\documentclass[aps,pra,twocolumn,superscriptaddress,amsmath,amssymb,showpacs,notitlepage]{revtex4-1}
\pdfoutput=1
\usepackage{graphicx}
\usepackage{dcolumn}
\usepackage{bm}
\usepackage{hyperref}
\usepackage{tikz}
\usepackage{mathtools}
\usepackage{physics}
\usetikzlibrary{matrix,calc}
\usepackage{karnaugh-map}
\usepackage{kvmap}
\usepackage{subfigure}
\usepackage{graphicx}
\usepackage{braket}
\usepackage{hyperref}
\usepackage{multirow}
\usepackage{stackengine}
\usepackage{array}
\usepackage{xcolor,color}
\usepackage{lipsum}
\usepackage{multirow}
\usepackage{stackengine}
\usepackage{enumitem}
\usepackage{algorithm}
\usepackage{algpseudocode}
\usepackage{algorithmicx}

\usetikzlibrary{quantikz}
\newcolumntype{P}[1]{>{\centering\arraybackslash}p{#1}}

\usepackage{array}
\newsavebox{\boxA}
\newsavebox{\boxB}
\newsavebox{\boxC}
\newsavebox{\boxF}
\newsavebox{\boxG}
\newsavebox{\boxH}
\newsavebox{\boxD}
\newsavebox{\boxE}
\newsavebox{\boxI}
\newsavebox{\boxJ}
\newsavebox{\boxK}
\newsavebox{\boxL}
\newsavebox{\boxM}
\newcolumntype{?}{!{\vrule width 1pt}}
\newcommand\xrowht[2][0]{\addstackgap[.5\dimexpr#2\relax]{\vphantom{#1}}}

\begin{document}


\title{Resource-efficient loss-aware photonic graph state preparation using atomic emitters}
\author{Eneet Kaur}
\thanks{These authors contributed equally.}
\affiliation{Wyant College of Optical Sciences, University of Arizona, \\1630 E University Blvd, Tucson, AZ, 85719, USA}
\affiliation{Cisco Quantum Lab, Los Angeles, USA}

\author{Ashlesha Patil}
\thanks{These authors contributed equally.}
\affiliation{Wyant College of Optical Sciences, University of Arizona, \\1630 E University Blvd, Tucson, AZ, 85719, USA}

\author{Saikat Guha}
\affiliation{Wyant College of Optical Sciences, University of Arizona, \\1630 E University Blvd, Tucson, AZ, 85719, USA}
\affiliation{Department of Electrical and Computer Engineering, University of Maryland, \\ 8228 Paint Branch Drive, College Park, MD, 20742, USA}

\begin{abstract}
Multi-qubit entangled photonic graph states are an important ingredient for all-photonic quantum computing, repeaters and networking. Preparing them using probabilistic stitching of single photons using linear optics presents a formidable resource challenge due to multiplexing needs. Quantum emitters provide a viable solution to prepare photonic graph states as they enable deterministic production of photons entangled with emitter qubits, and deterministic two-qubit interactions among emitters. A handful of emitters often suffice to generate useful-size graph states that would otherwise require millions of emitters used as single photon sources, using the linear-optics method. Photon loss however impedes the emitter method due to a large circuit depth, and hence loss accrual on the photons of the graph state produced, given the typically large number of slow two-qubit CNOT gates between emitters. We propose an algorithm that can trade the number of emitters with the graph-state depth, while minimizing the number of emitter CNOTs. We apply our algorithm to generate a repeater graph state (RGS) for a new all-photonic repeater protocol, which achieves a far superior rate-distance tradeoff compared to using the least number of emitters needed to generate the RGS. Yet, it needs five orders of magnitude fewer emitters than the multiplexed linear-optics method---with each emitter used as a photon source---to achieve a desired rate-distance performance.
\end{abstract}

\maketitle


\section{Introduction}
\label{sec:intro}

Graph states are a class of highly entangled multi-qubit states that are a key resource for quantum information applications spanning cryptography, computation~ \cite{bartolucci2021fusionbased,Kitaev2003,Briegel}, communications \cite{RevModPhys.83.33,Azuma2015} and sensing \cite{Damian2020}. A graph state $|G\rangle$ of $n$ qubits can be arrived at by laying one qubit each, prepared in the $|+\rangle = (\ket{0}+\ket{1})/\sqrt{2}$ state, on each of the $n$ vertices of a graph $G$, and applying two-qubit controlled-phase (CZ) gates on qubit pairs lying on each edge of $G$. The measurement-based model of quantum computing (MQBC)---which relies on a continual preparation of a long-range-connected graph state on a 3D regular-lattice-shaped graph~\cite{Raussendorf2007} (known as a {\em cluster state})---is particularly suited to photonic qubits~\cite{Kieling2007,GimenoSegovia2015,Pant2019}. Graph states of photonic qubits, therefore, are of particular importance and the subject of interest in this paper. Furthermore, we focus on the dual-rail photonic qubit, which encodes the two computational states of a qubit by exciting one of two orthogonal optical (spatial, temporal or polarization) modes in a single-photon Fock state with the other mode in vacuum. The dual-rail photonic qubit is easy to prepare using single-photon sources: either probabilistic ones that use nonlinear optical processes such as spontaneous parametric downconversion (SPDC) or spontaneous four-wave mixing (SFWM) integrated with photonic waveguides~\cite{silverstone2016silicon,meyer2020single,mosley2008heralded,alexander2024manufacturable}, or deterministic sources of single photons using quantum dots, color centers and other artificial atom quantum emitters~\cite{senellart2017high,aharonovich2016solid}. Dual-rail qubits admit arbitrary (deterministic) single-qubit gates using a two-input two-output beamsplitter, and arbitrary two-qubit gates also using linear optical (LO) circuit elements, i.e., beamsplitters and phase shifters, but in a probabilistic (heralded) fashion~\cite{Knill2001}. In other words, the realization of a two-qubit LO gate succeeds only with a sub-unity probability, but one knows, e.g., with the occurrence of a particular detector-click pattern, that the gate did succeed. Another well-studied application of photonic graph states is in quantum repeaters for long-distance high-rate quantum communications, but without the need of matter-based quantum memories. Here, one mimics the action of a quantum memory and processor at quantum repeater nodes by a special class of photonic graph states known as the repeater graph state (RGS). The multi-photon RGS helps store faithfully a logical qubit at a repeater---by acting as an error-correction code for photon loss---for however long it takes to generate heralded entanglement across a link between that repeater and a neighboring repeater across a link. The RGS also serves as a substrate for measurement-based realization of multiplexed Bell-state measurements (BSMs) to connect successful link-entanglement attempts across the repeater's neighboring nodes~\cite{Azuma2015, Mihir2017}. There have also been studies exploring the use of photonic graph states for quantum-enhanced sensing~\cite{Wang2020}. Unheralded photon loss after state preparation is the primary source of noise that corrupts a photonic graph state. Error correction for loss adds an overhead of additional physical photonic qubits, but this also results in an increased {\em circuit depth}, i.e., an increased time lag between the oldest and the youngest photonic qubit in the graph state produced, which causes higher physical-qubit loss accrual in turn. This poses an important resource-efficiency problem that is pertinent to all the aforesaid applications of photonic graph states in computing, communications, and sensing. The challenge is to develop a resource-efficient method to prepare a photonic graph state, viz., one that requires the smallest number of sources and detectors to prepare a graph state of a desired number of photonic qubits of a target fidelity desired by the underlying application.

\begin{figure}
    \centering
    \includegraphics[scale = 0.7]{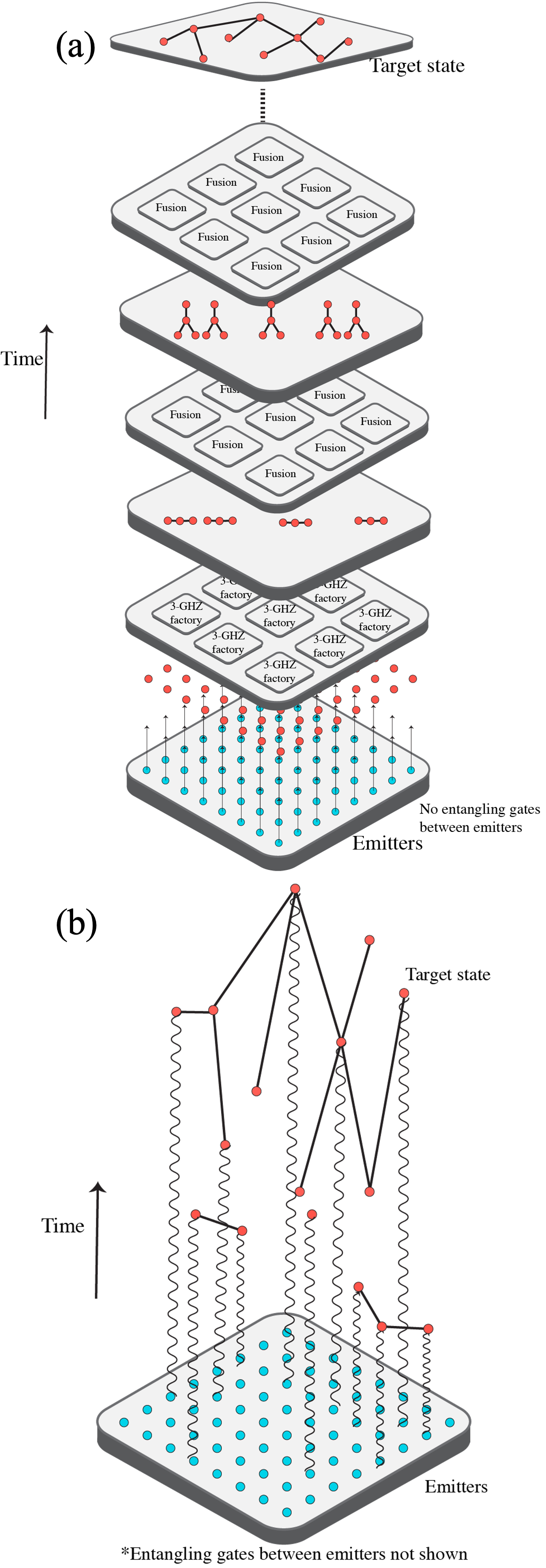}
    \caption{(a) The linear optical method simultaneously generates single photons (red circles) entangled with (squiggly lines) quantum emitters (cyan circles). Emitters are measured to unentangle the photons (not shown). The photons then go through probabilistic linear optical circuits to form the target graph state. (b) The quantum emitters emit entangled photons of the target graph state separated in time.}
    \label{fig:schematic}
\end{figure}

There are two genres of methods to prepare graph states of dual-rail photonic qubits, each of which could be considered one of two extremes, in a sense that will become clear in what follows. The first method uses an array of sources---either emitters or nonlinear-optics-based heralded sources---emitting a stream of single photons in every $\tau$-second time-step. This is followed by a linear-optical (LO) heralded assembly of the target graph state, which usually happens in a few steps (see  Fig.~\ref{fig:schematic}(a) for an illustration). The first step uses LO circuits to prepare small (e.g., $3$ to $6$ photonic-qubit) {\em resource states} such as GHZ states in a probabilistic (heralded) fashion. With a sufficient number of sources emitting photons in parallel, despite the probabilistic failures of the LO circuits preparing the resource states, one can ensure that sufficiently many resource states are prepared, near-deterministically, at every time-step. We call this a multiplexed resource-state {\em factory}. From hereon, parallel attempts of two-qubit (entangling, but destructive) BSMs, and their unitary-rotated variants known as {\em fusion} circuits, are used to progressively grow the graph-state size. LO fusion circuits are probabilistic and heralded, a feature they inherit from the underlying two-qubit LO entangling gates. Each step in above is multiplexed sufficiently to ensure that one produces, with probability close to $1$, at least one copy of the final target graph state, at each time step $\tau$. As the size of the target graph state grows, the probabilistic nature of the LO circuits for resource-state preparation and fusion causes a steep increase in the required number of photon sources, making this a very resource-intensive process~\cite{Varnava_2008,Ying2015,Mihir2017}. Assuming that the graph state fragments produced during all the failed fusion attempts in the above method are discarded~\footnote{There have been studies on recycling graph states generated from failed fusion attempts and optimizing the space-time multiplexing for fully-LO preparation of photonic graph states~\cite{Gimeno-Segovia2017}.}, the final target state is prepared all-at-once at time $K\tau$ and available for use, where $K$, the number of sequential steps in the above method grows logarithmically with the size of the target state, and is hence typically small. Further, all the photons in the target graph state accrue the same amount of loss proportional to the loss-rate (dB per second of propagation in the waveguide) times $\tau$ times $K$, which is also small if the repetition period $\tau$ can be engineered to be small. 

The second well-known method for graph state preparation~\cite{lindner2009proposal,Li2022} uses the least number $n_{\rm min}$ of emitters possible to prepare a photonic graph state $\ket{G}$, where $n_{\rm min}$ is the rank of the matrix product representation of $\ket{G}$~\cite{Schon2005,Schon2007}. Ref.~\cite{Li2022} characterized $n_{\rm min}$ as the {\em height function}, $h(G)$. This method uses the quantum emitters not just as emitters of single photons but also as atomic {\em qubits}. The optically-active emitter qubit is excited (deterministically) to generate a dual-rail photonic qubit entangled with the emitter qubit in a Bell state or a maximally-entangled two-qubit graph state. In addition, one can perform two-qubit entangling, e.g., CZ or CNOT gates deterministically among two distinct emitter qubits~\cite{Stas2022}. This is done by leveraging a shelving qubit (such as a host-lattice atomic nuclear spin in a color center) into which the optically-active emitter qubit (e.g., electronic-spin of a color center) is deterministically transferred to, followed by repeat-until-success generation of photonic-BSM-mediated entanglement among the freed-up optically-active emitter qubits~\cite{Dhara2023}, and another round of entangling gates and single-qubit measurements among the respective optically-active and shelving qubits at each emitter-qubit site. Appendix~\ref{sec:apx} describes the above process in detail (see Fig.~\ref{fig:enSWAP} for a circuit schematic of performing a single emitter-emitter CNOT using the above method). Taking into account all the relevant timescales---viz., qubit initialization, photonic-qubit  emission, single- and two-qubit gates, and single-qubit readout---the time $t_{{\rm CNOT}_{\rm{e,e}}}$ it takes to perform this CNOT is typically much larger than the repetition period $\tau$ of single-photon emissions possible with the same emitters if they were just used as photon emitters (rather than qubits)~\cite{Stas2022,inlek2017multispecies,tan2015multi}. Therefore, there is a risk that despite the deterministic entangling gates among emitter qubits, the feature that makes this method of photonic graph-state preparation attractive and enables minimizing the number of emitters used, could also potentially result in a large CNOT-depth, i.e., the number of time-consuming sequential steps, leading to long time delays between the constituent photonic qubits of the graph state prepared in subsequent time steps, unlike the LO method (see  FIG.~\ref{fig:schematic}(b)). This in turn may render the photonic graph state useless for the application in question. In recent work, Ref.~\cite{Ghanbari2024}---while continuing to use the minimum required $h(G)$ number of emitters---reduced the CNOT depth by about a half, by allowing one to produce a graph state of smaller depth that is local-Clifford equivalent to the target graph state. 

In this paper, we develop a general algorithm for emitter-based preparation of photonic graph states, where we can reduce the CNOT depth further at the expense of an increase in the number of emitters, in a controlled manner (see Fig.~\ref{fig:emitter_vs_depth}). The underlying innovation is a time-reversed decomposition of a quantum circuit (see Table~\ref{tab:SFE_cases}) involving a user-provided number of available emitter qubits and the photonic qubits of the desired photonic graph state, in term of two-qubit gates between emitter and photonic qubits, single- and two-qubit Clifford gates on emitter qubits, and single-qubit computational-basis measurements on emitter qubits. We evaluate the performance of our algorithm in terms of the rate-vs.-distance achievable with a new-variant of all-photonic RGS protocol adapted to our algorithm, which measures off photons as they are created to mitigate errors due to loss accumulation. We show that our protocol works far better compared to both extremes: (1) using millions of emitters at each repeater, purely as photon emitters, followed by multiplexed-LO circuitry to generate an RGS $G$ every $\tau$ seconds~\cite{Mihir2017}, and (2) using $t_{{\rm CNOT}_{\rm{e,e}}}/\tau$ banks of $h(G)$ emitters each, used as emitter-qubits, emitting photons in a staggered manner, so as to again generate an RGS every $\tau$ seconds~\cite{Buterakos2017}. For a chosen set of parameter regimes of a generic emitter platform, we show that method (2) does not act as a viable repeater, i.e., it does not outperform the repeaterless rate-distance bound~\cite{Pirandola2017}. Our method however---programmed to operate at the right emitter-number vs. CNOT-depth tradeoff point---can achieve the same rate-distance performance as method (1) does, but with $5$ orders of magnitude fewer emitters per repeater node (see Fig.~\ref{fig:ratePlot}).

\section{Notation}
\label{sec:notation}
Consider a graph $G \equiv (V,E)$, where $V$ represents the set of vertices and $E$ represents the set of edges. To define a graph state, we associate each vertex of the graph with a qubit. The graph's edges are associated with the action of controlled-phase $(\operatorname{CZ})$ gates. Mathematically, for a given graph $G$, the graph state is defined as:

\begin{equation}
\ket{G} = \prod_{\substack{i,j \in E}} \operatorname{CZ}_{i,j} \ket{+}^{\otimes |V|},
\end{equation}

where $\operatorname{CZ}_{i,j}$ represents a $\operatorname{CZ}$ gate with vertex $i$ as the control qubit and vertex $j$ as the target qubit. Here, $\ket{+}^{\otimes |V|}$ denotes the initialization of $|V|$ qubits in the $\ket{+}$ state.

Alternatively, a graph state can be described using the stabilizer formalism. For each vertex $j$, we define an operator $S_j = X_j \prod_{\substack{k \in \textbf{N}(j)}} Z_k$, where $\textbf{N}(j)$ denotes the neighborhood of vertex $j$ in the graph $G$. Then, the graph state is defined as the simultaneous eigenstate with eigenvalue +1 of the operators $\left\{S_j\right\}_{j \in V}$.

We note that the graph state can accommodate photonic or matter qubits. We use dual-rail encoding for the photonic qubits. We name the qubits in the graph state as follows: the notation $k_p$ represents photonic qubits, where $k$ is a positive integer. We use the notation $j_e$ to represent emitter qubits, where $j$ is a positive integer. 

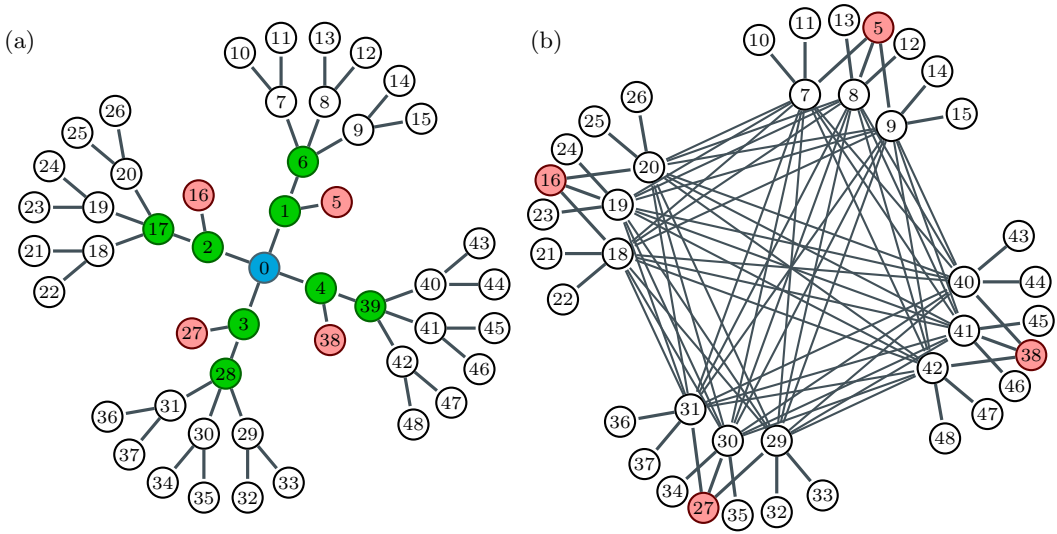
\begin{figure*}
    \centering
    \begin{tikzpicture}[scale=1,shorten >=1pt, auto, node distance=5cm,
    unode/.style = {
    circle, 
    draw = black, 
    thick,
    fill=white, 
    inner sep=0pt,
    minimum size=0.4cm,font=\scriptsize},
    unodec/.style = {
    circle, 
    draw = cyan!30!black, 
    thick,
    fill=cyan!90!black, 
    inner sep=0pt,
    minimum size=0.4cm,font=\scriptsize},
     unodeg/.style = {
    circle, 
    draw = green!40!black, 
    thick,
    fill=green!80!black, 
    inner sep=0pt,
    minimum size=0.4cm,font=\scriptsize},
     unodep/.style = {
    circle, 
    draw = red!40!black, 
    thick,
    fill=red!40!white, 
    inner sep=0pt,
    minimum size=0.4cm,font=\scriptsize},
    unode_i/.style = {
    circle, 
    draw = white, 
    thick,
    fill=white, 
    inner sep=0pt,
    minimum size=0.4cm,font=\scriptsize, text = white},
    unode2/.style = {
    rectangle, 
    draw=black, 
    thick,
    fill=white,
    inner sep=0pt,
    minimum size=0.4cm,font=\scriptsize },
    uedge/.style = {
    draw=cyan!20!black, 
    very thick},
    uedge2/.style = {
    draw=cyan!20!black, thick}]
  \node at(-3.25,3) {(a)};
  \node[unodec] (t) at (0,0) {$0$}; 
  \node[unodeg] (t1) at ([shift={(-20:0.8)}]t) {$4$};
  \node[unodeg] (t2) at ([shift={(-110:.8)}]t) {$3$};
  \node[unodeg] (t3) at ([shift={(70:0.8)}]t) {$1$};
  \node[unodeg] (t4) at ([shift={(160:0.8)}]t) {$2$};
  \path[uedge] (t)  edge (t1) edge (t2) edge (t3) edge (t4);

  \node[unodep] (t11) at ([shift={(-80:0.7)}]t1) {$38$};
  \path[uedge] (t1) edge (t11);
  \node[unodep] (t21) at ([shift={(-170:0.7)}]t2) {$27$};
  \path[uedge] (t2) edge (t21);
  \node[unodep] (t31) at ([shift={(10:0.7)}]t3) {$5$};
  \path[uedge] (t3) edge (t31);
  \node[unodep] (t41) at ([shift={(100:0.7)}]t4) {$16$};
  \path[uedge] (t4) edge (t41);

  \node[unodeg] (t111) at ([shift={(-20:.7)}]t1) {$39$};
   \path[uedge] (t111) edge (t1);
   \node[unode] (t1111) at ([shift={(10+10:0.85)}]t111) {$40$};
   \path[uedge] (t111) edge (t1111);
   \node[unode] (t1112) at ([shift={(-20:0.85)}]t111) {$41$};
   \path[uedge] (t111) edge (t1112);
   \node[unode] (t1113) at ([shift={(-50-10:0.85)}]t111) {$42$};
   \path[uedge] (t111) edge (t1113);

    \node[unode] (t11111) at ([shift={(40:0.85)}]t1111) {$43$};
    \path[uedge] (t11111) edge (t1111);
    \node[unode] (t11112) at ([shift={(0:0.85)}]t1111) {$44$};
    \path[uedge] (t11112) edge (t1111);

    \node[unode] (t11121) at ([shift={(-0:0.85)}]t1112) {$45$};
   \path[uedge] (t1112) edge (t11121);
    \node[unode] (t11121) at ([shift={(-40:0.85)}]t1112) {$46$};
   \path[uedge] (t1112) edge (t11121);
   
   \node[unode] (t11131) at ([shift={(-40:0.85)}]t1113) {$47$};
   \path[uedge] (t1113) edge (t11131);
   \node[unode] (t11132) at ([shift={(-80:0.85)}]t1113) {$48$};
   \path[uedge] (t1113) edge (t11132);

   \node[unodeg] (t222) at ([shift={(-110:.7)}]t2) {$28$};
   \path[uedge] (t222) edge (t2);
   \node[unode] (t2221) at ([shift={(10-90+10:0.85)}]t222) {$29$};
   \path[uedge] (t222) edge (t2221);
   \node[unode] (t2222) at ([shift={(-20-90:0.85)}]t222) {$30$};
   \path[uedge] (t222) edge (t2222);
   \node[unode] (t2223) at ([shift={(-50-90-10:0.85)}]t222) {$31$};
   \path[uedge] (t222) edge (t2223);
   
    \node[unode] (t22212) at ([shift={(10-90-10:0.85)}]t2221) {$32$};
   \path[uedge] (t2221) edge (t22212);
   \node[unode] (t22211) at ([shift={(10-90+30:0.85)}]t2221) {$33$};
   \path[uedge] (t2221) edge (t22211);
   \node[unode] (t22221) at ([shift={(-40-90:0.85)}]t2222) {$34$};
   \path[uedge] (t2222) edge (t22221);
   \node[unode] (t22222) at ([shift={(-90:0.85)}]t2222) {$35$};
   \path[uedge] (t2222) edge (t22222);
   \node[unode] (t22231) at ([shift={(-50-90-30:0.85)}]t2223) {$36$};
   \path[uedge] (t2223) edge (t22231);
   \node[unode] (t22232) at ([shift={(-50-90+10:0.85)}]t2223) {$37$};
   \path[uedge] (t2223) edge (t22232);

\node[unodeg] (t333) at ([shift={(70:.7)}]t3) {$6$};
   \path[uedge] (t333) edge (t3);
   \node[unode] (t3331) at ([shift={(10+90+10:0.85)}]t333) {$7$};
   \path[uedge] (t333) edge (t3331);
   \node[unode] (t3332) at ([shift={(-20+90:0.85)}]t333) {$8$};
   \path[uedge] (t333) edge (t3332);
   \node[unode] (t3333) at ([shift={(-50+90-10:0.85)}]t333) {$9$};
   \path[uedge] (t333) edge (t3333);
   
   \node[unode] (t33311) at ([shift={(10+90+30:0.85)}]t3331) {$10$};
   \path[uedge] (t3331) edge (t33311);
   \node[unode] (t33312) at ([shift={(10+90-10:0.85)}]t3331) {$11$};
   \path[uedge] (t3331) edge (t33312);
   \node[unode] (t33321) at ([shift={(-40+90:0.85)}]t3332) {$12$};
   \path[uedge] (t3332) edge (t33321);
    \node[unode] (t33322) at ([shift={(+90:0.85)}]t3332) {$13$};
   \path[uedge] (t3332) edge (t33322);
   \node[unode] (t33331) at ([shift={(-50+90+10:0.85)}]t3333) {$14$};
   \path[uedge] (t3333) edge (t33331);
    \node[unode] (t33332) at ([shift={(-50+90-30:0.85)}]t3333) {$15$};
   \path[uedge] (t3333) edge (t33332);

\node[unodeg] (t444) at ([shift={(160:.7)}]t4) {$17$};
   \path[uedge] (t444) edge (t4);

    \node[unode] (t4441) at ([shift={(10-90-90+10:0.85)}]t444) {$18$};
   \path[uedge] (t444) edge (t4441);
   \node[unode] (t4442) at ([shift={(-20-90-90:0.85)}]t444) {$19$};
   \path[uedge] (t444) edge (t4442);
   \node[unode] (t4443) at ([shift={(-50-90-90-10:0.85)}]t444) {$20$};
   \path[uedge] (t444) edge (t4443);

    \node[unode] (t44411) at ([shift={(-90-90:0.85)}]t4441) {$21$};
   \path[uedge] (t4441) edge (t44411);
   \node[unode] (t44412) at ([shift={(10-90-90+30:0.85)}]t4441) {$22$};
   \path[uedge] (t4441) edge (t44412);
   \node[unode] (t44421) at ([shift={(-90-90:0.85)}]t4442) {$23$};
   \path[uedge] (t4442) edge (t44421);
   \node[unode] (t44422) at ([shift={(-40-90-90:0.85)}]t4442) {$24$};
   \path[uedge] (t4442) edge (t44422);
   \node[unode] (t44431) at ([shift={(-50-90-90+10:0.85)}]t4443) {$25$};
   \path[uedge] (t4443) edge (t44431);
   \node[unode] (t44432) at ([shift={(-50-90-90-30:0.85)}]t4443) {$26$};
   \path[uedge] (t4443) edge (t44432);

     \node at(-3.25+7,3) {(b)};
  \node[unode_i] (t) at (0+7,0) {$0$}; 
  \node[unode_i] (t1) at ([shift={(-20:0.8)}]t) {$4$};
  \node[unode_i] (t2) at ([shift={(-110:.8)}]t) {$3$};
  \node[unode_i] (t3) at ([shift={(70:0.8)}]t) {$1$};
  \node[unode_i] (t4) at ([shift={(160:0.8)}]t) {$2$};

  \
  
  

  \node[unode_i] (t111) at ([shift={(-20:.7)}]t1) {$39$};
   \node[unode] (t1111) at ([shift={(10+10:0.95)}]t111) {$40$};
   \node[unode] (t1112) at ([shift={(-20:0.95)}]t111) {$41$};
   \node[unode] (t1113) at ([shift={(-50-10:0.95)}]t111) {$42$};

   \node[unodep] (t11) at ([shift={(-20:0.95)}]t1112) {$38$};
   \path[uedge] (t11)  edge (t1111) edge (t1112) edge (t1113);

    \node[unode] (t11111) at ([shift={(40:0.95)}]t1111) {$43$};
    \path[uedge] (t11111) edge (t1111);
    \node[unode] (t11112) at ([shift={(0:0.95)}]t1111) {$44$};
    \path[uedge] (t11112) edge (t1111);

    \node[unode] (t11121) at ([shift={(7.5:1)}]t1112) {$45$};
   \path[uedge] (t1112) edge (t11121);
    \node[unode] (t11121) at ([shift={(-47.5:1)}]t1112) {$46$};
   \path[uedge] (t1112) edge (t11121);
   
   \node[unode] (t11131) at ([shift={(-40:0.95)}]t1113) {$47$};
   \path[uedge] (t1113) edge (t11131);
   \node[unode] (t11132) at ([shift={(-80:0.95)}]t1113) {$48$};
   \path[uedge] (t1113) edge (t11132);

   \node[unode_i] (t222) at ([shift={(-110:.7)}]t2) {$28$};
   \node[unode] (t2221) at ([shift={(10-90+10:0.95)}]t222) {$29$};
   \node[unode] (t2222) at ([shift={(-20-90:0.95)}]t222) {$30$};
   \node[unode] (t2223) at ([shift={(-50-90-10:0.95)}]t222) {$31$};
   
    \node[unode] (t22212) at ([shift={(10-90-10:0.95)}]t2221) {$32$};
   \path[uedge] (t2221) edge (t22212);
   \node[unode] (t22211) at ([shift={(10-90+30:0.95)}]t2221) {$33$};
   \path[uedge] (t2221) edge (t22211);
   \node[unode] (t22221) at ([shift={(-40-97.5:1)}]t2222) {$34$};
   \path[uedge] (t2222) edge (t22221);
   \node[unode] (t22222) at ([shift={(-90+7.5:1)}]t2222) {$35$};
   \path[uedge] (t2222) edge (t22222);
   \node[unode] (t22231) at ([shift={(-50-90-30:0.95)}]t2223) {$36$};
   \path[uedge] (t2223) edge (t22231);
   \node[unode] (t22232) at ([shift={(-50-90+10:0.95)}]t2223) {$37$};
   \path[uedge] (t2223) edge (t22232);

   \node[unodep] (t21) at ([shift={(-110:0.95)}]t2222) {$27$};
   \path[uedge] (t21) edge (t2222) edge (t2221) edge (t2223);

\node[unode_i] (t333) at ([shift={(70:.7)}]t3) {$6$};
   \node[unode] (t3331) at ([shift={(10+90+10:0.95)}]t333) {$7$};
   \node[unode] (t3332) at ([shift={(-20+90:0.95)}]t333) {$8$};
   \node[unode] (t3333) at ([shift={(-50+90-10:0.95)}]t333) {$9$};

   \node[unodep] (t31) at ([shift={(70:0.95)}]t3332) {$5$};
    \path[uedge] (t31)  edge (t3333) edge (t3331) edge (t3332);
   
   \node[unode] (e) at ([shift={(10+90+30:0.95)}]t3331) {$10$};
   \path[uedge] (t3331) edge (e);
   \node[unode] (t33312) at ([shift={(10+90-10:0.95)}]t3331) {$11$};
   \path[uedge] (t3331) edge (t33312);
   \node[unode] (t33321) at ([shift={(-47.5+90:1)}]t3332) {$12$};
   \path[uedge] (t3332) edge (t33321);
    \node[unode] (t33322) at ([shift={(+97.5:1)}]t3332) {$13$};
   \path[uedge] (t3332) edge (t33322);
   \node[unode] (t33331) at ([shift={(-50+90+10:0.95)}]t3333) {$14$};
   \path[uedge] (t3333) edge (t33331);
    \node[unode] (t33332) at ([shift={(-50+90-30:0.95)}]t3333) {$15$};
   \path[uedge] (t3333) edge (t33332);

\node[unode_i] (t444) at ([shift={(160:.7)}]t4) {$17$};

    \node[unode] (t4441) at ([shift={(10-90-90+10:0.95)}]t444) {$18$};
   \node[unode] (t4442) at ([shift={(-20-90-90:0.95)}]t444) {$19$};
   \node[unode] (t4443) at ([shift={(-50-90-90-10:0.95)}]t444) {$20$};

    \node[unode] (t44411) at ([shift={(-90-90:0.95)}]t4441) {$21$};
   \path[uedge] (t4441) edge (t44411);
   \node[unode] (t44412) at ([shift={(10-90-90+30:0.95)}]t4441) {$22$};
   \path[uedge] (t4441) edge (t44412);
   \node[unode] (t44421) at ([shift={(-90-90+7.5:1)}]t4442) {$23$};
   \path[uedge] (t4442) edge (t44421);
   \node[unode] (t44422) at ([shift={(-40-90-97.5:1)}]t4442) {$24$};
   \path[uedge] (t4442) edge (t44422);
   \node[unode] (t44431) at ([shift={(-50-90-90+10:0.95)}]t4443) {$25$};
   \path[uedge] (t4443) edge (t44431);
   \node[unode] (t44432) at ([shift={(-50-90-90-30:0.95)}]t4443) {$26$};
   \path[uedge] (t4443) edge (t44432);
   \node[unodep] (t41) at ([shift={(160:0.95)}]t4442) {$16$};
   \path[uedge] (t41) edge (t4443) edge (t4442) edge (t4441);

   \path[uedge2] (t3331) edge (t2221) edge (t1111) edge (t4441);
   \path[uedge2] (t3332) edge (t2221) edge (t1111) edge (t4441);
   \path[uedge2] (t3333) edge (t2221) edge (t1111) edge (t4441);
   \path[uedge2] (t3331) edge (t2222) edge (t1112) edge (t4442);
   \path[uedge2] (t3332) edge (t2222) edge (t1112) edge (t4442);
   \path[uedge2] (t3333) edge (t2222) edge (t1112) edge (t4442);
   \path[uedge2] (t3331) edge (t2223) edge (t1113) edge (t4443);
   \path[uedge2] (t3332) edge (t2223) edge (t1113) edge (t4443);
   \path[uedge2] (t3333) edge (t2223) edge (t1113) edge (t4443);

   \path[uedge2] (t2221) edge (t1111) edge (t4441);
   \path[uedge2] (t2221) edge (t1112) edge (t4442);
   \path[uedge2] (t2221) edge (t1113) edge (t4443);
   \path[uedge2] (t2222) edge (t1111) edge (t4441);
   \path[uedge2] (t2222) edge (t1112) edge (t4442);
   \path[uedge2] (t2222) edge (t1113) edge (t4443);
   \path[uedge2] (t2223) edge (t1111) edge (t4441);
   \path[uedge2] (t2223) edge (t1112) edge (t4442);
   \path[uedge2] (t2223) edge (t1113) edge (t4443);

    \path[uedge2] (t1111) edge (t4441);
   \path[uedge2] (t1111) edge (t4442);
   \path[uedge2] (t1111) edge (t4443);
   \path[uedge2] (t1112) edge (t4441);
   \path[uedge2] (t1112) edge (t4442);
   \path[uedge2] (t1112) edge (t4443);
   \path[uedge2] (t1113) edge (t4441);
   \path[uedge2] (t1113) edge (t4442);
   \path[uedge2](t1113) edge (t4443);

  \end{tikzpicture}
  \caption{This figure shows construction of a RGS with $m=2$, and $b = [3,2]$. (a) Start with a star graph state with $2m+1$ qubits. Attach a qubit (pink qubits) and a tree with branching vector $b$ to every qubit of the clique graph state. (b) Performing $X$ measurements on the green nodes and $Y$ measurement the central blue node on the graph state in (a) gives the RGS. The pink qubits depict the flying qubits, which are sent to the two adjacent nodes, and the white nodes remain at the repeaters and mimic quantum memories.}
\label{fig:RSG_Const}
  \end{figure*}

We utilize the algorithm developed in this work to analyze the properties of repeater graph states \cite{Azuma2015,Varnava2007}. The repeater graph states (RGS) were primarily introduced to give way to an all-photonic quantum repeater architecture. To achieve this Ref.~\cite{Azuma2015} replaced the matter-based quantum memories in~\cite{Sinclair2014}, by the optical graph states along the lines of~\cite{Varnava2007}. These optical graph states, can be thought of as photonic quantum memories, paving the way for an all photonic repeater architecture. The RGS are then characterized by two parameters $m$ and $b$. Here, $m$ is connected to the multiplexing in the architecture, i.e. $m$ physical qubits are sent to the two nodes on either side of the repeaters. The physical qubit is loss-error protected by a regular tree described by the branching vector $b \equiv [b_0,b_1, \cdots,b_d]$, which signifies that the root of the tree has $b_0$ children nodes, and each of those nodes
have $b_1$ children nodes, till we reach the $(d+1)^{\textrm{th}}$ level. An illustrative example is depicted in FIG.~\ref{fig:RSG_Const}. 

\section{Overview of the algorithm}
\label{sec:emitters}
The target photonic graph state $\ket{G}$, represented by graph $G$, has $m$ qubits. We are given $n$ quantum emitter qubits, each of which can be initialized in any state we wish. There are no photonic qubits to start with. When a quantum emitter qubit (henceforth referred to simply as {\em emitter} for brevity), is made to {\em emit} a dual-rail photonic qubit (henceforth referred to simply as {\em photon} for brevity) in an entangled emitter-photon Bell state, we model this as a CNOT gate ($\operatorname{CNOT}_{e,p}$) acting between the emitter (control qubit) and a {\em fresh} photonic (target) qubit prepared in the $\ket{0}$ state. This is the {\em only} allowed two-qubit interaction between an emitter and a photon. For example, a CNOT between an emitter (as control) and an already-emitted photonic qubit (as target) is not allowed. A CNOT with any photon (as control) and an emitter (as target) is also not allowed. Also, a CNOT between two photons is not allowed. We do however have access to the following deterministically-realizable operations: an arbitrary single-qubit (including Pauli, Hadamard and non-Clifford) gate on either an emitter or a photon, CNOT gates between two emitters ($\operatorname{CNOT}_{e,e}$) (deterministic, yet time-consuming: see Appendix~\ref{sec:apx}), and arbitrary single-qubit measurements of any (emitter or photon) qubit. A single qubit measurement on an emitters unentangles it from the emitted photon(s) and stops further emission till the emitter is initialized again. With these set of constraints, we pose the following question: 
\begin{quote}
  Given $n$ emitters, a desired photonic graph state $\ket{G}$ of $m$ photons, a preferred order of emission of photons in $\ket{G}$, what is the {\em optimal} sequence of above-mentioned allowed (deterministic) operations to generate $\ket{G}$ using the emitters?
\end{quote}

We refer to the sequence in which the allowed operations are performed to generate $\ket{G}$ an \textit{algorithm} and the preferred order of emission of photons in $\ket{G}$ as an \textit{initial condition}. For a given initial condition, the optimality of the algorithm depends on the objective function. One could for example seek to minimize the circuit depth of (parallel) emitter-emitter $\operatorname{CNOT}_{e,e}$ gates in the circuit (relevant for all-photonic repeater as it minimizes unheralded-loss accrual), minimize the total number of $\operatorname{CNOT}$ gates, or minimize the total number of emitters employed (as done in \cite{Li2022}). One could also optimize another application-driven objective function driven, for instance, by the loss-error threshold for preparing a fault-tolerant photonic cluster state for MBQC, or creating a loss-protected resource for entanglement-assisted sensor probe.

We introduce three primitives and then combine the three primitives to yield the algorithm. The algorithm might use these primitives multiple times. Note that, as the algorithm takes the target graph state to single photons, it is considered ``time-reversed", i.e., we reverse both the order of primitives in the algorithm and the order of quantum gates in the primitives themselves to generate the target photonic graph state from emitters.

First, we observe that with the set of available operations listed above, all entanglement among the photons in the final photonic graph state $\ket{G}$, i.e., all the edges of $\ket{G}$, must come from the emitter-emitter CNOTs or by the allowed two-qubit interaction between photon and the emitter. So, it is not surprising that the CNOT depth is the longest when one uses the fewest number (i.e., $n = h(G)$) of emitters possible, since it is hard to parallelize the CNOTs. Our algorithm enables for such CNOT parallelization by allowing for $n > h(G)$ emitters, in a structured way. Rather than minimizing a chosen objective function, we provide an intuitive algorithm that is shown to be able to trade the CNOT depth with $n$ well, and yield a dramatic improvement in the rate-vs.-distance performance of all-photonic repeaters compared to all known RGS-preparation methods.

\section{Building blocks of the algorithm}\label{sec:algo_primitives}
The goal of the algorithm is to convert the graph state $\ket{G}$ with $m$ photon qubits and $n$ emitters prepared in the state $\ket{0}^{\otimes n}$ to $\ket{0}^{\otimes n+m}$. The algorithm's output is the sequence of operations that achieve the goal. We can then reverse the output to obtain the sequence of operations which starting from $\ket{0}^{\otimes n+m}$ gives a state $\ket{G} \otimes \ket{0}^{\otimes m}$. We divide the sequence of allowed operations into three primitives:
\begin{itemize}
    \item \textbf{Swapping with a free emitter (SFE)} - Absorb a photon using an emitter in the 
 (unentangled) state $\ket{0}$. This primitive replaces the vertex of the absorbed photon in the graph $G$ with the emitter. The algorithm starts with this process. The hardware implementation of this process translates to the emission of a photon, followed by computational basis measurement on the emitter. The emitter needs to be reinitialized to the $\ket{0}$ state after measurement to emit more photons.
 \item \textbf{Absorption by an entangled emitter (AEE)} - Absorb a photon in the graph state with an emitter that is entangled with photons and emitters in $\ket{G}$ under certain conditions explained in Section~\ref{sec:photon_absorption}. The hardware implementation of this process translates to the emission of a photon. Here, the emitter does not need re-initialization and keeps emitting more photons.
 \item \textbf{Unentangle emitters} - The hardware implementation of this process entangles emitters using CNOT gates. This is typically the most time-consuming step.
\end{itemize}
 In the following sections, we discuss these processes along with their quantum circuits for these processes. These circuits are time-reversed. We also give the ``time-forward" quantum circuits or simply, quantum circuits for these processes, which are the hardware implementation of these processes to generate photons. 

\subsection{Absorption by an entangled emitter (AEE)} \label{sec:photon_absorption}

As described earlier, absorbing a photon with an emitter is a fundamental process in our algorithm. This process converts an entangled photon in the graph state to a single photon in $\ket{0}$ state. The emitter used for absorption then inherits neighbors of the photon in the graph state. In this section, we describe one of the special cases of photon absorption, where the emitter to be swapped is entangled with other emitters and photons.

\noindent\textbf{Inputs:}  (1) an $m'$ qubit graph state $\ket{G'}$ between $m'_{e}\geq 1$  emitters and $m'_{p}\geq 1$ photons s.t. $m'_{e}+m'_{p}=m'$ and $m'\leq (m+n)$, and (2) an emitter $e$ in graph state $\ket{G'}$ to be used for photon absorption.

Only a subset of photons in $\ket{G'}$ can be absorbed with $e$. We now describe three conditions to identify the photons that can be absorbed in Table~\ref{tab:AEE_cases} along with the corresponding quantum circuits and the graph-theoretic rules to describe the actions of these three cases on $\ket{G'}$. We pictorially depict AEE in Figure~\ref{fig:AEE_12} and Figure~\ref{fig:AEE_3}. We justify the graph-theoretic rules and the choice of corresponding quantum circuits in Appendix~\ref{app:stabilizer_form}.

\savebox{\boxA}{\begin{quantikz}
    \lstick{$p$}&\qw & \targ{}&\qw\rstick{$\ket{0}$}\\
    \lstick{$e$}&\gate{H}& \ctrl{-1}&\qw\\
\end{quantikz}}%
\savebox{\boxF}{\begin{quantikz}
    \lstick{$p$}& \targ{}&\qw &\qw\\
    \lstick{$e$}& \ctrl{-1}&\gate{H}&\qw\\
\end{quantikz}}%
\savebox{\boxB}{\begin{quantikz}
    \lstick{$p$}&\gate{H} & \targ{}&\qw\rstick{$\ket{0}$}\\
    \lstick{$e$}& \qw & \ctrl{-1}&\qw\\
\end{quantikz}}%
\savebox{\boxG}{\begin{quantikz}
    \lstick{$p$} & \targ{}&\gate{H}&\qw\\
    \lstick{$e$} & \ctrl{-1}&\qw &\qw\\
\end{quantikz}}%
\savebox{\boxC}{\begin{quantikz}
     \lstick{$p$}&\gate{H} & \targ{}&\qw &\qw\rstick{$\ket{0}$}\\
    \lstick{$e$}& \gate{H} & \ctrl{-1}&\gate{H} &\qw\\
\end{quantikz}}
\savebox{\boxH}{\begin{quantikz}
     \lstick{$p$}&\qw & \targ{}&\gate{H}&\qw \\
    \lstick{$e$}& \gate{H} & \ctrl{-1}&\gate{H} &\qw\\
\end{quantikz}}

\begin{table*}[htb]
    \centering
  \resizebox{\textwidth}{!}{\begin{tabular}{?P{1cm}|P{4cm}|P{4cm}|P{5cm}?}
      \hline\xrowht{10pt}
     &\textbf{Condition} & \textbf{Output} & \textbf{Circuit} \\
    \noalign{\hrule height 1pt}\xrowht{20pt}
    Case 1 & A photon $p$ can be absorbed by emitter $e$ if $\textbf{N}_{G'}(e) = \{p\}$, i.e., $p$ is the only neighbor of $e$ in $G'$. &$\ket{G'}\rightarrow\ket{G'_1}\otimes \ket{0}_p$.
 To obtain $G_1'$ from $G'$ \begin{itemize}
     \item Add edges from $\mathbf{N}_{G'}(p)$ to $e$ in $G'$.
     \item $G_1' = G'\setminus\{p\}$
 \end{itemize} & Time-reversed:  \usebox\boxA
 \newline Time-forward: \usebox\boxF
\\
   \hline\xrowht{20pt}
   Case 2 & A photon $p$ can be absorbed by emitter $e$ if $\textbf{N}_{G'}(p) = \{e\}$, i.e., $e$ is the only neighbor of $p$ in $G'$.  & $\ket{G'}\rightarrow\ket{G'_1}\otimes \ket{0}_p$.
$G_1'=G'\setminus\{p\}$. & Time-reversed: \usebox\boxB \newline Time-forward: \usebox\boxG\\
   \hline\xrowht{20pt}
   Case 3 & A photon $p$ can be absorbed by emitter $e$ if $\textbf{N}_{G'}(p) = \textbf{N}_{G'}(e)$, i.e., $p$ and $e$ have the same neighborhood in $G'$. 
 &$\ket{G'}\rightarrow\ket{G'_1}\otimes \ket{0}_p$.
$G_1'=G'\setminus\{p\}$. &Time-reversed: \usebox\boxC
\newline Time-forward: \usebox\boxH\\
    \noalign{\hrule height 1pt}
  \end{tabular}}
\caption{This table summarizes the three cases to identify a photon $p$ in graph state $\ket{G'}$ that can be absorbed by an emitter $e$ in $\ket{G'}$. The graph state after photon absorption is $\ket{G_1'}$. $\mathbf{N}_{G'}(i)$ is the set of neighbors of vertex $i$ in $G'$.} 
    \label{tab:AEE_cases}
\end{table*}

\begin{figure*}[htb]
\centering
\begin{tikzpicture}[shorten >=1pt, auto, node distance=5cm,
    unode/.style = {
    circle, 
    draw = black, 
    thick,
    fill=white, 
    inner sep=0pt,
    minimum size=0.45cm,font=\scriptsize},
    unode2/.style = {
    rectangle, 
    draw=black, 
    thick,
    fill=gray!40!white,,
    inner sep=0pt,
    minimum size=0.45cm,font=\scriptsize },
    uedge/.style = {
    draw=cyan!20!black, 
    very thick}]
 \node at(-4.2,3) {(a)};
  \node[unode] (t) at (-2,0) {$0$}; 
  \node[unode] (t1) at ([shift={(-45:0.8)}]t) {$2$};
  \node[unode] (t4) at ([shift={(135:0.8)}]t) {$1$};
  \path[uedge] (t)  edge (t1) edge (t4);

   \node[unode] (t1111) at ([shift={(0:0.85)}]t1) {$5$};
   \path[uedge] (t1) edge (t1111);
   \node[unode] (t1113) at ([shift={(-50-70:0.85)}]t1) {$6$};
   \path[uedge] (t1) edge (t1113);

    \node[unode] (t11111) at ([shift={(45:0.85)}]t1111) {$11$};
    \path[uedge] (t11111) edge (t1111);
    \node[unode] (t11112) at ([shift={(-45:0.85)}]t1111) {$12$};
    \path[uedge] (t11112) edge (t1111);

   \node[unode] (t11131) at ([shift={(-45:0.85)}]t1113) {$13$};
   \path[uedge] (t1113) edge (t11131);
   \node[unode] (t11132) at ([shift={(-135:0.85)}]t1113) {$14$};
   \path[uedge] (t1113) edge (t11132);

    \node[unode] (t4441) at ([shift={(180:0.85)}]t4) {$4$};
   \path[uedge] (t4) edge (t4441);
  
   \node[unode] (t4443) at ([shift={(60:0.85)}]t4) {$3$};
   \path[uedge] (t4) edge (t4443);

    \node[unode2] (t44411) at ([shift={(135:0.85)}]t4441) {$1$};
   \path[uedge] (t4441) edge (t44411);
   \node[unode] (t44412) at ([shift={(225:0.85)}]t4441) {$10$};
   \path[uedge] (t4441) edge (t44412);
   
   \node[unode] (t44431) at ([shift={(105:0.85)}]t4443) {$8$};
   \path[uedge] (t4443) edge (t44431);
   \node[unode] (t44432) at ([shift={(15:0.85)}]t4443) {$7$};
   \path[uedge] (t4443) edge (t44432);

 \node at(-4.2+5.5,3) {(b)};
  \node[unode] (e1) at (-1+5.5,0.5) {$4$};
  \node[unode] (t) at (-2+5.5,0) {$0$}; 
  \node[unode] (t1) at ([shift={(-45:0.8)}]t) {$2$};
  \node[unode] (t4) at ([shift={(135:0.8)}]t) {$1$};
  \path[uedge] (t)  edge (t1) edge (t4);

   \node[unode] (t1111) at ([shift={(0:0.85)}]t1) {$5$};
   \path[uedge] (t1) edge (t1111);
   \node[unode] (t1113) at ([shift={(-50-70:0.85)}]t1) {$6$};
   \path[uedge] (t1) edge (t1113);

    \node[unode] (t11111) at ([shift={(45:0.85)}]t1111) {$11$};
    \path[uedge] (t11111) edge (t1111);
    \node[unode] (t11112) at ([shift={(-45:0.85)}]t1111) {$12$};
    \path[uedge] (t11112) edge (t1111);

   \node[unode] (t11131) at ([shift={(-45:0.85)}]t1113) {$13$};
   \path[uedge] (t1113) edge (t11131);
   \node[unode] (t11132) at ([shift={(-135:0.85)}]t1113) {$14$};
   \path[uedge] (t1113) edge (t11132);

    \node[unode2] (t4441) at ([shift={(180:0.85)}]t4) {$1$};
   \path[uedge] (t4) edge (t4441);
  
   \node[unode] (t4443) at ([shift={(60:0.85)}]t4) {$3$};
   \path[uedge] (t4) edge (t4443);

   \node[unode] (t44412) at ([shift={(225:0.85)}]t4441) {$10$};
   \path[uedge] (t4441) edge (t44412);
   
   \node[unode] (t44431) at ([shift={(105:0.85)}]t4443) {$8$};
   \path[uedge] (t4443) edge (t44431);
   \node[unode] (t44432) at ([shift={(15:0.85)}]t4443) {$7$};
   \path[uedge] (t4443) edge (t44432);
   
   \node at(-4.2+11,3) {(c)};
  \node[unode] (e1) at (-1+11,0.5) {$4$};
  \node[unode] (e2) at (-1+8.5,-0.25) {$10$};
  \node[unode] (t) at (-2+11,0) {$0$}; 
  \node[unode] (t1) at ([shift={(-45:0.8)}]t) {$2$};
  \node[unode] (t4) at ([shift={(135:0.8)}]t) {$1$};
  \path[uedge] (t)  edge (t1) edge (t4);

   \node[unode] (t1111) at ([shift={(0:0.85)}]t1) {$5$};
   \path[uedge] (t1) edge (t1111);
   \node[unode] (t1113) at ([shift={(-50-70:0.85)}]t1) {$6$};
   \path[uedge] (t1) edge (t1113);

    \node[unode] (t11111) at ([shift={(45:0.85)}]t1111) {$11$};
    \path[uedge] (t11111) edge (t1111);
    \node[unode] (t11112) at ([shift={(-45:0.85)}]t1111) {$12$};
    \path[uedge] (t11112) edge (t1111);

   \node[unode] (t11131) at ([shift={(-45:0.85)}]t1113) {$13$};
   \path[uedge] (t1113) edge (t11131);
   \node[unode] (t11132) at ([shift={(-135:0.85)}]t1113) {$14$};
   \path[uedge] (t1113) edge (t11132);

    \node[unode2] (t4441) at ([shift={(180:0.85)}]t4) {$1$};
   \path[uedge] (t4) edge (t4441);
  
   \node[unode] (t4443) at ([shift={(60:0.85)}]t4) {$3$};
   \path[uedge] (t4) edge (t4443);
   
   \node[unode] (t44431) at ([shift={(105:0.85)}]t4443) {$8$};
   \path[uedge] (t4443) edge (t44431);
   \node[unode] (t44432) at ([shift={(15:0.85)}]t4443) {$7$};
   \path[uedge] (t4443) edge (t44432);
  \end{tikzpicture}
  
\caption{Example to demonstrate AEE. Circles and squares are photons and emitters, respectively. In this and all subsequent figures, we use shapes to differentiate emitters and photons instead of the subscripts $e$ and $p$ used in the main text. (a) Input state to Case 1. The emitter $1_e$ absorbed photon $4_p$ to get (b). (b) is the input state for Case 2.  $1_e$ absorbed $10_p$ to get (c). }
\label{fig:AEE_12}
\end{figure*}
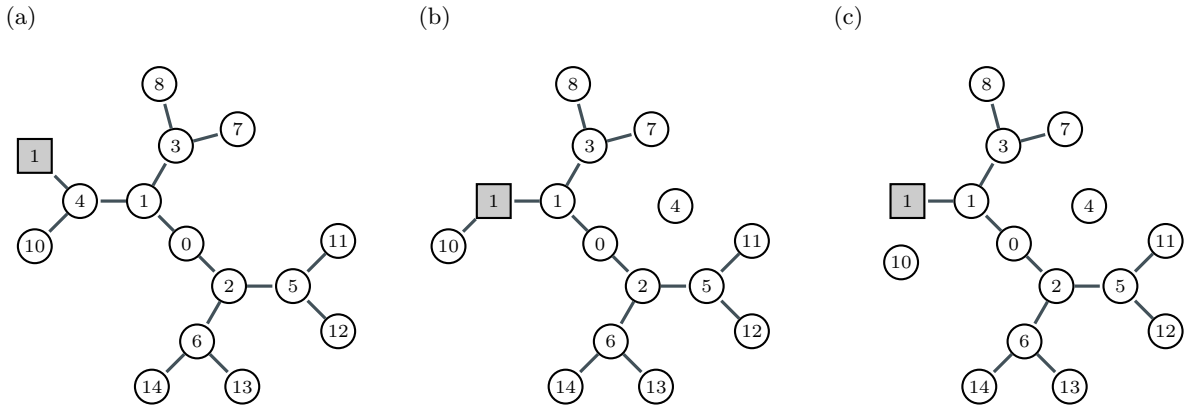


\begin{figure*}[htb]
\centering
\begin{tikzpicture}[shorten >=1pt, auto, node distance=5cm,
    unode/.style = {
    circle, 
    draw = black, 
    thick,
    fill=white, 
    inner sep=0pt,
    minimum size=0.45cm,font=\scriptsize},
    unode2/.style = {
    rectangle, 
    draw=black, 
    thick,
    fill=gray!40!white,,
    inner sep=0pt,
    minimum size=0.45cm,font=\scriptsize },
    uedge/.style = {
    draw=cyan!20!black, 
    very thick}]
 \node at(-1.5,2.75) {(a)};
 \node[unode2] (t) at (-1,1) {$1$}; 
  \node[unode] (e1) at ([shift={(90:1.5)}]t) {$1$};
  
  \node[unode] (t1) at ([shift={(0:1.5)}]t) {$2$};
 
    \node[unode] (t4441) at ([shift={(90:1.5)}]t1) {$3$};
     \path[uedge] (t)  edge (t1) edge (t4441);
      \path[uedge] (e1)  edge (t1) edge (t4441);

 \node at(-1.5+3,2.75) {(b)};
 \node[unode2] (t) at (-1+3,1) {$1$}; 
  \node[unode] (e1) at ([shift={(90:1.5)}]t) {$1$};
  
  \node[unode] (t1) at ([shift={(0:1.5)}]t) {$2$};
 
    \node[unode] (t4441) at ([shift={(90:1.5)}]t1) {$3$};
     \path[uedge] (t)  edge (t1) edge (t4441);

  \end{tikzpicture}
  
\caption{Example to demonstrate AEE Case 3. Circles and squares are photons and emitters, respectively. (a) Input state to Case 3. The emitter $1_e$ absorbed photon $1_p$ to get (b).}
\label{fig:AEE_3}
\end{figure*}
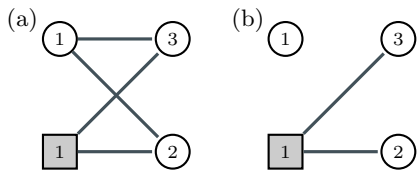

\subsection{Swapping with the free emitter (SFE)}\label{sec:SFE}

\noindent\textbf{Inputs:}  (1) An $m'$ qubit graph state $\ket{G'}$ between $m'_{e}\geq 0$ emitters and $m'_{p}\geq 1$ photons s.t. $m'\leq (m+n)$, (2) a free emitter $e$ in the state $\ket{0}_{e}$, and (3) a photon $p$ in $\ket{G'}$ that is to be swapped with $e$.


As the name suggests, this step swaps a photon in the graph state with a free emitter, i.e., an emitter that is not entangled with the graph state. We break down SFE into two steps (see the Circuit column of TABLE~\ref{tab:SFE_cases}). The first step entangles the emitter $e$ with qubits (photons and emitters) of $\ket{G'}$ to create $\ket{G'_1}$, such that $\mathbf{N_{G'_1}}(p)=\mathbf{N_{G'_1}}(e)$. As a result, in step two, $p$ is swapped with $e$ in $\ket{G'_1}$ according to Case 3 of AEE. The final graph state $\ket{G'_2}$ is obtained by replacing the vertex $p$ with $e$ in $\ket{G'}$.

The time-forward implementation of SFE is given in TABLE~\ref{tab:SFE_cases}. The first CNOT gate corresponds to the emitter emitting the photon. To implement the second CNOT gate, we would require an entangling operation between the photon and the emitter after the photon has been emitted. However, this operation is not allowed according to the architecture constraints. TABLE~\ref{tab:SFE_cases} also shows a hardware-compatible implementation of the time-forward circuit. Note that after the completion of the time-forward circuit, the emitter is in the $\ket{0}$ state. This constraint allows us to model the $CNOT_{e,p} \otimes H_{e}$ by a computation measurement on the emitter followed by a measurement-result dependent X rotation on the photon. If the emitter measurement outcome is one, apply a Pauli-X on the photon. 
The Pauli-X gate on the photonic qubit affects only the phase of the stabilizer of $\ket{G}$. As a result, it is not necessary to physically apply the Pauli-X gate if we keep track of the phase of the stabilizers using classical post-processing. This implementation is similar to the time-reversed measurement introduced in \cite{Li2022}. We justify the graph-theoretic rules and the choice of corresponding quantum circuits in Appendix~\ref{app:stabilizer_form}.

To implement this primitive in the algorithm, we need to specify the qubit in the graph that the emitter needs to replace. This specification is given to the algorithm under the initial conditions. The initial condition is a list of photons in $\ket{G}$, ordered by preference for selecting photons for SFE. Finding the optimal initial conditions for the desired objective function are beyond the scope of this work. 
\savebox{\boxK}{\begin{quantikz}
    \lstick{$p$}&\qw & \targ{}&\gate{H}&\targ{} &\qw&\qw\\
    \lstick{$e$}&\gate{H}& \ctrl{-1}&\gate{H}&\ctrl{-1}&\gate{H}&\qw\\
\end{quantikz}}%
\savebox{\boxL}{\begin{quantikz}
    \lstick{$p$}&\qw & \targ{}&\gate{H}&\targ{} &\qw&\qw\\
    \lstick{$e$}&\gate{H}& \ctrl{-1}&\gate{H}&\ctrl{-1}&\gate{H}&\qw\\
\end{quantikz}}%
\savebox{\boxM}{\begin{quantikz}
    \lstick{$p$}&\qw & \targ{}&\gate{H} &\gate{X}\\
    \lstick{$e$}&\gate{H}& \ctrl{-1}&\gate{H}&\meter{}\\
\end{quantikz}}%
\begin{table*}[ht]
    \centering
  \resizebox{\textwidth}{!}{\begin{tabular}{?P{0.75cm}|P{3cm}|P{3cm}|P{7.25cm}?}
      \hline\xrowht{10pt}
     &\textbf{Condition} & \textbf{Output} & \textbf{Circuit} \\
    \noalign{\hrule height 1pt}\xrowht{20pt}

   Case 1 & Emitter $e$ in state $\ket{0}$ and a photon $p$ in $\ket{G'}$
 & $\ket{G''}\rightarrow\ket{G'}\otimes \ket{0}_{p}$.
To get $G''$ replace the vertex $p$ with $e$ in $G'$. & Time reversed circuit: 
\begin{tikzpicture}[roundnode/.style={circle, draw = black, very thick,  minimum size=1mm,font=\sffamily\small\bfseries},edgestyle/.style={draw=black, thick}]
\path [rounded corners,draw=black, dotted, thick] (-12.4, 4.4) rectangle (-10.25, 2.65) {};
       \path [rounded corners,draw=black, dashed, thick] (-10.1, 4.4) rectangle (-7.2, 2.65) {};
    \node at (-9.9,3.25){\usebox\boxK};
    \end{tikzpicture}
\newline Time forward circuit: \begin{tikzpicture}[roundnode/.style={circle, draw = black, very thick,  minimum size=1mm,font=\sffamily\small\bfseries},edgestyle/.style={draw=black, thick}]

    \path [rounded corners,draw=black, thick] (-3.9, 2.78) rectangle (-3.2, 4.35) {};
    \node at (-2.8,3.25){\usebox\boxL};
    \end{tikzpicture}
\newline Hardware compatible implementation: \begin{tikzpicture}[roundnode/.style={circle, draw = black, very thick,  minimum size=1mm,font=\sffamily\small\bfseries},edgestyle/.style={draw=black, thick}]

    \node at (4,3.25){\usebox\boxM};
    \draw [thick] (6,3.25 ) -- (6,3.9);
         \draw [thick] (6.1,3.25 ) -- (6.1,3.9);
         \node at (6,2.4) {$Z$};
    \end{tikzpicture}\\
    \noalign{\hrule height 1pt}
  \end{tabular}}
\caption{Swapping the photon $p$ in the graph state with a free emitter $e$.  In the time-reversed circuit for SFE, the circuit inside the dotted rectangle entangles $e$ with $\ket{G'}$. The circuit inside the dashed rectangle represents Case 3 of AEE. The CNOT inside the rectangle corresponds to the photon emission process in the time-forward circuit. (c) Hardware-compatible implementation of SFE.} 
    \label{tab:SFE_cases}
\end{table*}

The impact of the replacement by the free emitter on the graph can be represented pictorially as well, as given in Figure~\ref{fig:pic_rep_SFE_1}.

\begin{figure*}[htb]
\centering
\begin{tikzpicture}[shorten >=1pt, auto, node distance=5cm,
    unode/.style = {
    circle, 
    draw = black, 
    thick,
    fill=white, 
    inner sep=0pt,
    minimum size=0.45cm,font=\scriptsize},
    unode2/.style = {
    rectangle, 
    draw=black, 
    thick,
    fill=gray!40!white,,
    inner sep=0pt,
    minimum size=0.45cm,font=\scriptsize },
    uedge/.style = {
    draw=cyan!20!black, 
    very thick}]
 \node at(-4.2,3) {(a)};
  \node[unode2] (e1) at (-1,0.5) {$1$};
  \node[unode] (t) at (-2,0) {$0$}; 
  \node[unode] (t1) at ([shift={(-45:0.8)}]t) {$2$};
  \node[unode] (t4) at ([shift={(135:0.8)}]t) {$1$};
  \path[uedge] (t)  edge (t1) edge (t4);

   \node[unode] (t1111) at ([shift={(0:0.85)}]t1) {$5$};
   \path[uedge] (t1) edge (t1111);
   \node[unode] (t1113) at ([shift={(-50-70:0.85)}]t1) {$6$};
   \path[uedge] (t1) edge (t1113);

    \node[unode] (t11111) at ([shift={(45:0.85)}]t1111) {$11$};
    \path[uedge] (t11111) edge (t1111);
    \node[unode] (t11112) at ([shift={(-45:0.85)}]t1111) {$12$};
    \path[uedge] (t11112) edge (t1111);

   \node[unode] (t11131) at ([shift={(-45:0.85)}]t1113) {$13$};
   \path[uedge] (t1113) edge (t11131);
   \node[unode] (t11132) at ([shift={(-135:0.85)}]t1113) {$14$};
   \path[uedge] (t1113) edge (t11132);

    \node[unode] (t4441) at ([shift={(180:0.85)}]t4) {$3$};
   \path[uedge] (t4) edge (t4441);
  
   \node[unode] (t4443) at ([shift={(60:0.85)}]t4) {$4$};
   \path[uedge] (t4) edge (t4443);

    \node[unode] (t44411) at ([shift={(135:0.85)}]t4441) {$9$};
   \path[uedge] (t4441) edge (t44411);
   \node[unode] (t44412) at ([shift={(225:0.85)}]t4441) {$10$};
   \path[uedge] (t4441) edge (t44412);
   
   \node[unode] (t44431) at ([shift={(105:0.85)}]t4443) {$8$};
   \path[uedge] (t4443) edge (t44431);
   \node[unode] (t44432) at ([shift={(15:0.85)}]t4443) {$7$};
   \path[uedge] (t4443) edge (t44432);

 \node at(-4.2+6,3) {(b)};
  \node[unode] (e1) at (-1+6,0.5) {$0$};
  \node[unode2] (t) at (-2+6,0) {$1$}; 
  \node[unode] (t1) at ([shift={(-45:0.8)}]t) {$2$};
  \node[unode] (t4) at ([shift={(135:0.8)}]t) {$1$};
  \path[uedge] (t)  edge (t1) edge (t4);

   \node[unode] (t1111) at ([shift={(0:0.85)}]t1) {$5$};
   \path[uedge] (t1) edge (t1111);
   \node[unode] (t1113) at ([shift={(-50-70:0.85)}]t1) {$6$};
   \path[uedge] (t1) edge (t1113);

    \node[unode] (t11111) at ([shift={(45:0.85)}]t1111) {$11$};
    \path[uedge] (t11111) edge (t1111);
    \node[unode] (t11112) at ([shift={(-45:0.85)}]t1111) {$12$};
    \path[uedge] (t11112) edge (t1111);

   \node[unode] (t11131) at ([shift={(-45:0.85)}]t1113) {$13$};
   \path[uedge] (t1113) edge (t11131);
   \node[unode] (t11132) at ([shift={(-135:0.85)}]t1113) {$14$};
   \path[uedge] (t1113) edge (t11132);

    \node[unode] (t4441) at ([shift={(180:0.85)}]t4) {$3$};
   \path[uedge] (t4) edge (t4441);
  
   \node[unode] (t4443) at ([shift={(60:0.85)}]t4) {$4$};
   \path[uedge] (t4) edge (t4443);

    \node[unode] (t44411) at ([shift={(135:0.85)}]t4441) {$9$};
   \path[uedge] (t4441) edge (t44411);
   \node[unode] (t44412) at ([shift={(225:0.85)}]t4441) {$10$};
   \path[uedge] (t4441) edge (t44412);
   
   \node[unode] (t44431) at ([shift={(105:0.85)}]t4443) {$8$};
   \path[uedge] (t4443) edge (t44431);
   \node[unode] (t44432) at ([shift={(15:0.85)}]t4443) {$7$};
   \path[uedge] (t4443) edge (t44432);
  \end{tikzpicture}
  
\caption{Example to demonstrate SFE. CCircles and squares are photons and emitters, respectively. (a) Input state to Case 1. The emitter $1_e$ swapped photon $0_p$ to get (b).}
\label{fig:pic_rep_SFE_1}
\end{figure*}

\subsection{Unentangle emitters} \label{sec:emitter-disent}
The goal of this primitive is to disentangle as many emitters as possible. This can be achieved by performing $\operatorname{CNOT}$ gates between the emitters and Hadamard gates on the emitters. We now outline the conditions under which the two-qubit gates can be performed. 

\noindent\textbf{Inputs:}  (1) An $m''$ qubit graph state $\ket{G''}$ between $m''_{e}\geq 2$ emitters and $m''_{p}\geq 0$ photons and s.t. $m''_{e}
+m''_{p}=m''$ and $m''\leq (m+n)$, and (2) emitters $e_1, e_2$ in graph state $\ket{G''}$ to be unentanlged. 

\savebox{\boxD}{\begin{quantikz}
    \lstick{$e_1$}& \ctrl{}{}&\qw\\
    \lstick{$e_2$}& \ctrl{-1}&\qw\\
\end{quantikz}}%
\savebox{\boxE}{\begin{quantikz}
    \lstick{$e_1$} & \targ{}&\qw\\
    \lstick{$e_2$}& \ctrl{-1}&\qw\\
\end{quantikz}}%

\begin{table*}[ht]
    \centering
  \resizebox{\textwidth}{!}{\begin{tabular}{?P{1cm}|P{4cm}|P{4cm}|P{5cm}?}
      \hline\xrowht{10pt}
     &\textbf{Condition} & \textbf{Output} & \textbf{Circuit} \\
    \noalign{\hrule height 1pt}\xrowht{20pt}
    Case 1 & If $e_2\in \mathbf{N}_{G_1''}(e_1)$, i.e., if an edge exists between $e_1$ and $e_2$ & Remove the between $e_1$ and $e_2$ & \usebox\boxD
\\
   \hline\xrowht{20pt}
   Case 2 & If $\textbf{N}_{G_1''}(e_1) = \textbf{N}_{G_1''}(e_2)$. 
 & $\ket{G''}\rightarrow\ket{G''_1}\otimes \ket{+}_{e_2}$.
$G_1''=G''\setminus\{e_2\}$. &\usebox\boxE\\
    \noalign{\hrule height 1pt}
  \end{tabular}}
\caption{This table summarizes the two cases to unentangle emitters $e_1$ and $e_2$ in graph state $\ket{G''}$. The graph state after photon absorption is $\ket{G_1''}$. $\mathbf{N}_{G_1''}(i)$ is the set of neighbors of vertex $i$ in $G''$. Here, the hardware implementation is same as the time-reversed circuits.} 
    \label{tab:EE_cases}
\end{table*}

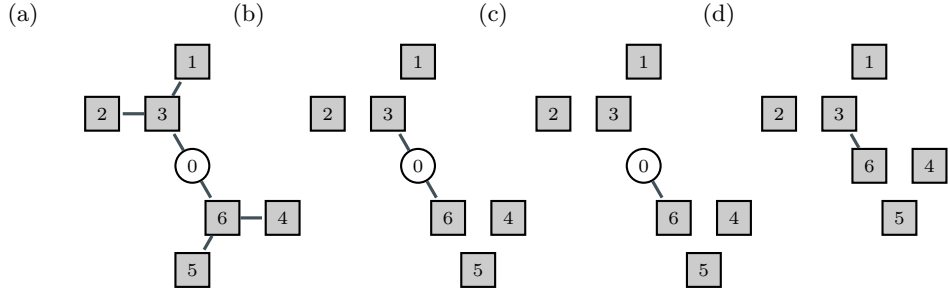
\begin{figure*}[htb]
\centering
\begin{tikzpicture}[shorten >=1pt, auto, node distance=5cm,
    unode/.style = {
    circle, 
    draw = black, 
    thick,
    fill=white, 
    inner sep=0pt,
    minimum size=0.45cm,font=\scriptsize},
    unode2/.style = {
    rectangle, 
    draw=black, 
    thick,
    fill=gray!40!white,,
    inner sep=0pt,
    minimum size=0.45cm,font=\scriptsize },
    uedge/.style = {
    draw=cyan!20!black, 
    very thick}]
  \node at(-3.25,3) {(a)};
  \node[unode] (t) at (-1,1) {$0$}; 
  \node[unode2] (t1) at ([shift={(-60:0.8)}]t) {$6$};
  \node[unode2] (t2) at ([shift={(120:.8)}]t) {$3$};
  \node[unode2] (t3) at ([shift={(60:0.8)}]t2) {$1$};
  \node[unode2] (t4) at ([shift={(180:0.8)}]t2) {$2$};
  \node[unode2] (t5) at ([shift={(0:0.8)}]t1) {$4$};
  \node[unode2] (t6) at ([shift={(-120:0.8)}]t1) {$5$};
  \path[uedge] (t)  edge (t1) edge (t2);
  \path[uedge] (t2) edge (t3) edge (t4);
 \path[uedge] (t1) edge (t5) edge (t6);

 \node at(-3.25+3,3) {(b)};
  \node[unode] (t) at (-1+3,1) {$0$}; 
  \node[unode2] (t1) at ([shift={(-60:0.8)}]t) {$6$};
  \node[unode2] (t2) at ([shift={(120:.8)}]t) {$3$};
  \node[unode2] (t3) at ([shift={(60:0.8)}]t2) {$1$};
  \node[unode2] (t4) at ([shift={(180:0.8)}]t2) {$2$};
  \node[unode2] (t5) at ([shift={(0:0.8)}]t1) {$4$};
  \node[unode2] (t6) at ([shift={(-60:0.8)}]t1) {$5$};
  \path[uedge] (t)  edge (t1) edge (t2);

 \node at(-3+6,3) {(c)};
  \node[unode] (t) at (-1+6,1) {$0$}; 
  \node[unode2] (t1) at ([shift={(-60:0.8)}]t) {$6$};
  \node[unode2] (t2) at ([shift={(120:.8)}]t) {$3$};
  \node[unode2] (t3) at ([shift={(60:0.8)}]t2) {$1$};
  \node[unode2] (t4) at ([shift={(180:0.8)}]t2) {$2$};
  \node[unode2] (t5) at ([shift={(0:0.8)}]t1) {$4$};
  \node[unode2] (t6) at ([shift={(-60:0.8)}]t1) {$5$};
  \path[uedge] (t)  edge (t1);

\node at(-3.25+9.25,3) {(d)};
  \node[unode2] (t1) at (8,1) {$6$};
  \node[unode2] (t2) at ([shift={(120:.8)}]t1) {$3$};
  \node[unode2] (t3) at ([shift={(60:0.8)}]t2) {$1$};
  \node[unode2] (t4) at ([shift={(180:0.8)}]t2) {$2$};
  \node[unode2] (t5) at ([shift={(0:0.8)}]t1) {$4$};
  \node[unode2] (t6) at ([shift={(-60:0.8)}]t1) {$5$};
  \path[uedge]  (t1) edge (t2);
  
  \end{tikzpicture}
  
\caption{Pictorial representation of unentangling emitters, Case 1. (a) Initial graph, Case 1 (b) Final graph, Case 1. Initial case for Case 2 (c) Final graph for Case 2 (d) Final graph if the photon $0_p$ is absorbed first by one of the emitters.}
\label{fig:EE}
\end{figure*}

We give an example of unentangling emitters in Figure~\ref{fig:EE}. Figure~\ref{fig:EE}(b) depicts the initial state for Case 2 of unentangling emitters with identical neighborhoods of $3_e$ and $6_e$. Given Figure~\ref{fig:EE}(b) also satisfies Case 1 of AEE such that the emitters $3_e$ and $6_e$ have only one photonic qubit as a neighbor, one of the emitters can absorb the photon $0_p$ first, resulting in Figure~\ref{fig:EE}(d). The objective function determines whether the algorithm performs photon absorption or unentangling emitters first.

\subsection{Building an Algorithm}\label{sec:final_algo}
We combine the three primitives outlined above to obtain the final algorithms. We note here that, in principle, various ways exist to combine the above primitives to absorb the graph state completely. Our numerics show that intending to minimize the CNOT depth of the circuit for repeater graph states, the following two algorithms yield the minimal CNOT depth. The choice of the algorithm depends on the initial number of emitters. Further optimization on the ordering of the primitives is left for future work.


\subsubsection{Algorithm 1}
\textbf{Input:} the number of emitters $n_e$, initial conditions, and the target graph state $\ket{G}$

\begin{figure}
\begin{minipage}{\linewidth}
\begin{algorithm}[H]
\caption{}\label{alg:alg1}
\textbf{Input:} the number of emitters $n_e$, the target graph state $\ket{G}$ with $m$ photons, and initial conditions represented by an array of length $m$\\
\textbf{Output:} The algorithm to generate $\ket{G}$
\vspace{1em}
\begin{algorithmic}[1]

\Procedure{getAlgorithm1}{$n_e$,$G$, \textrm{initial conditions}}
    \For {$i\gets 1:n_e$}
    \State Apply SFE using $i$-th emitter to $i$-th photon in initial conditions.
    \State Remove $i$-th photon in initial conditions and update initial conditions.
    \State Apply AEE using $i$-th emitter to as many photons as possible.
    \State Remove the photons undergone AEE from initial conditions.
\EndFor
\State Unentangle emitters.
\If {$\textrm{length(initial conditions)}>0 $} \Comment{If all photons are not absorbed}
\State {getAlgorithm1}($n_e$,$G$, \textrm{initial conditions})
\EndIf

\EndProcedure

\end{algorithmic}
\end{algorithm}
\end{minipage}
\end{figure}

In Algorithm~1, we repeatedly remove the absorbed photons from the initial conditions. The reason being that once the photon has been absorbed, it cannot be swapped with a free emitter. 

We also note that with this algorithm, there is an upper limit on the number of emitters that can be used. This algorithm gives a preference to photon absorption over SFE. This implies that the emitter absorbs as many photons as possible. Then, after a point, there are no more photons left for the emitter to replace. To incorporate more emitters, we would need a different ordering of the primitives, as depicted in Algorithm 2. The Algorithm 2 gives preference to replacing the photons with quantum emitters over photon absorption.

\subsubsection{Algorithm 2}
\begin{figure}
\begin{minipage}{\linewidth}
\begin{algorithm}[H]
\caption{}\label{alg:alg2}
\textbf{Input:} the number of emitters $n_e$, the target graph state $\ket{G}$ with $m$ photons, and initial conditions represented by an array of length $m$\\
\textbf{Output:} The algorithm to generate $\ket{G}$
\vspace{1em}
\begin{algorithmic}[1]

\Procedure{getAlgorithm2}{$n_e$,$G$, \textrm{initial conditions}}
    
    \State Apply SFE simultaneously using all $n_e$ emitters to the first $n_e$ photons in initial conditions.
    \State Remove the first $n_e$ photons in initial conditions and update initial conditions.
    \State Apply AEE simultaneously using $n_e$ emitters to as many photons as possible.
    \State Remove the photons undergone AEE from initial conditions.

\State Unentangle emitters.
\If {$\textrm{length(initial conditions)}>0 $}
\State {getAlgorithm2}($n_e$,$G$, \textrm{initial conditions})
\EndIf

\EndProcedure

\end{algorithmic}
\end{algorithm}
\end{minipage}
\end{figure}


The main point of difference between Algorithm 1 and Algorithm 2 is as follows: In Algorithm 1, after swapping the photon with an emitter, the emitter tries to absorb as many photons as possible. The algorithm tries to swap another photon with a free emitter if no more photons can be absorbed by the existing emitters in the graph. The Algorithm 1 prefers absorbing photons over swapping with free emitter. In Algorithm 2, the preference is given to swapping photons with free emitters. The Algorithm will first swap photons with all available free emitters and then try to absorb photons. 

Now, consider that the number of emitters equals the number of photons in the graph state. Then, the Algorithm 2, first prepares the graph state on the emitter and transduces the state to the photons. In this case, all the photons are emitted in one time step. 
\subsubsection{Example for Algorithm 1}
We illustrate Algorithm 1 with the following example. Two parameters $m$ and the branching vector $b$ describe a repeater graph state. For this example, we choose $m=2$ and $b = [3,2]$. We choose the number of emitters as twelve. The initial conditions are given as a list of the end nodes of the attached tree, implying that the emitters can replace these photons if the other emitters have not absorbed them. 

\begin{figure*}
    \centering
    \resizebox{\textwidth}{!}{
    \begin{tikzpicture}[scale=1,shorten >=1pt, auto, node distance=5cm,
    unode/.style = {
    circle, 
    draw = black, 
    thick,
    fill=white, 
    inner sep=0pt,
    minimum size=0.45cm,font=\scriptsize},
    unode2/.style = {
    rectangle, 
    draw=black, 
    thick,
    fill=gray!40!white,,
    inner sep=0pt,
    minimum size=0.45cm,font=\scriptsize },
    uedge/.style = {
    draw=cyan!20!black, 
    very thick}]
  \node at(-3.25,3) {(a)};
  \node[unode2] (e1) at (1.4,0.2) {$1$};
  \node[unode] (t) at (0,0) {$0$}; 
  \node[unode] (t1) at ([shift={(-20:0.8)}]t) {$4$};
  \node[unode] (t2) at ([shift={(-110:.8)}]t) {$3$};
  \node[unode] (t3) at ([shift={(70:0.8)}]t) {$1$};
  \node[unode] (t4) at ([shift={(160:0.8)}]t) {$2$};
  \path[uedge] (t)  edge (t1) edge (t2) edge (t3) edge (t4);

  \node[unode] (t11) at ([shift={(-80:0.7)}]t1) {$38$};
  \path[uedge] (t1) edge (t11);
  \node[unode] (t21) at ([shift={(-170:0.7)}]t2) {$27$};
  \path[uedge] (t2) edge (t21);
  \node[unode] (t31) at ([shift={(10:0.7)}]t3) {$5$};
  \path[uedge] (t3) edge (t31);
  \node[unode] (t41) at ([shift={(100:0.7)}]t4) {$16$};
  \path[uedge] (t4) edge (t41);

  \node[unode] (t111) at ([shift={(-20:.7)}]t1) {$39$};
   \path[uedge] (t111) edge (t1);
   \node[unode] (t1111) at ([shift={(10+10:0.85)}]t111) {$40$};
   \path[uedge] (t111) edge (t1111);
   \node[unode] (t1112) at ([shift={(-20:0.85)}]t111) {$41$};
   \path[uedge] (t111) edge (t1112);
   \node[unode] (t1113) at ([shift={(-50-10:0.85)}]t111) {$42$};
   \path[uedge] (t111) edge (t1113);

    \node[unode] (t11111) at ([shift={(40:0.85)}]t1111) {$43$};
    \path[uedge] (t11111) edge (t1111);
    \node[unode] (t11112) at ([shift={(0:0.85)}]t1111) {$44$};
    \path[uedge] (t11112) edge (t1111);

    \node[unode] (t11121) at ([shift={(-0:0.85)}]t1112) {$45$};
   \path[uedge] (t1112) edge (t11121);
    \node[unode] (t11121) at ([shift={(-40:0.85)}]t1112) {$46$};
   \path[uedge] (t1112) edge (t11121);
   
   \node[unode] (t11131) at ([shift={(-40:0.85)}]t1113) {$47$};
   \path[uedge] (t1113) edge (t11131);
   \node[unode] (t11132) at ([shift={(-80:0.85)}]t1113) {$48$};
   \path[uedge] (t1113) edge (t11132);

   \node[unode] (t222) at ([shift={(-110:.7)}]t2) {$28$};
   \path[uedge] (t222) edge (t2);
   \node[unode] (t2221) at ([shift={(10-90+10:0.85)}]t222) {$29$};
   \path[uedge] (t222) edge (t2221);
   \node[unode] (t2222) at ([shift={(-20-90:0.85)}]t222) {$30$};
   \path[uedge] (t222) edge (t2222);
   \node[unode] (t2223) at ([shift={(-50-90-10:0.85)}]t222) {$31$};
   \path[uedge] (t222) edge (t2223);
   
    \node[unode] (t22212) at ([shift={(10-90-10:0.85)}]t2221) {$32$};
   \path[uedge] (t2221) edge (t22212);
   \node[unode] (t22211) at ([shift={(10-90+30:0.85)}]t2221) {$33$};
   \path[uedge] (t2221) edge (t22211);
   \node[unode] (t22221) at ([shift={(-40-90:0.85)}]t2222) {$34$};
   \path[uedge] (t2222) edge (t22221);
   \node[unode] (t22222) at ([shift={(-90:0.85)}]t2222) {$35$};
   \path[uedge] (t2222) edge (t22222);
   \node[unode] (t22231) at ([shift={(-50-90-30:0.85)}]t2223) {$36$};
   \path[uedge] (t2223) edge (t22231);
   \node[unode] (t22232) at ([shift={(-50-90+10:0.85)}]t2223) {$37$};
   \path[uedge] (t2223) edge (t22232);

\node[unode] (t333) at ([shift={(70:.7)}]t3) {$6$};
   \path[uedge] (t333) edge (t3);
   \node[unode] (t3331) at ([shift={(10+90+10:0.85)}]t333) {$7$};
   \path[uedge] (t333) edge (t3331);
   \node[unode] (t3332) at ([shift={(-20+90:0.85)}]t333) {$8$};
   \path[uedge] (t333) edge (t3332);
   \node[unode] (t3333) at ([shift={(-50+90-10:0.85)}]t333) {$9$};
   \path[uedge] (t333) edge (t3333);
   
   \node[unode] (t33311) at ([shift={(10+90+30:0.85)}]t3331) {$10$};
   \path[uedge] (t3331) edge (t33311);
   \node[unode] (t33312) at ([shift={(10+90-10:0.85)}]t3331) {$11$};
   \path[uedge] (t3331) edge (t33312);
   \node[unode] (t33321) at ([shift={(-40+90:0.85)}]t3332) {$12$};
   \path[uedge] (t3332) edge (t33321);
    \node[unode] (t33322) at ([shift={(+90:0.85)}]t3332) {$13$};
   \path[uedge] (t3332) edge (t33322);
   \node[unode] (t33331) at ([shift={(-50+90+10:0.85)}]t3333) {$14$};
   \path[uedge] (t3333) edge (t33331);
    \node[unode] (t33332) at ([shift={(-50+90-30:0.85)}]t3333) {$15$};
   \path[uedge] (t3333) edge (t33332);

\node[unode] (t444) at ([shift={(160:.7)}]t4) {$17$};
   \path[uedge] (t444) edge (t4);

    \node[unode] (t4441) at ([shift={(10-90-90+10:0.85)}]t444) {$18$};
   \path[uedge] (t444) edge (t4441);
   \node[unode] (t4442) at ([shift={(-20-90-90:0.85)}]t444) {$19$};
   \path[uedge] (t444) edge (t4442);
   \node[unode] (t4443) at ([shift={(-50-90-90-10:0.85)}]t444) {$20$};
   \path[uedge] (t444) edge (t4443);

    \node[unode] (t44411) at ([shift={(-90-90:0.85)}]t4441) {$21$};
   \path[uedge] (t4441) edge (t44411);
   \node[unode] (t44412) at ([shift={(10-90-90+30:0.85)}]t4441) {$22$};
   \path[uedge] (t4441) edge (t44412);
   \node[unode] (t44421) at ([shift={(-90-90:0.85)}]t4442) {$23$};
   \path[uedge] (t4442) edge (t44421);
   \node[unode] (t44422) at ([shift={(-40-90-90:0.85)}]t4442) {$24$};
   \path[uedge] (t4442) edge (t44422);
   \node[unode] (t44431) at ([shift={(-50-90-90+10:0.85)}]t4443) {$25$};
   \path[uedge] (t4443) edge (t44431);
   \node[unode] (t44432) at ([shift={(-50-90-90-30:0.85)}]t4443) {$26$};
   \path[uedge] (t4443) edge (t44432);

     \node at(-3.25+7,3) {(b)};
  \node[unode] (t) at (0+7,0) {$0$}; 
  \node[unode] (t1) at ([shift={(-20:0.8)}]t) {$4$};
  \node[unode] (t2) at ([shift={(-110:.8)}]t) {$3$};
  \node[unode] (t3) at ([shift={(70:0.8)}]t) {$1$};
  \node[unode] (t4) at ([shift={(160:0.8)}]t) {$2$};
  \path[uedge] (t)  edge (t1) edge (t2) edge (t3) edge (t4);

  \node[unode] (t11) at ([shift={(-80:0.7)}]t1) {$38$};
  \path[uedge] (t1) edge (t11);
  \node[unode] (t21) at ([shift={(-170:0.7)}]t2) {$27$};
  \path[uedge] (t2) edge (t21);
  \node[unode] (t31) at ([shift={(10:0.7)}]t3) {$5$};
  \path[uedge] (t3) edge (t31);
  \node[unode] (t41) at ([shift={(100:0.7)}]t4) {$16$};
  \path[uedge] (t4) edge (t41);

  \node[unode] (t111) at ([shift={(-20:.7)}]t1) {$39$};
   \path[uedge] (t111) edge (t1);
   \node[unode] (t1111) at ([shift={(10+10:0.85)}]t111) {$40$};
   \path[uedge] (t111) edge (t1111);
   \node[unode] (t1112) at ([shift={(-20:0.85)}]t111) {$41$};
   \path[uedge] (t111) edge (t1112);
   \node[unode] (t1113) at ([shift={(-50-10:0.85)}]t111) {$42$};
   \path[uedge] (t111) edge (t1113);

    \node[unode] (t11111) at ([shift={(40:0.85)}]t1111) {$43$};
    \path[uedge] (t11111) edge (t1111);
    \node[unode] (t11112) at ([shift={(0:0.85)}]t1111) {$44$};
    \path[uedge] (t11112) edge (t1111);

    \node[unode] (t11121) at ([shift={(-0:0.85)}]t1112) {$45$};
   \path[uedge] (t1112) edge (t11121);
    \node[unode] (t11121) at ([shift={(-40:0.85)}]t1112) {$46$};
   \path[uedge] (t1112) edge (t11121);
   
   \node[unode] (t11131) at ([shift={(-40:0.85)}]t1113) {$47$};
   \path[uedge] (t1113) edge (t11131);
   \node[unode] (t11132) at ([shift={(-80:0.85)}]t1113) {$48$};
   \path[uedge] (t1113) edge (t11132);

   \node[unode] (t222) at ([shift={(-110:.7)}]t2) {$28$};
   \path[uedge] (t222) edge (t2);
   \node[unode] (t2221) at ([shift={(10-90+10:0.85)}]t222) {$29$};
   \path[uedge] (t222) edge (t2221);
   \node[unode] (t2222) at ([shift={(-20-90:0.85)}]t222) {$30$};
   \path[uedge] (t222) edge (t2222);
   \node[unode] (t2223) at ([shift={(-50-90-10:0.85)}]t222) {$31$};
   \path[uedge] (t222) edge (t2223);
   
    \node[unode] (t22212) at ([shift={(10-90-10:0.85)}]t2221) {$32$};
   \path[uedge] (t2221) edge (t22212);
   \node[unode] (t22211) at ([shift={(10-90+30:0.85)}]t2221) {$33$};
   \path[uedge] (t2221) edge (t22211);
   \node[unode] (t22221) at ([shift={(-40-90:0.85)}]t2222) {$34$};
   \path[uedge] (t2222) edge (t22221);
   \node[unode] (t22222) at ([shift={(-90:0.85)}]t2222) {$35$};
   \path[uedge] (t2222) edge (t22222);
   \node[unode] (t22231) at ([shift={(-50-90-30:0.85)}]t2223) {$36$};
   \path[uedge] (t2223) edge (t22231);
   \node[unode] (t22232) at ([shift={(-50-90+10:0.85)}]t2223) {$37$};
   \path[uedge] (t2223) edge (t22232);

\node[unode] (t333) at ([shift={(70:.7)}]t3) {$6$};
   \path[uedge] (t333) edge (t3);
   \node[unode] (t3331) at ([shift={(10+90+10:0.85)}]t333) {$7$};
   \path[uedge] (t333) edge (t3331);
   \node[unode] (t3332) at ([shift={(-20+90:0.85)}]t333) {$8$};
   \path[uedge] (t333) edge (t3332);
   \node[unode] (t3333) at ([shift={(-50+90-10:0.85)}]t333) {$9$};
   \path[uedge] (t333) edge (t3333);
   
   \node[unode2] (e) at ([shift={(10+90+30:0.85)}]t3331) {$1$};
   \path[uedge] (t3331) edge (e);
   \node[unode] (t33312) at ([shift={(10+90-10:0.85)}]t3331) {$11$};
   \path[uedge] (t3331) edge (t33312);
   \node[unode] (t33321) at ([shift={(-40+90:0.85)}]t3332) {$12$};
   \path[uedge] (t3332) edge (t33321);
    \node[unode] (t33322) at ([shift={(+90:0.85)}]t3332) {$13$};
   \path[uedge] (t3332) edge (t33322);
   \node[unode] (t33331) at ([shift={(-50+90+10:0.85)}]t3333) {$14$};
   \path[uedge] (t3333) edge (t33331);
    \node[unode] (t33332) at ([shift={(-50+90-30:0.85)}]t3333) {$15$};
   \path[uedge] (t3333) edge (t33332);

\node[unode] (t444) at ([shift={(160:.7)}]t4) {$17$};
   \path[uedge] (t444) edge (t4);

    \node[unode] (t4441) at ([shift={(10-90-90+10:0.85)}]t444) {$18$};
   \path[uedge] (t444) edge (t4441);
   \node[unode] (t4442) at ([shift={(-20-90-90:0.85)}]t444) {$19$};
   \path[uedge] (t444) edge (t4442);
   \node[unode] (t4443) at ([shift={(-50-90-90-10:0.85)}]t444) {$20$};
   \path[uedge] (t444) edge (t4443);

    \node[unode] (t44411) at ([shift={(-90-90:0.85)}]t4441) {$21$};
   \path[uedge] (t4441) edge (t44411);
   \node[unode] (t44412) at ([shift={(10-90-90+30:0.85)}]t4441) {$22$};
   \path[uedge] (t4441) edge (t44412);
   \node[unode] (t44421) at ([shift={(-90-90:0.85)}]t4442) {$23$};
   \path[uedge] (t4442) edge (t44421);
   \node[unode] (t44422) at ([shift={(-40-90-90:0.85)}]t4442) {$24$};
   \path[uedge] (t4442) edge (t44422);
   \node[unode] (t44431) at ([shift={(-50-90-90+10:0.85)}]t4443) {$25$};
   \path[uedge] (t4443) edge (t44431);
   \node[unode] (t44432) at ([shift={(-50-90-90-30:0.85)}]t4443) {$26$};
   \path[uedge] (t4443) edge (t44432);

     \node at(-3.25+14,3) {(c)};
  \node[unode] (t) at (0+14,0) {$0$}; 
  \node[unode] (t1) at ([shift={(-20:0.8)}]t) {$4$};
  \node[unode] (t2) at ([shift={(-110:.8)}]t) {$3$};
  \node[unode] (t3) at ([shift={(70:0.8)}]t) {$1$};
  \node[unode] (t4) at ([shift={(160:0.8)}]t) {$2$};
  \path[uedge] (t)  edge (t1) edge (t2) edge (t3) edge (t4);

  \node[unode] (t11) at ([shift={(-80:0.7)}]t1) {$38$};
  \path[uedge] (t1) edge (t11);
  \node[unode] (t21) at ([shift={(-170:0.7)}]t2) {$27$};
  \path[uedge] (t2) edge (t21);
  \node[unode] (t31) at ([shift={(10:0.7)}]t3) {$5$};
  \path[uedge] (t3) edge (t31);
  \node[unode] (t41) at ([shift={(100:0.7)}]t4) {$16$};
  \path[uedge] (t4) edge (t41);

  \node[unode] (t111) at ([shift={(-20:.7)}]t1) {$39$};
   \path[uedge] (t111) edge (t1);
   \node[unode] (t1111) at ([shift={(10+10:0.85)}]t111) {$40$};
   \path[uedge] (t111) edge (t1111);
   \node[unode] (t1112) at ([shift={(-20:0.85)}]t111) {$41$};
   \path[uedge] (t111) edge (t1112);
   \node[unode] (t1113) at ([shift={(-50-10:0.85)}]t111) {$42$};
   \path[uedge] (t111) edge (t1113);

    \node[unode] (t11111) at ([shift={(40:0.85)}]t1111) {$43$};
    \path[uedge] (t11111) edge (t1111);
    \node[unode] (t11112) at ([shift={(0:0.85)}]t1111) {$44$};
    \path[uedge] (t11112) edge (t1111);

    \node[unode] (t11121) at ([shift={(-0:0.85)}]t1112) {$45$};
   \path[uedge] (t1112) edge (t11121);
    \node[unode] (t11121) at ([shift={(-40:0.85)}]t1112) {$46$};
   \path[uedge] (t1112) edge (t11121);
   
   \node[unode] (t11131) at ([shift={(-40:0.85)}]t1113) {$47$};
   \path[uedge] (t1113) edge (t11131);
   \node[unode] (t11132) at ([shift={(-80:0.85)}]t1113) {$48$};
   \path[uedge] (t1113) edge (t11132);

   \node[unode] (t222) at ([shift={(-110:.7)}]t2) {$28$};
   \path[uedge] (t222) edge (t2);
   \node[unode] (t2221) at ([shift={(10-90+10:0.85)}]t222) {$29$};
   \path[uedge] (t222) edge (t2221);
   \node[unode] (t2222) at ([shift={(-20-90:0.85)}]t222) {$30$};
   \path[uedge] (t222) edge (t2222);
   \node[unode] (t2223) at ([shift={(-50-90-10:0.85)}]t222) {$31$};
   \path[uedge] (t222) edge (t2223);
   
    \node[unode] (t22212) at ([shift={(10-90-10:0.85)}]t2221) {$32$};
   \path[uedge] (t2221) edge (t22212);
   \node[unode] (t22211) at ([shift={(10-90+30:0.85)}]t2221) {$33$};
   \path[uedge] (t2221) edge (t22211);
   \node[unode] (t22221) at ([shift={(-40-90:0.85)}]t2222) {$34$};
   \path[uedge] (t2222) edge (t22221);
   \node[unode] (t22222) at ([shift={(-90:0.85)}]t2222) {$35$};
   \path[uedge] (t2222) edge (t22222);
   \node[unode] (t22231) at ([shift={(-50-90-30:0.85)}]t2223) {$36$};
   \path[uedge] (t2223) edge (t22231);
   \node[unode] (t22232) at ([shift={(-50-90+10:0.85)}]t2223) {$37$};
   \path[uedge] (t2223) edge (t22232);

\node[unode] (t333) at ([shift={(70:.7)}]t3) {$6$};
   \path[uedge] (t333) edge (t3);
   \node[unode2] (e) at ([shift={(10+90+10:0.85)}]t333) {$1$};
   \path[uedge] (t333) edge (e);
   \node[unode] (t3332) at ([shift={(-20+90:0.85)}]t333) {$8$};
   \path[uedge] (t333) edge (t3332);
   \node[unode] (t3333) at ([shift={(-50+90-10:0.85)}]t333) {$9$};
   \path[uedge] (t333) edge (t3333);

   \node[unode] (t33312) at ([shift={(10+90-10:0.85)}]e) {$11$};
   \path[uedge] (e) edge (t33312);
   \node[unode] (t33321) at ([shift={(-40+90:0.85)}]t3332) {$12$};
   \path[uedge] (t3332) edge (t33321);
    \node[unode] (t33322) at ([shift={(+90:0.85)}]t3332) {$13$};
   \path[uedge] (t3332) edge (t33322);
   \node[unode] (t33331) at ([shift={(-50+90+10:0.85)}]t3333) {$14$};
   \path[uedge] (t3333) edge (t33331);
    \node[unode] (t33332) at ([shift={(-50+90-30:0.85)}]t3333) {$15$};
   \path[uedge] (t3333) edge (t33332);

\node[unode] (t444) at ([shift={(160:.7)}]t4) {$17$};
   \path[uedge] (t444) edge (t4);

    \node[unode] (t4441) at ([shift={(10-90-90+10:0.85)}]t444) {$18$};
   \path[uedge] (t444) edge (t4441);
   \node[unode] (t4442) at ([shift={(-20-90-90:0.85)}]t444) {$19$};
   \path[uedge] (t444) edge (t4442);
   \node[unode] (t4443) at ([shift={(-50-90-90-10:0.85)}]t444) {$20$};
   \path[uedge] (t444) edge (t4443);

    \node[unode] (t44411) at ([shift={(-90-90:0.85)}]t4441) {$21$};
   \path[uedge] (t4441) edge (t44411);
   \node[unode] (t44412) at ([shift={(10-90-90+30:0.85)}]t4441) {$22$};
   \path[uedge] (t4441) edge (t44412);
   \node[unode] (t44421) at ([shift={(-90-90:0.85)}]t4442) {$23$};
   \path[uedge] (t4442) edge (t44421);
   \node[unode] (t44422) at ([shift={(-40-90-90:0.85)}]t4442) {$24$};
   \path[uedge] (t4442) edge (t44422);
   \node[unode] (t44431) at ([shift={(-50-90-90+10:0.85)}]t4443) {$25$};
   \path[uedge] (t4443) edge (t44431);
   \node[unode] (t44432) at ([shift={(-50-90-90-30:0.85)}]t4443) {$26$};
   \path[uedge] (t4443) edge (t44432);

     \node at(-3.25+0,3-7) {(d)};
  \node[unode] (t) at (0,0-7) {$0$}; 
  \node[unode] (t1) at ([shift={(-20:0.8)}]t) {$4$};
  \node[unode] (t2) at ([shift={(-110:.8)}]t) {$3$};
  \node[unode] (t3) at ([shift={(70:0.8)}]t) {$1$};
  \node[unode] (t4) at ([shift={(160:0.8)}]t) {$2$};
  \path[uedge] (t)  edge (t1) edge (t2) edge (t3) edge (t4);

  \node[unode] (t11) at ([shift={(-80:0.7)}]t1) {$38$};
  \path[uedge] (t1) edge (t11);
  \node[unode] (t21) at ([shift={(-170:0.7)}]t2) {$27$};
  \path[uedge] (t2) edge (t21);
  \node[unode] (t31) at ([shift={(10:0.7)}]t3) {$5$};
  \path[uedge] (t3) edge (t31);
  \node[unode] (t41) at ([shift={(100:0.7)}]t4) {$16$};
  \path[uedge] (t4) edge (t41);

  \node[unode] (t111) at ([shift={(-20:.7)}]t1) {$39$};
   \path[uedge] (t111) edge (t1);
   \node[unode] (t1111) at ([shift={(10+10:0.85)}]t111) {$40$};
   \path[uedge] (t111) edge (t1111);
   \node[unode] (t1112) at ([shift={(-20:0.85)}]t111) {$41$};
   \path[uedge] (t111) edge (t1112);
   \node[unode] (t1113) at ([shift={(-50-10:0.85)}]t111) {$42$};
   \path[uedge] (t111) edge (t1113);

    \node[unode] (t11111) at ([shift={(40:0.85)}]t1111) {$43$};
    \path[uedge] (t11111) edge (t1111);
    \node[unode] (t11112) at ([shift={(0:0.85)}]t1111) {$44$};
    \path[uedge] (t11112) edge (t1111);

    \node[unode] (t11121) at ([shift={(-0:0.85)}]t1112) {$45$};
   \path[uedge] (t1112) edge (t11121);
    \node[unode] (t11121) at ([shift={(-40:0.85)}]t1112) {$46$};
   \path[uedge] (t1112) edge (t11121);
   
   \node[unode] (t11131) at ([shift={(-40:0.85)}]t1113) {$47$};
   \path[uedge] (t1113) edge (t11131);
   \node[unode] (t11132) at ([shift={(-80:0.85)}]t1113) {$48$};
   \path[uedge] (t1113) edge (t11132);

   \node[unode] (t222) at ([shift={(-110:.7)}]t2) {$28$};
   \path[uedge] (t222) edge (t2);
   \node[unode] (t2221) at ([shift={(10-90+10:0.85)}]t222) {$29$};
   \path[uedge] (t222) edge (t2221);
   \node[unode] (t2222) at ([shift={(-20-90:0.85)}]t222) {$30$};
   \path[uedge] (t222) edge (t2222);
   \node[unode] (t2223) at ([shift={(-50-90-10:0.85)}]t222) {$31$};
   \path[uedge] (t222) edge (t2223);
   
    \node[unode] (t22212) at ([shift={(10-90-10:0.85)}]t2221) {$32$};
   \path[uedge] (t2221) edge (t22212);
   \node[unode] (t22211) at ([shift={(10-90+30:0.85)}]t2221) {$33$};
   \path[uedge] (t2221) edge (t22211);
   \node[unode] (t22221) at ([shift={(-40-90:0.85)}]t2222) {$34$};
   \path[uedge] (t2222) edge (t22221);
   \node[unode] (t22222) at ([shift={(-90:0.85)}]t2222) {$35$};
   \path[uedge] (t2222) edge (t22222);
   \node[unode] (t22231) at ([shift={(-50-90-30:0.85)}]t2223) {$36$};
   \path[uedge] (t2223) edge (t22231);
   \node[unode] (t22232) at ([shift={(-50-90+10:0.85)}]t2223) {$37$};
   \path[uedge] (t2223) edge (t22232);

\node[unode] (t333) at ([shift={(70:.7)}]t3) {$6$};
   \path[uedge] (t333) edge (t3);
   \node[unode2] (e) at ([shift={(10+90+10:0.85)}]t333) {$1$};
   \path[uedge] (t333) edge (e);
   \node[unode] (t3332) at ([shift={(-20+90:0.85)}]t333) {$8$};
   \path[uedge] (t333) edge (t3332);
   \node[unode] (t3333) at ([shift={(-50+90-10:0.85)}]t333) {$9$};
   \path[uedge] (t333) edge (t3333);

   \node[unode] (t33321) at ([shift={(-40+90:0.85)}]t3332) {$12$};
   \path[uedge] (t3332) edge (t33321);
    \node[unode] (t33322) at ([shift={(+90:0.85)}]t3332) {$13$};
   \path[uedge] (t3332) edge (t33322);
   \node[unode] (t33331) at ([shift={(-50+90+10:0.85)}]t3333) {$14$};
   \path[uedge] (t3333) edge (t33331);
    \node[unode] (t33332) at ([shift={(-50+90-30:0.85)}]t3333) {$15$};
   \path[uedge] (t3333) edge (t33332);

\node[unode] (t444) at ([shift={(160:.7)}]t4) {$17$};
   \path[uedge] (t444) edge (t4);

    \node[unode] (t4441) at ([shift={(10-90-90+10:0.85)}]t444) {$18$};
   \path[uedge] (t444) edge (t4441);
   \node[unode] (t4442) at ([shift={(-20-90-90:0.85)}]t444) {$19$};
   \path[uedge] (t444) edge (t4442);
   \node[unode] (t4443) at ([shift={(-50-90-90-10:0.85)}]t444) {$20$};
   \path[uedge] (t444) edge (t4443);

    \node[unode] (t44411) at ([shift={(-90-90:0.85)}]t4441) {$21$};
   \path[uedge] (t4441) edge (t44411);
   \node[unode] (t44412) at ([shift={(10-90-90+30:0.85)}]t4441) {$22$};
   \path[uedge] (t4441) edge (t44412);
   \node[unode] (t44421) at ([shift={(-90-90:0.85)}]t4442) {$23$};
   \path[uedge] (t4442) edge (t44421);
   \node[unode] (t44422) at ([shift={(-40-90-90:0.85)}]t4442) {$24$};
   \path[uedge] (t4442) edge (t44422);
   \node[unode] (t44431) at ([shift={(-50-90-90+10:0.85)}]t4443) {$25$};
   \path[uedge] (t4443) edge (t44431);
   \node[unode] (t44432) at ([shift={(-50-90-90-30:0.85)}]t4443) {$26$};
   \path[uedge] (t4443) edge (t44432);

     \node at(-3.25+7,3-7) {(e)};
  \node[unode] (t) at (7,0-7) {$0$}; 
  \node[unode] (t1) at ([shift={(-20:0.8)}]t) {$4$};
  \node[unode] (t2) at ([shift={(-110:.8)}]t) {$3$};
  \node[unode] (t3) at ([shift={(70:0.8)}]t) {$1$};
  \node[unode] (t4) at ([shift={(160:0.8)}]t) {$2$};
  \path[uedge] (t)  edge (t1) edge (t2) edge (t3) edge (t4);

  \node[unode] (t11) at ([shift={(-80:0.7)}]t1) {$38$};
  \path[uedge] (t1) edge (t11);
  \node[unode] (t21) at ([shift={(-170:0.7)}]t2) {$27$};
  \path[uedge] (t2) edge (t21);
  \node[unode] (t31) at ([shift={(10:0.7)}]t3) {$5$};
  \path[uedge] (t3) edge (t31);
  \node[unode] (t41) at ([shift={(100:0.7)}]t4) {$16$};
  \path[uedge] (t4) edge (t41);

  \node[unode] (t111) at ([shift={(-20:.7)}]t1) {$39$};
   \path[uedge] (t111) edge (t1);
   \node[unode] (t1111) at ([shift={(10+10:0.85)}]t111) {$40$};
   \path[uedge] (t111) edge (t1111);
   \node[unode] (t1112) at ([shift={(-20:0.85)}]t111) {$41$};
   \path[uedge] (t111) edge (t1112);
   \node[unode] (t1113) at ([shift={(-50-10:0.85)}]t111) {$42$};
   \path[uedge] (t111) edge (t1113);

    \node[unode] (t11111) at ([shift={(40:0.85)}]t1111) {$43$};
    \path[uedge] (t11111) edge (t1111);
    \node[unode] (t11112) at ([shift={(0:0.85)}]t1111) {$44$};
    \path[uedge] (t11112) edge (t1111);

    \node[unode] (t11121) at ([shift={(-0:0.85)}]t1112) {$45$};
   \path[uedge] (t1112) edge (t11121);
    \node[unode] (t11121) at ([shift={(-40:0.85)}]t1112) {$46$};
   \path[uedge] (t1112) edge (t11121);
   
   \node[unode] (t11131) at ([shift={(-40:0.85)}]t1113) {$47$};
   \path[uedge] (t1113) edge (t11131);
   \node[unode] (t11132) at ([shift={(-80:0.85)}]t1113) {$48$};
   \path[uedge] (t1113) edge (t11132);

   \node[unode] (t222) at ([shift={(-110:.7)}]t2) {$28$};
   \path[uedge] (t222) edge (t2);
   \node[unode] (t2221) at ([shift={(10-90+10:0.85)}]t222) {$29$};
   \path[uedge] (t222) edge (t2221);
   \node[unode] (t2222) at ([shift={(-20-90:0.85)}]t222) {$30$};
   \path[uedge] (t222) edge (t2222);
   \node[unode] (t2223) at ([shift={(-50-90-10:0.85)}]t222) {$31$};
   \path[uedge] (t222) edge (t2223);
   
    \node[unode] (t22212) at ([shift={(10-90-10:0.85)}]t2221) {$32$};
   \path[uedge] (t2221) edge (t22212);
   \node[unode] (t22211) at ([shift={(10-90+30:0.85)}]t2221) {$33$};
   \path[uedge] (t2221) edge (t22211);
   \node[unode] (t22221) at ([shift={(-40-90:0.85)}]t2222) {$34$};
   \path[uedge] (t2222) edge (t22221);
   \node[unode] (t22222) at ([shift={(-90:0.85)}]t2222) {$35$};
   \path[uedge] (t2222) edge (t22222);
   \node[unode] (t22231) at ([shift={(-50-90-30:0.85)}]t2223) {$36$};
   \path[uedge] (t2223) edge (t22231);
   \node[unode] (t22232) at ([shift={(-50-90+10:0.85)}]t2223) {$37$};
   \path[uedge] (t2223) edge (t22232);

\node[unode2] (t333) at ([shift={(70:.7)}]t3) {$1$};
   \path[uedge] (t333) edge (t3);
   \node[unode] (t3332) at ([shift={(-20+90:0.85)}]t333) {$8$};
   \path[uedge] (t333) edge (t3332);
   \node[unode] (t3333) at ([shift={(-50+90-10:0.85)}]t333) {$9$};
   \path[uedge] (t333) edge (t3333);

   \node[unode] (t33321) at ([shift={(-40+90:0.85)}]t3332) {$12$};
   \path[uedge] (t3332) edge (t33321);
    \node[unode] (t33322) at ([shift={(+90:0.85)}]t3332) {$13$};
   \path[uedge] (t3332) edge (t33322);
   \node[unode] (t33331) at ([shift={(-50+90+10:0.85)}]t3333) {$14$};
   \path[uedge] (t3333) edge (t33331);
    \node[unode] (t33332) at ([shift={(-50+90-30:0.85)}]t3333) {$15$};
   \path[uedge] (t3333) edge (t33332);

\node[unode] (t444) at ([shift={(160:.7)}]t4) {$17$};
   \path[uedge] (t444) edge (t4);

    \node[unode] (t4441) at ([shift={(10-90-90+10:0.85)}]t444) {$18$};
   \path[uedge] (t444) edge (t4441);
   \node[unode] (t4442) at ([shift={(-20-90-90:0.85)}]t444) {$19$};
   \path[uedge] (t444) edge (t4442);
   \node[unode] (t4443) at ([shift={(-50-90-90-10:0.85)}]t444) {$20$};
   \path[uedge] (t444) edge (t4443);

    \node[unode] (t44411) at ([shift={(-90-90:0.85)}]t4441) {$21$};
   \path[uedge] (t4441) edge (t44411);
   \node[unode] (t44412) at ([shift={(10-90-90+30:0.85)}]t4441) {$22$};
   \path[uedge] (t4441) edge (t44412);
   \node[unode] (t44421) at ([shift={(-90-90:0.85)}]t4442) {$23$};
   \path[uedge] (t4442) edge (t44421);
   \node[unode] (t44422) at ([shift={(-40-90-90:0.85)}]t4442) {$24$};
   \path[uedge] (t4442) edge (t44422);
   \node[unode] (t44431) at ([shift={(-50-90-90+10:0.85)}]t4443) {$25$};
   \path[uedge] (t4443) edge (t44431);
   \node[unode] (t44432) at ([shift={(-50-90-90-30:0.85)}]t4443) {$26$};
   \path[uedge] (t4443) edge (t44432);

     \node at(-3.25+14,3-7) {(f)};
  \node[unode] (t) at (14,0-7) {$0$}; 
  \node[unode] (t1) at ([shift={(-20:0.8)}]t) {$4$};
  \node[unode] (t2) at ([shift={(-110:.8)}]t) {$3$};
  \node[unode] (t3) at ([shift={(70:0.8)}]t) {$1$};
  \node[unode] (t4) at ([shift={(160:0.8)}]t) {$2$};
  \path[uedge] (t)  edge (t1) edge (t2) edge (t3) edge (t4);

  \node[unode] (t11) at ([shift={(-80:0.7)}]t1) {$38$};
  \path[uedge] (t1) edge (t11);
  \node[unode] (t21) at ([shift={(-170:0.7)}]t2) {$27$};
  \path[uedge] (t2) edge (t21);
  \node[unode] (t31) at ([shift={(10:0.7)}]t3) {$5$};
  \path[uedge] (t3) edge (t31);
  \node[unode] (t41) at ([shift={(100:0.7)}]t4) {$16$};
  \path[uedge] (t4) edge (t41);

  \node[unode] (t111) at ([shift={(-20:.7)}]t1) {$39$};
   \path[uedge] (t111) edge (t1);
   \node[unode] (t1111) at ([shift={(10+10:0.85)}]t111) {$40$};
   \path[uedge] (t111) edge (t1111);
   \node[unode] (t1112) at ([shift={(-20:0.85)}]t111) {$41$};
   \path[uedge] (t111) edge (t1112);
   \node[unode] (t1113) at ([shift={(-50-10:0.85)}]t111) {$42$};
   \path[uedge] (t111) edge (t1113);

    \node[unode] (t11111) at ([shift={(40:0.85)}]t1111) {$43$};
    \path[uedge] (t11111) edge (t1111);
    \node[unode] (t11112) at ([shift={(0:0.85)}]t1111) {$44$};
    \path[uedge] (t11112) edge (t1111);

    \node[unode] (t11121) at ([shift={(-0:0.85)}]t1112) {$45$};
   \path[uedge] (t1112) edge (t11121);
    \node[unode] (t11121) at ([shift={(-40:0.85)}]t1112) {$46$};
   \path[uedge] (t1112) edge (t11121);
   
   \node[unode] (t11131) at ([shift={(-40:0.85)}]t1113) {$47$};
   \path[uedge] (t1113) edge (t11131);
   \node[unode] (t11132) at ([shift={(-80:0.85)}]t1113) {$48$};
   \path[uedge] (t1113) edge (t11132);

   \node[unode] (t222) at ([shift={(-110:.7)}]t2) {$28$};
   \path[uedge] (t222) edge (t2);
   \node[unode] (t2221) at ([shift={(10-90+10:0.85)}]t222) {$29$};
   \path[uedge] (t222) edge (t2221);
   \node[unode] (t2222) at ([shift={(-20-90:0.85)}]t222) {$30$};
   \path[uedge] (t222) edge (t2222);
   \node[unode] (t2223) at ([shift={(-50-90-10:0.85)}]t222) {$31$};
   \path[uedge] (t222) edge (t2223);
   
    \node[unode] (t22212) at ([shift={(10-90-10:0.85)}]t2221) {$32$};
   \path[uedge] (t2221) edge (t22212);
   \node[unode] (t22211) at ([shift={(10-90+30:0.85)}]t2221) {$33$};
   \path[uedge] (t2221) edge (t22211);
   \node[unode] (t22221) at ([shift={(-40-90:0.85)}]t2222) {$34$};
   \path[uedge] (t2222) edge (t22221);
   \node[unode] (t22222) at ([shift={(-90:0.85)}]t2222) {$35$};
   \path[uedge] (t2222) edge (t22222);
   \node[unode] (t22231) at ([shift={(-50-90-30:0.85)}]t2223) {$36$};
   \path[uedge] (t2223) edge (t22231);
   \node[unode] (t22232) at ([shift={(-50-90+10:0.85)}]t2223) {$37$};
   \path[uedge] (t2223) edge (t22232);

\node[unode2] (t333) at ([shift={(70:.7)}]t3) {$1$};
   \path[uedge] (t333) edge (t3);
   \node[unode] (t3332) at ([shift={(-20+90:0.85)}]t333) {$8$};
   \path[uedge] (t333) edge (t3332);
   \node[unode] (t3333) at ([shift={(-50+90-10:0.85)}]t333) {$9$};
   \path[uedge] (t333) edge (t3333);

   \node[unode2] (t33321) at ([shift={(-40+90:0.85)}]t3332) {$2$};
   \path[uedge] (t3332) edge (t33321);
    \node[unode] (t33322) at ([shift={(+90:0.85)}]t3332) {$13$};
   \path[uedge] (t3332) edge (t33322);
   \node[unode] (t33331) at ([shift={(-50+90+10:0.85)}]t3333) {$14$};
   \path[uedge] (t3333) edge (t33331);
    \node[unode] (t33332) at ([shift={(-50+90-30:0.85)}]t3333) {$15$};
   \path[uedge] (t3333) edge (t33332);

\node[unode] (t444) at ([shift={(160:.7)}]t4) {$17$};
   \path[uedge] (t444) edge (t4);

    \node[unode] (t4441) at ([shift={(10-90-90+10:0.85)}]t444) {$18$};
   \path[uedge] (t444) edge (t4441);
   \node[unode] (t4442) at ([shift={(-20-90-90:0.85)}]t444) {$19$};
   \path[uedge] (t444) edge (t4442);
   \node[unode] (t4443) at ([shift={(-50-90-90-10:0.85)}]t444) {$20$};
   \path[uedge] (t444) edge (t4443);

    \node[unode] (t44411) at ([shift={(-90-90:0.85)}]t4441) {$21$};
   \path[uedge] (t4441) edge (t44411);
   \node[unode] (t44412) at ([shift={(10-90-90+30:0.85)}]t4441) {$22$};
   \path[uedge] (t4441) edge (t44412);
   \node[unode] (t44421) at ([shift={(-90-90:0.85)}]t4442) {$23$};
   \path[uedge] (t4442) edge (t44421);
   \node[unode] (t44422) at ([shift={(-40-90-90:0.85)}]t4442) {$24$};
   \path[uedge] (t4442) edge (t44422);
   \node[unode] (t44431) at ([shift={(-50-90-90+10:0.85)}]t4443) {$25$};
   \path[uedge] (t4443) edge (t44431);
   \node[unode] (t44432) at ([shift={(-50-90-90-30:0.85)}]t4443) {$26$};
   \path[uedge] (t4443) edge (t44432);

     \node at(-3.25,3-14) {(g)};
  \node[unode] (t) at (0,0-14) {$0$}; 
  \node[unode] (t1) at ([shift={(-20:0.8)}]t) {$4$};
  \node[unode] (t2) at ([shift={(-110:.8)}]t) {$3$};
  \node[unode] (t3) at ([shift={(70:0.8)}]t) {$1$};
  \node[unode] (t4) at ([shift={(160:0.8)}]t) {$2$};
  \path[uedge] (t)  edge (t1) edge (t2) edge (t3) edge (t4);

  \node[unode] (t11) at ([shift={(-80:0.7)}]t1) {$38$};
  \path[uedge] (t1) edge (t11);
  \node[unode] (t21) at ([shift={(-170:0.7)}]t2) {$27$};
  \path[uedge] (t2) edge (t21);
  \node[unode] (t31) at ([shift={(10:0.7)}]t3) {$5$};
  \path[uedge] (t3) edge (t31);
  \node[unode] (t41) at ([shift={(100:0.7)}]t4) {$16$};
  \path[uedge] (t4) edge (t41);

  \node[unode] (t111) at ([shift={(-20:.7)}]t1) {$39$};
   \path[uedge] (t111) edge (t1);
   \node[unode] (t1111) at ([shift={(10+10:0.85)}]t111) {$40$};
   \path[uedge] (t111) edge (t1111);
   \node[unode] (t1112) at ([shift={(-20:0.85)}]t111) {$41$};
   \path[uedge] (t111) edge (t1112);
   \node[unode] (t1113) at ([shift={(-50-10:0.85)}]t111) {$42$};
   \path[uedge] (t111) edge (t1113);

    \node[unode] (t11111) at ([shift={(40:0.85)}]t1111) {$43$};
    \path[uedge] (t11111) edge (t1111);
    \node[unode] (t11112) at ([shift={(0:0.85)}]t1111) {$44$};
    \path[uedge] (t11112) edge (t1111);

    \node[unode] (t11121) at ([shift={(-0:0.85)}]t1112) {$45$};
   \path[uedge] (t1112) edge (t11121);
    \node[unode] (t11121) at ([shift={(-40:0.85)}]t1112) {$46$};
   \path[uedge] (t1112) edge (t11121);
   
   \node[unode] (t11131) at ([shift={(-40:0.85)}]t1113) {$47$};
   \path[uedge] (t1113) edge (t11131);
   \node[unode] (t11132) at ([shift={(-80:0.85)}]t1113) {$48$};
   \path[uedge] (t1113) edge (t11132);

   \node[unode] (t222) at ([shift={(-110:.7)}]t2) {$28$};
   \path[uedge] (t222) edge (t2);
   \node[unode] (t2221) at ([shift={(10-90+10:0.85)}]t222) {$29$};
   \path[uedge] (t222) edge (t2221);
   \node[unode] (t2222) at ([shift={(-20-90:0.85)}]t222) {$30$};
   \path[uedge] (t222) edge (t2222);
   \node[unode] (t2223) at ([shift={(-50-90-10:0.85)}]t222) {$31$};
   \path[uedge] (t222) edge (t2223);
   
    \node[unode] (t22212) at ([shift={(10-90-10:0.85)}]t2221) {$32$};
   \path[uedge] (t2221) edge (t22212);
   \node[unode] (t22211) at ([shift={(10-90+30:0.85)}]t2221) {$33$};
   \path[uedge] (t2221) edge (t22211);
   \node[unode] (t22221) at ([shift={(-40-90:0.85)}]t2222) {$34$};
   \path[uedge] (t2222) edge (t22221);
   \node[unode] (t22222) at ([shift={(-90:0.85)}]t2222) {$35$};
   \path[uedge] (t2222) edge (t22222);
   \node[unode] (t22231) at ([shift={(-50-90-30:0.85)}]t2223) {$36$};
   \path[uedge] (t2223) edge (t22231);
   \node[unode] (t22232) at ([shift={(-50-90+10:0.85)}]t2223) {$37$};
   \path[uedge] (t2223) edge (t22232);

\node[unode2] (t333) at ([shift={(70:.7)}]t3) {$1$};
   \path[uedge] (t333) edge (t3);
   \node[unode2] (t3332) at ([shift={(-20+90:0.85)}]t333) {$2$};
   \path[uedge] (t333) edge (t3332);
   \node[unode] (t3333) at ([shift={(-50+90-10:0.85)}]t333) {$9$};
   \path[uedge] (t333) edge (t3333);

   \node[unode] (t33331) at ([shift={(-50+90+10:0.85)}]t3333) {$14$};
   \path[uedge] (t3333) edge (t33331);
    \node[unode] (t33332) at ([shift={(-50+90-30:0.85)}]t3333) {$15$};
   \path[uedge] (t3333) edge (t33332);

\node[unode] (t444) at ([shift={(160:.7)}]t4) {$17$};
   \path[uedge] (t444) edge (t4);

    \node[unode] (t4441) at ([shift={(10-90-90+10:0.85)}]t444) {$18$};
   \path[uedge] (t444) edge (t4441);
   \node[unode] (t4442) at ([shift={(-20-90-90:0.85)}]t444) {$19$};
   \path[uedge] (t444) edge (t4442);
   \node[unode] (t4443) at ([shift={(-50-90-90-10:0.85)}]t444) {$20$};
   \path[uedge] (t444) edge (t4443);

    \node[unode] (t44411) at ([shift={(-90-90:0.85)}]t4441) {$21$};
   \path[uedge] (t4441) edge (t44411);
   \node[unode] (t44412) at ([shift={(10-90-90+30:0.85)}]t4441) {$22$};
   \path[uedge] (t4441) edge (t44412);
   \node[unode] (t44421) at ([shift={(-90-90:0.85)}]t4442) {$23$};
   \path[uedge] (t4442) edge (t44421);
   \node[unode] (t44422) at ([shift={(-40-90-90:0.85)}]t4442) {$24$};
   \path[uedge] (t4442) edge (t44422);
   \node[unode] (t44431) at ([shift={(-50-90-90+10:0.85)}]t4443) {$25$};
   \path[uedge] (t4443) edge (t44431);
   \node[unode] (t44432) at ([shift={(-50-90-90-30:0.85)}]t4443) {$26$};
   \path[uedge] (t4443) edge (t44432);

     \node at(-3.25+7,3-14) {(h)};
  \node[unode] (t) at (0+7,0-14) {$0$}; 
  \node[unode] (t1) at ([shift={(-20:0.8)}]t) {$4$};
  \node[unode] (t2) at ([shift={(-110:.8)}]t) {$3$};
  \node[unode] (t3) at ([shift={(70:0.8)}]t) {$1$};
  \node[unode] (t4) at ([shift={(160:0.8)}]t) {$2$};
  \path[uedge] (t)  edge (t1) edge (t2) edge (t3) edge (t4);

  \node[unode] (t11) at ([shift={(-80:0.7)}]t1) {$38$};
  \path[uedge] (t1) edge (t11);
  \node[unode] (t21) at ([shift={(-170:0.7)}]t2) {$27$};
  \path[uedge] (t2) edge (t21);
  \node[unode] (t31) at ([shift={(10:0.7)}]t3) {$5$};
  \path[uedge] (t3) edge (t31);
  \node[unode] (t41) at ([shift={(100:0.7)}]t4) {$16$};
  \path[uedge] (t4) edge (t41);

  \node[unode] (t111) at ([shift={(-20:.7)}]t1) {$39$};
   \path[uedge] (t111) edge (t1);
   \node[unode] (t1111) at ([shift={(10+10:0.85)}]t111) {$40$};
   \path[uedge] (t111) edge (t1111);
   \node[unode] (t1112) at ([shift={(-20:0.85)}]t111) {$41$};
   \path[uedge] (t111) edge (t1112);
   \node[unode] (t1113) at ([shift={(-50-10:0.85)}]t111) {$42$};
   \path[uedge] (t111) edge (t1113);

    \node[unode] (t11111) at ([shift={(40:0.85)}]t1111) {$43$};
    \path[uedge] (t11111) edge (t1111);
    \node[unode] (t11112) at ([shift={(0:0.85)}]t1111) {$44$};
    \path[uedge] (t11112) edge (t1111);

    \node[unode] (t11121) at ([shift={(-0:0.85)}]t1112) {$45$};
   \path[uedge] (t1112) edge (t11121);
    \node[unode] (t11121) at ([shift={(-40:0.85)}]t1112) {$46$};
   \path[uedge] (t1112) edge (t11121);
   
   \node[unode] (t11131) at ([shift={(-40:0.85)}]t1113) {$47$};
   \path[uedge] (t1113) edge (t11131);
   \node[unode] (t11132) at ([shift={(-80:0.85)}]t1113) {$48$};
   \path[uedge] (t1113) edge (t11132);

   \node[unode] (t222) at ([shift={(-110:.7)}]t2) {$28$};
   \path[uedge] (t222) edge (t2);
   \node[unode] (t2221) at ([shift={(10-90+10:0.85)}]t222) {$29$};
   \path[uedge] (t222) edge (t2221);
   \node[unode] (t2222) at ([shift={(-20-90:0.85)}]t222) {$30$};
   \path[uedge] (t222) edge (t2222);
   \node[unode] (t2223) at ([shift={(-50-90-10:0.85)}]t222) {$31$};
   \path[uedge] (t222) edge (t2223);
   
    \node[unode] (t22212) at ([shift={(10-90-10:0.85)}]t2221) {$32$};
   \path[uedge] (t2221) edge (t22212);
   \node[unode] (t22211) at ([shift={(10-90+30:0.85)}]t2221) {$33$};
   \path[uedge] (t2221) edge (t22211);
   \node[unode] (t22221) at ([shift={(-40-90:0.85)}]t2222) {$34$};
   \path[uedge] (t2222) edge (t22221);
   \node[unode] (t22222) at ([shift={(-90:0.85)}]t2222) {$35$};
   \path[uedge] (t2222) edge (t22222);
   \node[unode] (t22231) at ([shift={(-50-90-30:0.85)}]t2223) {$36$};
   \path[uedge] (t2223) edge (t22231);
   \node[unode] (t22232) at ([shift={(-50-90+10:0.85)}]t2223) {$37$};
   \path[uedge] (t2223) edge (t22232);

\node[unode2] (t333) at ([shift={(70:.7)}]t3) {$1$};
   \path[uedge] (t333) edge (t3);
   \node[unode2] (t3332) at ([shift={(-20+90:0.85)}]t333) {$2$};
   \path[uedge] (t333) edge (t3332);
   \node[unode2] (t3333) at ([shift={(-50+90-10:0.85)}]t333) {$3$};
   \path[uedge] (t333) edge (t3333);

\node[unode] (t444) at ([shift={(160:.7)}]t4) {$17$};
   \path[uedge] (t444) edge (t4);

    \node[unode] (t4441) at ([shift={(10-90-90+10:0.85)}]t444) {$18$};
   \path[uedge] (t444) edge (t4441);
   \node[unode] (t4442) at ([shift={(-20-90-90:0.85)}]t444) {$19$};
   \path[uedge] (t444) edge (t4442);
   \node[unode] (t4443) at ([shift={(-50-90-90-10:0.85)}]t444) {$20$};
   \path[uedge] (t444) edge (t4443);

    \node[unode] (t44411) at ([shift={(-90-90:0.85)}]t4441) {$21$};
   \path[uedge] (t4441) edge (t44411);
   \node[unode] (t44412) at ([shift={(10-90-90+30:0.85)}]t4441) {$22$};
   \path[uedge] (t4441) edge (t44412);
   \node[unode] (t44421) at ([shift={(-90-90:0.85)}]t4442) {$23$};
   \path[uedge] (t4442) edge (t44421);
   \node[unode] (t44422) at ([shift={(-40-90-90:0.85)}]t4442) {$24$};
   \path[uedge] (t4442) edge (t44422);
   \node[unode] (t44431) at ([shift={(-50-90-90+10:0.85)}]t4443) {$25$};
   \path[uedge] (t4443) edge (t44431);
   \node[unode] (t44432) at ([shift={(-50-90-90-30:0.85)}]t4443) {$26$};
   \path[uedge] (t4443) edge (t44432);

     \node at(-3.25+14,3-14) {(i)};
  \node[unode] (t) at (0+14,0-14) {$0$}; 
  \node[unode] (t1) at ([shift={(-20:0.8)}]t) {$4$};
  \node[unode] (t2) at ([shift={(-110:.8)}]t) {$3$};
  \node[unode] (t3) at ([shift={(70:0.8)}]t) {$1$};
  \node[unode] (t4) at ([shift={(160:0.8)}]t) {$2$};
  \path[uedge] (t)  edge (t1) edge (t2) edge (t3) edge (t4);

  \node[unode] (t11) at ([shift={(-80:0.7)}]t1) {$38$};
  \path[uedge] (t1) edge (t11);
  \node[unode] (t21) at ([shift={(-170:0.7)}]t2) {$27$};
  \path[uedge] (t2) edge (t21);
  \node[unode] (t31) at ([shift={(10:0.7)}]t3) {$5$};
  \path[uedge] (t3) edge (t31);
  \node[unode] (t41) at ([shift={(100:0.7)}]t4) {$16$};
  \path[uedge] (t4) edge (t41);

  \node[unode2] (t111) at ([shift={(-20:.7)}]t1) {$10$};
   \path[uedge] (t111) edge (t1);
   \node[unode2] (t1111) at ([shift={(10+10:0.85)}]t111) {$11$};
   \path[uedge] (t111) edge (t1111);
   \node[unode2] (t1113) at ([shift={(-50-10:0.85)}]t111) {$12$};
   \path[uedge] (t111) edge (t1113);

   \node[unode2] (t222) at ([shift={(-110:.7)}]t2) {$7$};
   \path[uedge] (t222) edge (t2);
   \node[unode2] (t2221) at ([shift={(10-90+10:0.85)}]t222) {$8$};
   \path[uedge] (t222) edge (t2221);
   \node[unode2] (t2223) at ([shift={(-50-90-10:0.85)}]t222) {$9$};
   \path[uedge] (t222) edge (t2223);

\node[unode2] (t333) at ([shift={(70:.7)}]t3) {$1$};
   \path[uedge] (t333) edge (t3);
   \node[unode2] (t3332) at ([shift={(-20+90:0.85)}]t333) {$2$};
   \path[uedge] (t333) edge (t3332);
   \node[unode2] (t3333) at ([shift={(-50+90-10:0.85)}]t333) {$3$};
   \path[uedge] (t333) edge (t3333);

\node[unode2] (t444) at ([shift={(160:.7)}]t4) {$4$};
   \path[uedge] (t444) edge (t4);

    \node[unode2] (t4441) at ([shift={(10-90-90+10:0.85)}]t444) {$6$};
   \path[uedge] (t444) edge (t4441);
   \node[unode2] (t4443) at ([shift={(-50-90-90-10:0.85)}]t444) {$5$};
   \path[uedge] (t444) edge (t4443);

     \node at(-3.25,3-21) {(j)};
  \node[unode] (t) at (-1,0-20) {$0$}; 
  \node[unode] (t1) at ([shift={(-20:0.8)}]t) {$4$};
  \node[unode] (t2) at ([shift={(-110:.8)}]t) {$3$};
  \node[unode] (t3) at ([shift={(70:0.8)}]t) {$1$};
  \node[unode] (t4) at ([shift={(160:0.8)}]t) {$2$};
  \path[uedge] (t)  edge (t1) edge (t2) edge (t3) edge (t4);

  \node[unode] (t11) at ([shift={(-80:0.7)}]t1) {$38$};
  \path[uedge] (t1) edge (t11);
  \node[unode] (t21) at ([shift={(-170:0.7)}]t2) {$27$};
  \path[uedge] (t2) edge (t21);
  \node[unode] (t31) at ([shift={(10:0.7)}]t3) {$5$};
  \path[uedge] (t3) edge (t31);
  \node[unode] (t41) at ([shift={(100:0.7)}]t4) {$16$};
  \path[uedge] (t4) edge (t41);

  \node[unode2] (t111) at ([shift={(-20:.7)}]t1) {$10$};
   \path[uedge] (t111) edge (t1);

   \node[unode2] (t222) at ([shift={(-110:.7)}]t2) {$7$};
   \path[uedge] (t222) edge (t2);

\node[unode2] (t333) at ([shift={(70:.7)}]t3) {$1$};
   \path[uedge] (t333) edge (t3);

\node[unode2] (t444) at ([shift={(160:.7)}]t4) {$4$};
   \path[uedge] (t444) edge (t4);

     \node at(-3.25+6,3-21) {(k)};
  \node[unode] (t) at (0+3,0-20) {$0$}; 
  \node[unode] (t1) at ([shift={(-20:0.8)}]t) {$4$};
  \node[unode2] (t2) at ([shift={(-110:.8)}]t) {$7$};
  \node[unode] (t3) at ([shift={(70:0.8)}]t) {$1$};
  \node[unode] (t4) at ([shift={(160:0.8)}]t) {$2$};
  \path[uedge] (t)  edge (t1) edge (t2) edge (t3) edge (t4);

  \node[unode] (t11) at ([shift={(-80:0.7)}]t1) {$38$};
  \path[uedge] (t1) edge (t11);
  \node[unode] (t21) at ([shift={(-170:0.7)}]t2) {$27$};
  \path[uedge] (t2) edge (t21);
  \node[unode] (t31) at ([shift={(10:0.7)}]t3) {$5$};
  \path[uedge] (t3) edge (t31);
  \node[unode] (t41) at ([shift={(100:0.7)}]t4) {$16$};
  \path[uedge] (t4) edge (t41);

  \node[unode2] (t111) at ([shift={(-20:.7)}]t1) {$10$};
   \path[uedge] (t111) edge (t1);

\node[unode2] (t333) at ([shift={(70:.7)}]t3) {$1$};
   \path[uedge] (t333) edge (t3);

\node[unode2] (t444) at ([shift={(160:.7)}]t4) {$4$};
   \path[uedge] (t444) edge (t4);

     \node at(-3.25+9,3-21) {(l)};
  \node[unode2] (t) at (0+7,0-20) {$7$}; 
  \node[unode] (t1) at ([shift={(-20:0.8)}]t) {$4$};
  \node[unode] (t3) at ([shift={(70:0.8)}]t) {$1$};
  \node[unode] (t4) at ([shift={(160:0.8)}]t) {$2$};
  \path[uedge] (t)  edge (t1)  edge (t3) edge (t4);

  \node[unode] (t11) at ([shift={(-80:0.7)}]t1) {$38$};
  \path[uedge] (t1) edge (t11);
  \node[unode] (t31) at ([shift={(10:0.7)}]t3) {$5$};
  \path[uedge] (t3) edge (t31);
  \node[unode] (t41) at ([shift={(100:0.7)}]t4) {$16$};
  \path[uedge] (t4) edge (t41);

  \node[unode2] (t111) at ([shift={(-20:.7)}]t1) {$10$};
   \path[uedge] (t111) edge (t1);

\node[unode2] (t333) at ([shift={(70:.7)}]t3) {$1$};
   \path[uedge] (t333) edge (t3);

\node[unode2] (t444) at ([shift={(160:.7)}]t4) {$4$};
   \path[uedge] (t444) edge (t4);

     \node at(-3.25+13,3-21) {(m)};
  \node[unode2] (t) at (0+10,0-20) {$7$}; 
  \node[unode] (t1) at ([shift={(-20:0.8)}]t) {$4$};
  \node[unode] (t3) at ([shift={(70:0.8)}]t) {$1$};
  \node[unode2] (t4) at ([shift={(160:0.8)}]t) {$4$};
  \path[uedge] (t)  edge (t1)  edge (t3) edge (t4);

  \node[unode] (t11) at ([shift={(-80:0.7)}]t1) {$38$};
  \path[uedge] (t1) edge (t11);
  \node[unode] (t31) at ([shift={(10:0.7)}]t3) {$5$};
  \path[uedge] (t3) edge (t31);

  \node[unode2] (t111) at ([shift={(-20:.7)}]t1) {$10$};
   \path[uedge] (t111) edge (t1);

\node[unode2] (t333) at ([shift={(70:.7)}]t3) {$1$};
   \path[uedge] (t333) edge (t3);

     \node at(-3.25+16,3-21) {(n)};
  \node[unode2] (t) at (0+13,0-20) {$7$}; 
  \node[unode2] (t1) at ([shift={(-20:0.8)}]t) {$10$};
  \node[unode] (t3) at ([shift={(70:0.8)}]t) {$1$};
  \node[unode2] (t4) at ([shift={(160:0.8)}]t) {$4$};
  \path[uedge] (t)  edge (t1)  edge (t3) edge (t4);



\node[unode2] (t333) at ([shift={(70:.7)}]t3) {$1$};
   \path[uedge] (t333) edge (t3);

     \node at(-3.25+19,3-21) {(o)};
  \node[unode2] (t) at (0+16,0-20) {$7$}; 
  \node[unode2] (t1) at ([shift={(-20:0.8)}]t) {$10$};
  \node[unode2] (t3) at ([shift={(70:0.8)}]t) {$1$};
  \node[unode2] (t4) at ([shift={(160:0.8)}]t) {$4$};
  \path[uedge] (t)  edge (t1)  edge (t3) edge (t4);
  \end{tikzpicture}
  }
  \caption{In this figure, we show the progression of the graph state as the available emitters absorb the photons. The initial
state is given in (a).}
\label{fig:example}
  \end{figure*}
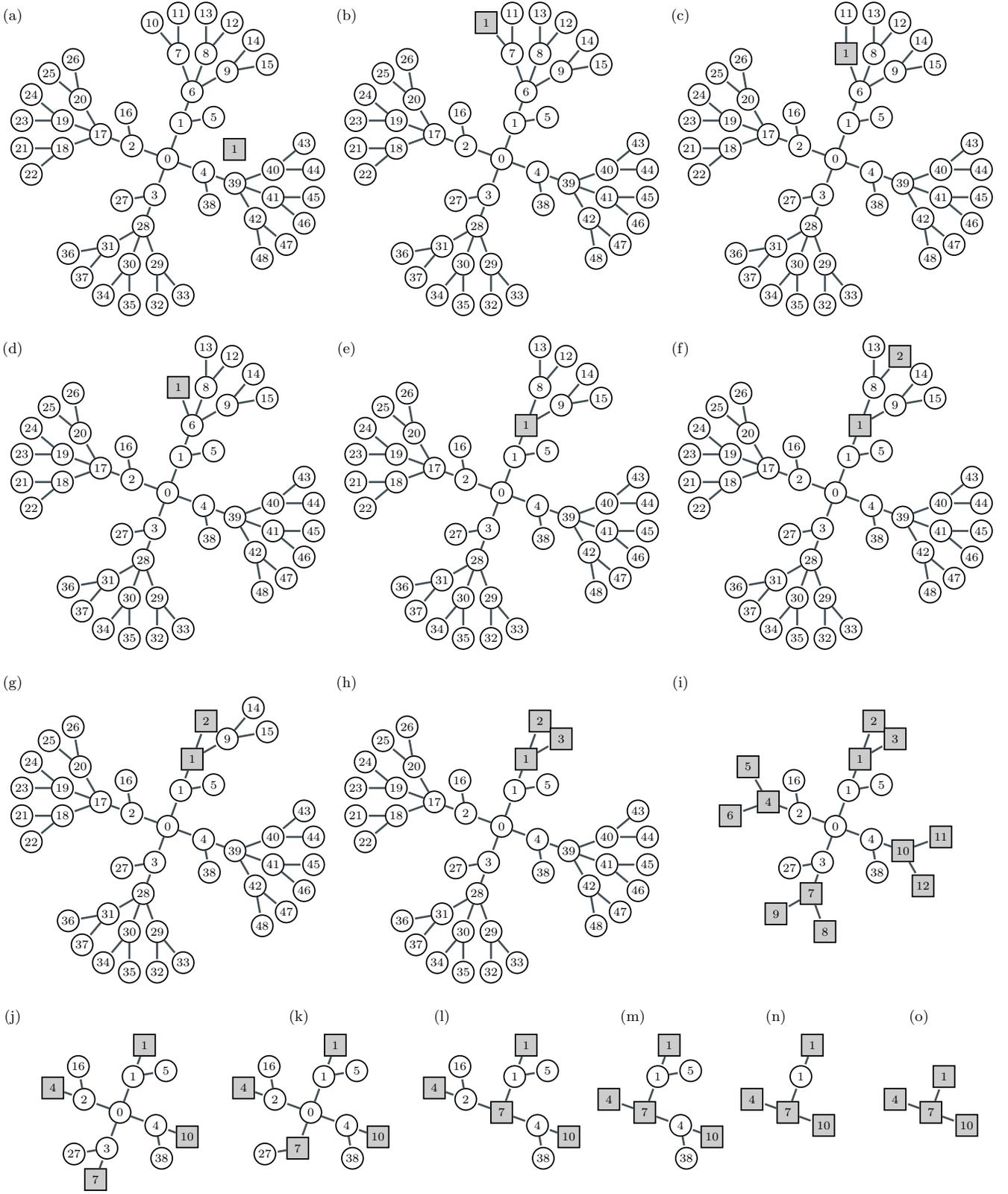

The initial graph state $\ket{G}$ is given in Figure~\ref{fig:example}(a). 
The emitter $1_e$ replaces $10_p$ via SFE to yield the graph in Figure.~\ref{fig:example}(b). The $\textbf{N}_{G}(1_e) = 7_p$, then the emitter $1_e$ can absorb $7_p$ to yield Figure~\ref{fig:example}(c). The $\textbf{N}_{G}(11_p) = 1_e$, then the emitter $1_e$ can absorb $11_p$ to yield Figure~\ref{fig:example}(d). Next, $\textbf{N}_{G}(1_e) = 6_p$, implying that the emitter $1_e$ can absorb $6_p$ to obtain the graph in Figure~\ref{fig:example}(e). We next see that the emitter $1_e$ can absorb no further. At this point, we use the initial conditions, and the emitter $2_e$ replace the photon $12_p$. We can then, by looking the photon absorption criterion given above, see that the emitter $2_e$ can absorb $8_p$ and $13_p$ to yield the graph given in Figure~\ref{fig:example}(g). Now, the emitter $1_e, 2_e$ can absorb no further photons. The emitter $3_e$ replaces the $14_p$ photons and absorbs $9_p$ and $15_p$ photons to yield the graph given in Figure~\ref{fig:example}(h). In this manner, we use the $12$ emitters to start by absorbing at the end nodes. This gives us the graph in Figure~\ref{fig:example}(i). Next, we no longer have any free emitters available, and we start with emitter disentanglement. All the edges between the emitters are removed by the application of $\operatorname{CNOT}$ gates between quantum emitters to yield the graph given in Figure~\ref{fig:example}(j). The absorption process continues until all the photons have been absorbed to yield the graph in Figure~\ref{fig:example}(o). Then, the emitters are disentangled to yield a completely unentangled state.

\subsection{Circuit Depth}\label{sec:circuit_depth}

In both algorithms presented above, the only step that involves $\operatorname{CNOT}$ operations between quantum emitters $${CNOT}{e,e}$$ is the emitter disentanglement step. The $\operatorname{CNOT}{e,e}$ gate is the most time-consuming resource in this architecture. Therefore, calculating the circuit depth in terms of $\operatorname{CNOT}{e,e}$ is critical.

In the algorithm specified above, the sequence in which $\operatorname{CNOT}{e,e}$ gates are applied during the emitter disentanglement step is not fixed. Ideally, we aim to parallelize these $\operatorname{CNOT}{e,e}$ operations to minimize the circuit depth. The following algorithm calculates the circuit depth without claiming optimality. Given the list of $\operatorname{CNOT}_{e,e}$ gates applied during the emitter disentanglement step, we assume that the gates can be applied in any sequence.

For each emitter disentanglement step, the algorithm proceeds as follows:

\begin{enumerate} \item From the list of $\operatorname{CNOT}{e,e}$ gates, select the gates that act on independent emitters. These gates can be applied in parallel within one time step. \item Remove the selected gates from the list of $\operatorname{CNOT}{e,e}$ gates. \item Repeat steps 1–2 until no gates remain in the list. \item The number of repetitions corresponds to the $\operatorname{CNOT}$ depth for the emitter disentanglement circuit. \end{enumerate}

The total $\operatorname{CNOT}$ circuit depth for the entire state generation circuit is then obtained by summing the $\operatorname{CNOT}$ circuit depths across all emitter disentanglement steps.

In step 1, there are multiple possible ways to choose independent $\operatorname{CNOT}_{e,e}$ gates. The algorithm does not attempt to optimize this selection.

\subsubsection{Example: Calculation for CNOT Depth}
\begin{figure}[htb]
    \centering
    \includegraphics[scale=0.35]{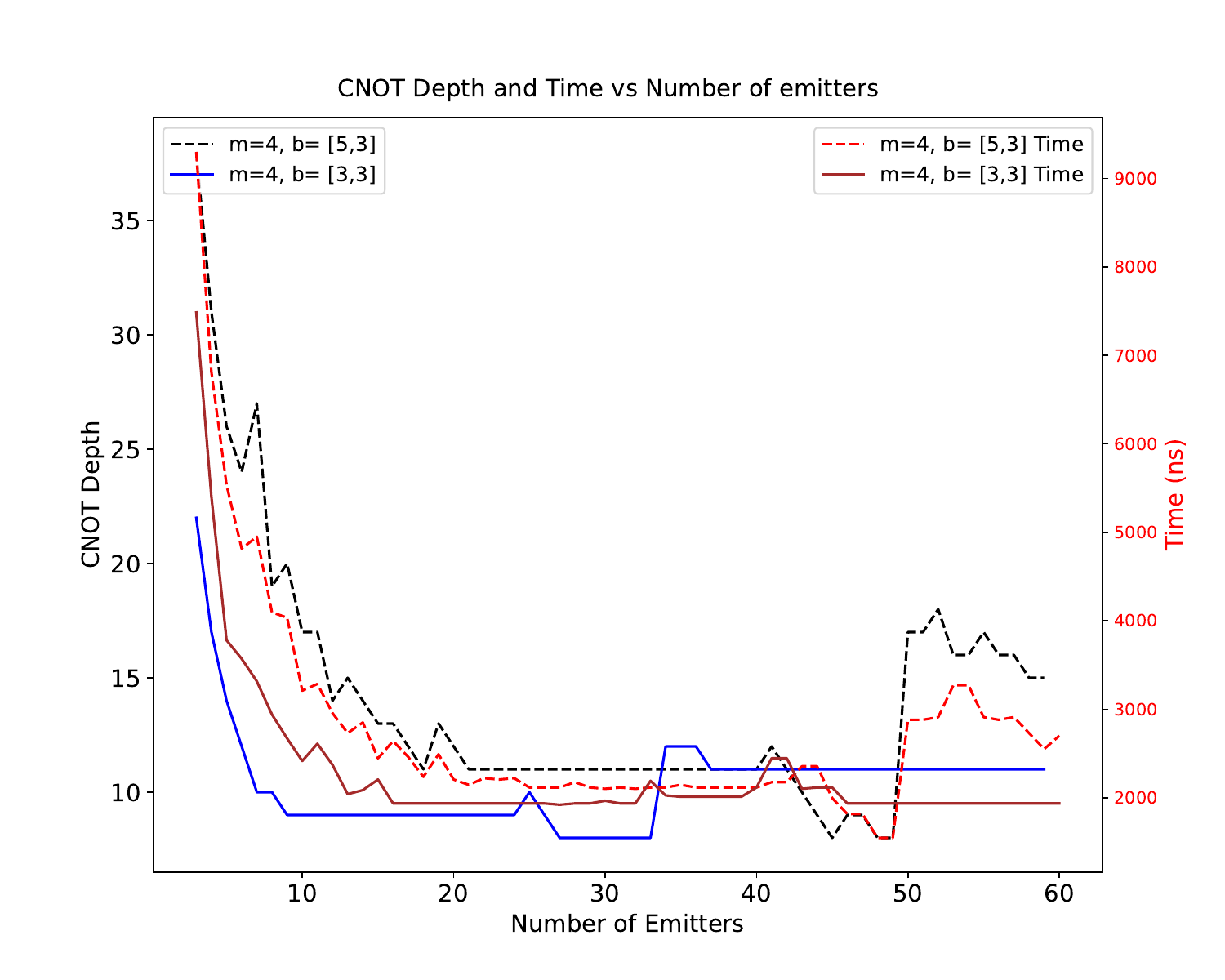}
    \caption{Emitter vs. CNOT depth for preparing regular-trees-on-clique repeater graph states from Refs.~\cite{Azuma2015, Mihir2017}, characterized by an integer $m$ and branching vector ${\vec b}$.}
    \label{fig:emitter_vs_depth}
\end{figure}

Figure~\ref{fig:emitter_vs_depth}, shows the $\operatorname{CNOT}_{e,e}$ depth and total generation time for the state vs. the number of emitters for various repeater graph states. Note that, in general, an increase in the number of emitters does not imply a decrease in the $\operatorname{CNOT}_{e,e}$ depth, as discussed in Appendix~\ref{sec:CNOT_depth}. For example, one can generate a $n$-qubit GHZ state with a single emitter, making the circuit's $\operatorname{CNOT}_{e,e}$ depth zero. However, when given $n$-emitters, we can first prepare the graph state on the quantum emitters, transfer the state to the photons, and then measure the emitters. The CNOT depth of this circuit is two. For further discussion on the behavior of the $\operatorname{CNOT}_{e,e}$ depth, refer to Appendix~\ref{sec:CNOT_depth}.
   In Appendix~\ref{appen: GHZ_generation}, we explore the dependence of the total generation time for a GHZ equivalent graph state with the total number of emitters used for generation. 
  
  In Figure~\ref{fig:emitter_vs_depth}, we have used both Algorithm 1 and Algorithm 2 for $n_e$ emitters and then plotted the minimal $\operatorname{CNOT}_{e,e}$ depth for each $n_e$. We have minimized over two initial conditions. For the first one, we begin with the outer level nodes of the attached trees, followed by the second level nodes of the tree, the root nodes, and finally, the clique nodes. To obtain the second initial condition, we reverse the first initial condition.

\section{Emission times of photons}
 To the algorithm given in Section~\ref{sec:final_algo}, we can associate a circuit composed of the following gate-set: 
\begin{itemize}
    \item Hadamard gate on the emitter ($\operatorname{H}_e$). Time required to perform $\operatorname{H}_e$ -  $t_{\textrm{H}_e}$.
    \item CNOT gate with the emitter as the control and the photonic qubit as the target ($\operatorname{CNOT}_{e,p}$). Time required - $t_{\textrm{CNOT}_{e,p}}$. This gate models the emission of the photon.
    \item CNOT gate between two emitters ($\operatorname{CNOT}_{e,e}$). Time required $t_{\textrm{CNOT}_{e,e}}$. 
    
    \item Measuring an emitter in the computational basis. Time taken $t_{\textrm{meas}}$.
    \item While not a part of gate set, after each measurement, we need to initialize the emitter in the $\ket{0}$ state. Time required - $t_{\textrm{init}}$.
\end{itemize}

We evaluate these timescales for SiV in Appendix~\ref{sec:apx}. We also outline the procedure for calculating the emission time of each photon in the graph state in Appendix~\ref{Emission times}. This information is useful in modeling the losses on the photons before they can be used for the desired application, as shown in Section~\ref{subsec:timing}. 


\section{All-photonic quantum repeater architecture}
\label{sec:repeater}
One of the primary applications of photonic graph states is as the resource state for the all-photonic quantum repeaters. Each repeater in this architecture is equipped with photonic dual-rail graph states. Each qubit of the graph state is a logical qubit encoded in a tree code~\cite{varnava2006loss, Mihir2017,Azuma2015}. In dual-rail photonic encoding, photon loss results in qubit loss error. The tree code protects the qubits of the graph state from losses, mimicking a quantum memory. Entanglement distribution rates and the maximum key rates over a lossy bosonic channel such as an optical fiber or free space link are known to drop exponentially with loss \cite{Takeoka2014}. The all-photonic quantum repeater architecture with tree-encoded graph state (referred to as repeater graph state (RGS) onwards) outperforms the maximum key rate obtained without using repeaters or the repeaterless rate given by $R_{\textrm{direct}}=-\log_2(1 - \eta)$ ebits per mode~\cite{Pirandola2017}, (see also \cite{Wilde2017} for a strong converse bound). Here, $\eta$ is the transmissivity of the optical fiber and is proportional to the length of the fiber, $L$ and its loss coefficient $\alpha $ $(\eta =e^{(-\alpha L)})$. 

In \cite{Mihir2017}, first, single photons are generated using emitters. Then, they are entangled using linear optical Bell state measurements (BSMs) that are probabilistic and multiplexed to create the RGS. The probabilistic entangling operation results in a massive overhead in the number of emitters required to produce one copy of the RGS. This section compares the performance of the RGS, created deterministically using our algorithm, with the RGS generated using probabilistic BSMs and multiplexing (hereafter referred to as the multiplexing method) in terms of (1) entanglement generation rate and (2) the number of emitters used per repeater.

\subsection{The protocol}
\begin{figure*}[htb]
    \centering
    \includegraphics[scale=0.7]{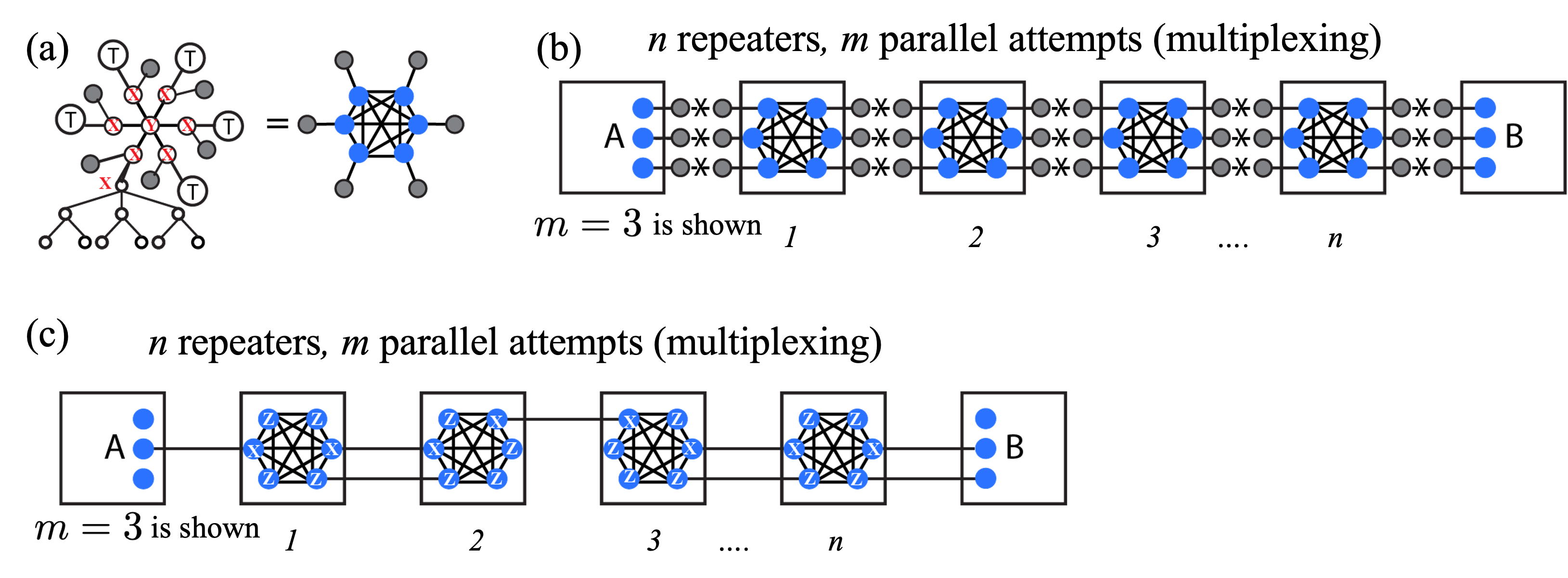}
    \caption{All-photonic quantum repeater architecture (a) The repeater graph state with tree encoded logical qubits (blue circles) and link qubits (grey circles) (b) A chain of $n$ repeater with multiplexing $m=3$, placed between the users - A and B. The link qubits are sent over the optical channel and they meet at the minor nodes (denoted by `x') to undergo a linear optical BSM (c) The solid black lines showing links generated from the successful BSMs at the minor nodes.}
    \label{fig:repSchem}
\end{figure*}

We begin by reviewing the all-photonic quantum repeater protocol. FIG.~\ref{fig:repSchem}(b) shows a chain of $n$ all-photonic quantum repeaters placed equidistant between the users Alice and Bob with $m$ parallel optical channels connecting each pair of repeaters. We refer to $m$ as the degree of multiplexing. In an all-photonic quantum repeater protocol, once the repeater graph state (RGS) (see FIG.~\ref{fig:repSchem}(a)) is generated at every repeater, the grey qubits or the \textit{link} qubits from the RGS are sent over the optical channels. The link qubits from the neighboring repeaters meet at the minor nodes, placed halfway between the neighboring repeaters, and undergo a photonic Bell state measurement (BSM). This measurement succeeds with probability $p$. If the users are placed distance $L$ apart, $p=\eta^{1/(n+1)}.p_{\textrm{BSM}}$, where $p_{\textrm{BSM}}$ is the success probability of the linear optical BSM and $\eta = \exp(-\alpha L)$. In other words, the BSM at the minor nodes succeeds if both the qubits undergoing BSM reach the minor node and the BSM itself is successful. For a simple linear optical system,  $p_{\textrm{BSM}}=50\%$, which can be boosted using ancilla single photons. If a BSM between the link qubits is successful, we say a \textit{link} was established. The success or failure outcomes of the BSMs are classically communicated to the respective neighboring repeaters. 

Once the repeaters receive the classical communication regarding the BSM outcomes, every repeater performs $X$ measurements on a pair of logical qubits with successfully heralded links on the opposite sides of the repeater and $Z$ measurements on the remaining $2m-2$ logical qubits in the graph state. These measurements are probabilistic as they are performed on lossy photonic qubits. If all measurements at every repeater are successful and at least one BSM succeeds at every minor node, users Alice and Bob end up with a shared Bell state. The entanglement generation rate is given by~\cite{Mihir2017}
\begin{align}
    R&=\frac{P_X^{2n}P_Z^{2{(m-1)n}}[1-(1-p)^m]^{(n+1)}}{2m\tau} \textrm{ebits/s}
    \label{eq:rate}
\end{align}
Here, $P_X$ and $P_Z$ are the probabilities of success of the logical single qubit $X$ and $Z$ measurements at the repeaters, respectively, and $\tau$ is the repetition time of the protocol. Eq.~\ref{eq:rate} assumes that qubits in all repeaters have identical $P_X$ and $P_Z$. The success probabilities of the Pauli measurements depend upon the shape and the size of the tree code used as we discuss in the following section.

\subsection{Tree code}
\label{sub:treeCode}
\begin{figure}[htb]
    \centering
    \includegraphics[scale = 0.5]{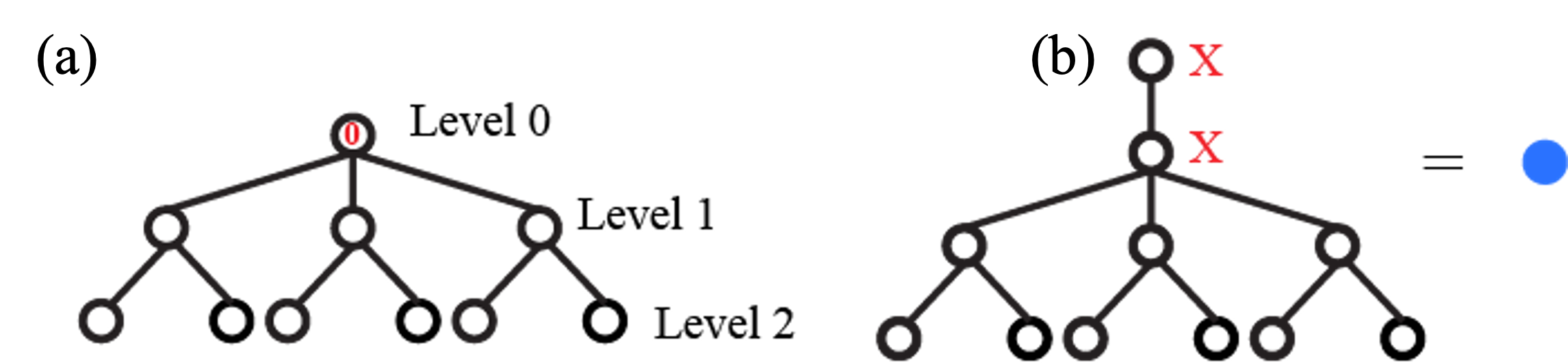}
    \caption{Tree code (a) Tree graph state with node 0 as the root and the branching vector $b = [3,2]$. (b) A logical qubit encoded in a tree code (shown as a blue circle) is created by first attaching the qubit to be encoded to a tree graph state using CZ gate, followed by $X$ measurements on the qubit to be encoded and the root of the tree. }
    \label{fig:treeEncoding}
\end{figure}

In this section, we derive the success probabilities of Pauli measurements on the logical qubit of a tree code assuming the qubits in the tree code have non-uniform loss probabilities. As discussed in Section~\ref{sec:notation}, we define a tree graph state using branching vector $b \equiv [b_0,b_1, \cdots,b_d]$. We define the root (labeled as qubit 0) of the tree as the qubit on level 0 of the tree as shown in FIG.~\ref{fig:treeEncoding}(a). Let $l_{i}$ be the probability of loss of qubit $i$ of the tree and $\mathcal{C}(i)$ be the set of \textit{children} of $i$, i.e., the qubits one level below $i$. Similarly, $i$ is the \textit{parent} of qubits in $\mathcal{C}(i)$. In order to encode a physical qubit into a tree code, a tree graph state is first attached to the physical qubit using a CZ gate as shown in FIG.~\ref{fig:treeEncoding}(b). $X$ measurements on the tree's root and the physical qubit encode the physical qubit into the logical qubit. 

The tree code protects the logical qubit from loss using the \textit{counterfactual} error correction scheme~\cite{varnava2006loss}. This scheme aims to infer a Pauli measurement result in the event a qubit of a graph state is lost by performing measurements on the other qubits in the graph state.
For example, consider a graph state stabilizer $Z_iX_j\prod_{k\in \mathcal{C}(j)}Z_k$. If the Pauli operators in this stabilizer are measured, the product of all measurement outcomes is 1. Using this property, if qubit $i$ is lost, the $Z$ measurement outcome of $i$ can be inferred from the outcomes of $X$ measurement on $j$ and $Z$ measurements on the set of qubits in $\mathcal{C}(j)$. This is an \textit{indirect}-$Z$ measurement. Note that, direct-$Z$ measurement succeeds if the qubit is not lost. For qubit $i$, the probability of success of direct or indirect-$Z$ measurement is given by ~\cite{Mihir2017,Azuma2015},
\begin{align}
   P_{Z_{i}} = (1-l_{i})+l_{i}\xi_{i}
    \label{eq:pz}
\end{align}
Here, $\xi_{i}$ is the success probability of indirect-$Z$ measurement. We perform indirect-$Z$ measurement on a qubit $i$ in a tree graph state using stabilizers of the form  $Z_iX_j\prod_{k\in\mathcal{C}(j)}Z_k$, $j\in\mathcal{C}(i)$. Out of the $\vert \mathcal{C}(i)\vert$ possible attempts of an indirect-$Z$ measurement on $i$, at least one must succeed. The success probability of an attempt is $(1-l_{j})\prod_{k\in \mathcal{C}(j)} P_{Z_{k}}$. Here, $(1-l_{j})$ is the success probability of $X$ measurement on $j$. If $i$ is not on the $(d+1)^{\textrm{th}}$ (last) level of the tree, using recursion, we can write: 
\begin{align} 
    \xi_{i} = 1-\prod_{j\in \mathcal{C}(i)}\Bigg[1-(1-l_{j})\prod_{k\in \mathcal{C}(j)} P_{Z_{k}} \Bigg]
\end{align}
We set $\xi_{i}=0$ if $i$ is on the $(d+1)^{\textrm{th}}$ level of the tree, as $i$ does not have any children, and indirect-$Z$ measurement cannot be performed without children. 

The probabilities of successful logical $Z$ and $X$ measurements are ~\cite{Mihir2017,Azuma2015},
\begin{align}
    P_Z =\prod_{i\textrm{ on level 1}}P_{Z_{i}}
    \label{eq:Pz}
\end{align}
\begin{align}
    P_X = \xi_{0}
    \label{eq:Px}
\end{align}
i.e., the $Z$ measurement probability of the logical qubit is the product of $Z$ measurement probabilities on all the level 1 qubits and the $X$ measurement probability of the logical qubit is the indirect-$Z$ measurement probability on the root qubit.

This paper considers tree code with a branching vector of length two. The following section discusses an all-photonic quantum repeater architecture with RGSs generated using the algorithm described in Section~\ref{sec:final_algo}.

\subsection{Emitter-based repeater architecture}
\label{subsec:timing}
As the entanglement rate is inversely proportional to qubit loss, designing a repeater architecture that reduces the losses becomes essential. When quantum emitters are used as single photon sources, the multiplexing method generates the RGS in time $t_{\rm init}$, in the limit that initializing emitters is much slower than the photonic chip that performs linear optical operations. The entire RGS is created in the same time step, resulting in identical loss probability for every qubit. The emitter-based method emits qubits of the RGS at different times. This section discusses the photonic qubit measurement sequence on the RGS that minimizes the loss and calculates the loss probability of the RGS's qubits for the emitter-based method. It also derives the number of emitters required for the emitter-based method to attain a given repetition time.

\subsubsection{Photonic qubit measurement sequence}
\label{subsec:measSeq}
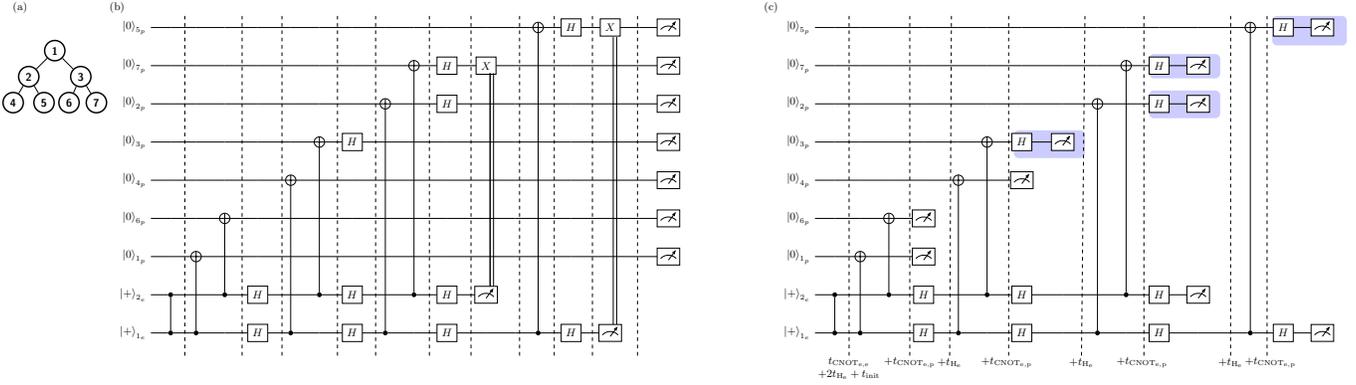
\begin{figure*}[htb]
    \centering
    \resizebox{\textwidth}{!}{%
    \begin{tikzpicture}[roundnode/.style={circle, draw = black, very thick,  minimum size=1mm,font=\sffamily\small\bfseries},edgestyle/.style={draw=black, thick}]
    \def \n {19}
     \path [rounded corners,fill=blue!20] (2.65+\n, 2.05) rectangle (4.7+\n, 2.85) {};
      \path [rounded corners,fill=blue!20] (2.65+\n, 3.05+0.15) rectangle (4.7+\n, 2.9+1) {};
       \path [rounded corners,fill=blue!20] (-1.25+\n, 2.05-.75-.4) rectangle (.8+\n, 2.1-.4) {};
    \path [rounded corners,fill=blue!20] (4+2.2+\n, 4.15) rectangle (5.85+2.5+\n, 5) {};
\node[roundnode]  (1)  at (-10+0.1,0+4) {1};
\node[roundnode]  (2)  at (-10.75+0.1,-.75+4) {2};
\node[roundnode]  (3)  at (-9.25+0.1,-.75+4) {3};
\node[roundnode]  (4)  at (-11.2+0.1,-1.5+4) {4};
\node[roundnode]  (5)  at (-10.31+0.1,-1.5+4) {5};
\node[roundnode]  (6)  at (-9.69+0.1+0.1,-1.5+4) {6};
\node[roundnode]  (7)  at (-8.8+0.1,-1.5+4) {7};
\draw[edgestyle] (1) edge (2);
\draw[edgestyle] (1) edge (3);
\draw[edgestyle] (4) edge (2);
\draw[edgestyle] (5) edge (2);
\draw[edgestyle] (6) edge (3);
\draw[edgestyle] (7) edge (3);
\node at (-10.9,5.25){(a)};
    \node at (-8.1,5.25){(b)};
    \node at (-8.25+\n,5.25){(c)};
        \node at (0,0) {
        \begin{quantikz}
        \lstick{$\ket{0}_{5_p}$} & \qw  & \qw  & \qw & \qw & \qw & \qw & \qw& \qw & \qw & \qw & \qw  & \qw  & \targ{} &\gate{H}&\gate{X}& \qw  &\meter{} \\
        \lstick{$\ket{0}_{7_p}$} & \qw  & \qw  & \qw & \qw & \qw& \qw & \qw & \qw  & \targ{} &\gate{H}&\gate{X} & \qw & \qw & \qw & \qw & \qw  &\meter{}\\
        \lstick{$\ket{0}_{2_p}$} & \qw & \qw & \qw & \qw & \qw & \qw & \qw & \targ{} & \qw  &\gate{H} & \qw & \qw & \qw & \qw & \qw & \qw &\meter{}\\
        \lstick{$\ket{0}_{3_p}$} & \qw  & \qw  & \qw & \qw & \qw & \targ{} &\gate{H}& \qw & \qw & \qw & \qw & \qw & \qw & \qw  & \qw  & \qw &\meter{} \\
        \lstick{$\ket{0}_{4_p}$} & \qw & \qw & \qw & \qw & \targ{} & \qw & \qw & \qw & \qw & \qw & \qw & \qw & \qw & \qw & \qw & \qw &\meter{} \\
        \lstick{$\ket{0}_{6_p}$} & \qw  & \qw & \targ{} & \qw & \qw & \qw & \qw & \qw & \qw & \qw & \qw & \qw & \qw & \qw & \qw & \qw &\meter{} \\
        \lstick{$\ket{0}_{1_p}$} & \qw & \targ{} & \qw & \qw & \qw & \qw & \qw & \qw& \qw & \qw & \qw & \qw & \qw & \qw & \qw & \qw &\meter{} \\
        \lstick{{$\ket{+}_{2_e}$}} & \ctrl{} & \qw & \ctrl{-2}&\gate{H}&\qw & \ctrl{-4}&\gate{H} &\qw & \ctrl{-6}&\gate{H} &\meter{} \\
	\lstick{{$\ket{+}_{1_e}$}} & \ctrl{-1}& \ctrl{-2} &\qw &\gate{H}  & \ctrl{-4}  &\qw &\gate{H}& \ctrl{-6}  &\qw &\gate{H} & \qw & \qw & \ctrl{-8}&\gate{H}&\meter{} \\
        \end{quantikz}
        };

        \draw [dashed] (-6.15,-4.8 ) -- (-6.15,5);
        \draw [dashed] (-4.5,-4.8 ) -- (-4.5,5);
        \draw [dashed] (-3.35,-4.8 ) -- (-3.35,5);
        \draw [dashed] (-1.75,-4.8 ) -- (-1.75,5);
        \draw [dashed] (-.65,-4.8 ) -- (-0.65,5);
        \draw [dashed] (.9,-4.8 ) -- (.9,5);
        \draw [dashed] (2.1,-4.8 ) -- (2.1,5);
        \draw [dashed] (3.5,-4.8 ) -- (3.5,5);
        \draw [dashed] (4.5,-4.8 ) -- (4.5,5);
        \draw [dashed] (5.6,-4.8 ) -- (5.6,5);
        \draw [dashed] (6.9,-4.8 ) -- (6.9,5);
        \draw  (2.65,-2.8 ) -- (2.65,3.35);
        \draw  (2.75,-2.8 ) -- (2.75,3.35);
        \draw  (2.65,-2.8 ) -- (2.65,3.35);
        \draw  (6.2,-3.9 ) -- (6.2,4.39);
        \draw  (6.3,-3.9 ) -- (6.3,4.39);

        \node at (\n,0) {
        \begin{quantikz}
        \lstick{$\ket{0}_{5_p}$} & \qw  & \qw  & \qw & \qw & \qw & \qw & \qw& \qw & \qw & \qw & \qw  & \qw  &\qw & \targ{} &\gate{H} &\meter{} \\
        \lstick{$\ket{0}_{7_p}$} & \qw  & \qw  & \qw & \qw & \qw& \qw & \qw & \qw &\qw  & \targ{} &\gate{H} &\meter{}\\
        \lstick{$\ket{0}_{2_p}$} & \qw & \qw & \qw & \qw & \qw & \qw & \qw& \qw & \targ{} & \qw  &\gate{H} &\meter{}\\
        \lstick{$\ket{0}_{3_p}$} & \qw  & \qw  & \qw & \qw & \qw & \targ{} &\gate{H} &\meter{} \\
        \lstick{$\ket{0}_{4_p}$} & \qw & \qw & \qw & \qw & \targ{}  & \qw &\meter{} \\
        \lstick{$\ket{0}_{6_p}$} & \qw  & \qw & \targ{}  &\meter{} \\
        \lstick{$\ket{0}_{1_p}$} & \qw & \targ{} &  \qw &\meter{} \\
        \lstick{{$\ket{+}_{2_e}$}} & \ctrl{} & \qw & \ctrl{-2}&\gate{H}&\qw & \ctrl{-4}&\gate{H} &\qw &\qw & \ctrl{-6}&\gate{H} &\meter{} \\
	\lstick{{$\ket{+}_{1_e}$}} & \ctrl{-1}& \ctrl{-2} & \qw  &\gate{H} & \ctrl{-4} & \qw  &\gate{H} & \qw &  \ctrl{-6}& \qw  &\gate{H} & \qw & \qw & \ctrl{-8}&\gate{H}&\meter{} \\
        \end{quantikz}
        };
        \draw [dashed] (-6.+\n,-4.8 ) -- (-6+\n,5);
        \node at (-6+\n,-5) {$t_{\rm CNOT_{e,e}}$};
        \node at (-6+\n,-5.35) {$+2t_{\rm H_e}+t_{\rm init}$};
        \draw [dashed] (-4.25+\n,-4.8 ) -- (-4.25+\n,5);
        \node at (-4.25+\n,-5) {$+t_{\rm CNOT_{e,p}}$};
        \draw [dashed] (-3.35+.25+\n,-4.8 ) -- (-3.3+.25+\n,5);
        \node at (-3.1+\n,-5) {$+t_{\rm H_{e}}$};
        \draw [dashed] (-1.65+.25+\n,-4.8 ) -- (-1.65+.25+\n,5);
         \node at (-1.45+\n,-5) {$+t_{\rm CNOT_{e,p}}$};
        \draw [dashed] (-.6+.55+.75+\n,-4.8 ) -- (-0.6+.65+.75+\n,5);
        \node at (-.6+.55+.75+\n,-5) {$+t_{\rm H_{e}}$};
        \draw [dashed] (2.+.5+\n,-4.8 ) -- (2.+.5+\n,5);
        \node at (2.45+\n,-5) {$+t_{\rm CNOT_{e,p}}$};
        \draw [dashed] (4.5+0.5+\n,-4.8 ) -- (4.5+0.5+\n,5);
        \node at (5+\n,-5) {$+t_{\rm H_{e}}$};
        \draw [dashed] (5.95+.1+\n,-4.8 ) -- (5.95+.1+\n,5);
        \node at (5.95+.2+\n,-5) {$+t_{\rm CNOT_{e,p}}$};
       
    \end{tikzpicture}
    }
    \caption{Quantum circuit for graph state generation (a) tree graph state with $b = [2,2]$ (b) the quantum circuit to generate tree graph state on (a) using two emitters followed by measurement of photonic qubits in arbitrary Pauli bases. Operations between two dashed vertical lines are concurrent. (c) Simplified quantum circuit in (b) to minimize the qubit losses. The blue rectangles denote Hadamard-rotated measurements on the emitted photons. These measurements are instantaneous. The time needed to perform operations on the emitters between two dashed lines is noted below the lines. }
    \label{fig:measSeq}
\end{figure*}

 In an all-photonic architecture, the photon loss probability is directly proportional to the lifetime of the photon, i.e., the time between photon generation and measurement. In this section, we outline some properties of the generation scheme that help mitigate the photon loss. Consider a quantum circuit to generate a graph state using two emitters as shown in FIG.~\ref{fig:measSeq}(b). From Section~\ref{sec:algo_primitives}, Hadamard, Pauli-$X$, and identity are the only operations performed on the photonic qubits of any graph state after they are emitted. In FIG.~\ref{fig:measSeq}(b), the photonic qubits are measured in the Pauli basis after generating the entire graph state. Since the measurements commute with operations on other qubits commute, the order of the measurements does not matter. FIG.~\ref{fig:measSeq}(c) shows an equivalent quantum circuit, s.t. the photonic qubits are measured as soon as the emitter emits the photons. We eliminate the conditional Pauli-$X$ gate as it only affects the phase of the generated state, which can be tracked using classical post-processing. Moreover,  we rotate the measurement bases of the photonic qubits instead of performing the Hadamard gates. 

Note that we have assumed above that the measurement bases of all qubits are pre-decided. If the measurements are adaptive, i.e., the measurement basis of a qubit depends upon the measurement outcome of another qubit, we must modify the measurement sequence accordingly. For example, consider qubits 2 and 4 from FIG.~\ref{fig:measSeq}(a) such that the measurement outcome of qubit 2 determines the measurement basis of qubit 4. In this case, we would have to measure qubit 4 after qubit 2, even if it is emitted earlier. This increases the loss probability of qubit 4. Consequently, an all-photonic repeater with fewer adaptive measurements performs better. In the following section, we outline the measurement sequence on the emitted RGS and calculate the loss introduced on each photon. 

\subsubsection{Loss calculations}
\label{subsec:loss}

At the beginning of our protocol, the quantum emitters in all repeaters simultaneously start generating the RGS. The repeaters send link qubits to the minor nodes as soon as they are emitted. Let $\tau_l = L/(2c_f(n+1))$ be the time for the link qubits to reach the minor nodes. Here, $c_f$ is the speed of light in the optical fiber, and $L$ is the distance between the end nodes. Each minor node performs BSM immediately after it receives a pair of qubits from both sides and classically communicates the BSM outcome to the neighboring repeaters. Let $T_l$ be the time the last link qubit is emitted. The repeaters have all BSM outcomes at time $T_l+2\tau_l$. This is when the repeaters know the measurement bases of all the logical qubits and, hence, all physical qubits in the tree codes. At this point, the logical $X$ and $Z$ measurements start at the repeaters. Note that, in our architecture, we assume that the indirect-$Z$ measurement is non-adaptive. If we require a $Z$ measurement outcome of a qubit at the repeater, irrespective of whether or not it is lost, we perform the indirect-$Z$ measurement sequence on its children. This design choice avoids the delays in measurement caused by adaptive measurements, as discussed in Section~\ref{subsec:measSeq}.

We now calculate the time at which the physical qubits of the tree code are measured. These times determine the loss probability of the qubits. Consider a qubit $i$ emitted at time $T_i$. If $T_l+2\tau_l<T_i$, the photon is measured immediately. This is because the basis in which the repeater needs to measure qubit $i$ is known prior to the emission of the photon. If $T_l+2\tau_l>T_i$, qubit $i$ has to wait for time $T_l+2\tau_l-T_i$ before measurement. In other words, it is measured at $T_{mi}=\max(T_l+2\tau_l, T_i)$. The amount of time the qubit waits before being measured or its \textit{measurement wait time} is $T_{wi}=T_{mi}-T_i$. 

The qubits emitted from the qubit chip with the spin emitters are first coupled into an optical fiber with efficiency $\eta_c$. These qubits keep undergoing losses in the optical fiber for their measurement wait times. The loss probability ($l$) of qubit $i$ with measurement wait time $T_{wi}$ is  $1-\eta_c\exp(-\alpha c_fT_{wi} )$. Note that, unlike the all-photonic repeater protocols studied earlier~\cite{Mihir2017,Azuma2015}, the qubits in the RGS generated using the algorithm in Section ~\ref{sec:final_algo} have different loss probabilities due to different wait times. We derive Pauli measurement success probabilities for tree code with non-uniform qubit loss probabilities using Section~\ref{sub:treeCode}.

\subsubsection{Resource requirements}
\label{subsub:resources}
The RGS generation event starts with initializing the emitters and ends after all the photonic qubits have been generated and measured. Let $T_{n_e}$ be the time required to generate one copy of the RGS using $n_e$ emitters. Let $\tau$ be the repetition rate required by the protocol. We now calculate the number of emitters one would require to support the repetition rate $\tau< T_{n_e}$. Here, we use the concept of staggered generation, wherein to maintain the repetition rate $\tau$, the repeater needs to start creating a new copy of the RGS at an interval of $\tau$ seconds.  

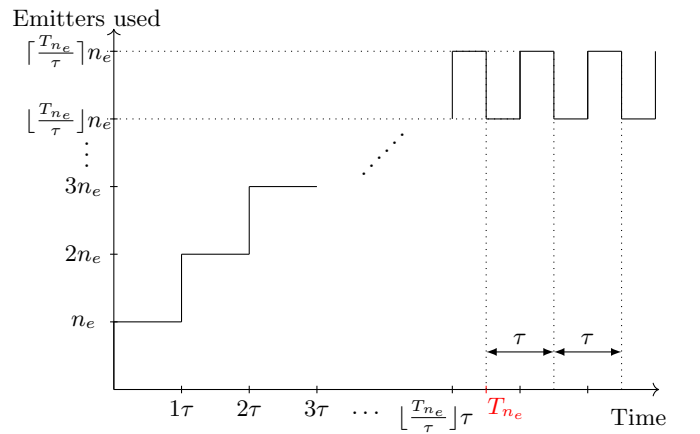
\begin{figure}[htb]
    \centering
    \begin{tikzpicture}
 \def \n {0.9}
 \draw[->] (0,0) -- (7.25,0);
 \draw[->] (0,0) -- (0,4.8);
\foreach \x in {1,2,3,5,6,7}
{
\draw (\n*\x,-.05) -- (\n*\x,.05);
}
\draw[red] (\n*5.5,-.05) -- (\n*5.5,.05);
\foreach \x in {1,2,3,4,5}
{
\draw (-.05,\n*\x) -- (.05,\n*\x);
}
\foreach \x in {1,2,3}
{
\node at (\n*\x,-.25) {\x$\tau$};
}
\foreach \x in {2,3}
{
\node at (-.4,\n*\x) {\x$n_e$};
}
\node at (-.4,\n) {$n_e$};
\node at (-.6,\n*5) {$\lceil \frac{T_{n_e}}{\tau}\rceil n_e$};
\node at (-.6,\n*4) {$\lfloor \frac{T_{n_e}}{\tau}\rfloor n_e$};
\foreach \x in {0,\n*1,\n*2}
{
\draw (\x,\x+\n*1) -- (\x+\n*1,\x+\n*1);
\draw (\x,\x) -- (\x,\x+\n*1);
}
\foreach \x in {2,3,1}
{
\draw (\n*4+\n*\x,\n*4 ) -- (\n*4+\n*\x,\n*5 );
\draw (\n*4+\n*\x,\n*5 ) -- (\n*4.5+\n*\x,\n*5 );
\draw (\n*4.5+\n*\x,\n*5 ) -- (\n*4.5+\n*\x,\n*4 );
\draw (\n*4.5+\n*\x,\n*4 ) -- (\n*5+\n*\x,\n*4 );
\draw (\n*5+\n*\x,\n*4 ) -- (\n*5+\n*\x,\n*5 );
}
\foreach \x in {1,2,3}
{
\draw [dotted] (\n*4.5+\n*\x,\n*0) -- (\n*4.5+\n*\x,\n*5);

}
\draw [dotted] (\n*-.1,\n*5) -- (\n*6,\n*5);
\draw [dotted] (\n*-.1,\n*4) -- (\n*6,\n*4);
\node[red] at (\n*5.8,-.25) {$T_{n_e}$};
\node at (\n*4.75,-.35) {$\lfloor \frac{T_{n_e}}{\tau}\rfloor\tau$};
\node [rotate=45] at (\n*4,\n*3.5) {$\dots\dots$};
\node [rotate=90] at (\n*-.4,\n*3.5) {$\dots$};
\node  at (\n*3.75,\n*-.35) {$\dots$};
\node  at (\n*7.75,\n*-.4) {Time};
\node  at (\n*-.4,\n*5.5) {Emitters used};
\node  at (\n*6,\n*.75) {$\tau$};
\node  at (\n*7,\n*.75) {$\tau$};
\draw[>=latex,<->] (\n*6.5,0.5) -- (\n*7.5,0.5);
\draw[>=latex,<->] (\n*6.5,0.5) -- (\n*5.5,0.5);
\end{tikzpicture}
    \caption{Timing diagram for the number of emitters used at a repeater. One copy of the RGS is generated every $\tau$ seconds. The time required to generate the RGS using $n_e$ emitters is $T_{n_e}>\tau$. }
    \label{fig:timing}
\end{figure}

The first RGS is generated using $n_e$ emitters and the quantum repeater protocol starts at $T_{n_e}$. Up to time $T_{n_e}$, the repeater needs to employ $N_e =\lceil\frac{T_{n_e}}{\tau}\rceil n_e$ number of emitters for the staggered generation of RGS. At $T_{n_e}$, $n_e$ emitters are measured, and the total number of emitters the repeater is actively using drops to $N_e-n_e =\lfloor\frac{T_{n_e}}{\tau}\rfloor n_e$. The measured $n_e$ emitters then begin another round of RGS generation at time $\lceil\frac{T_{n_e}}{\tau}\rceil\tau$, increasing the number of emitters being used to $N_e$. To summarize, the repeater requires $N_e$ emitters to produce an RGS every $T_{n_e}+k\tau$ seconds ($k\in\{0,1,2,\dots\}$), and attain the repetition time $\tau$. This process is outlined in FIG.~\ref{fig:timing}.

\subsection{Results and discussion}
In this section, we calculate the entanglement generation rate of our emitter-based all-photonic quantum repeater architecture and compare it with ~\cite{Mihir2017}. The RGS parameters are $b =[7,3]$ and $m=4$ and we set $\tau = t_{\rm init}$.  At the minor nodes, we use linear optical BSM whose success probability is boosted to 3/4 using ancilla qubits and assume that the detectors are perfect. The emitter parameters are $t_{\textrm{cnot}_{e,e}}=180 \textrm{ns},  t_{\textrm{meas}}=45 \textrm{ns},t_{\rm H_{e}} = 15 \textrm{ns}, $ and $t_{\rm init}= 15 \textrm{ns}$. It is worth noting that these values are not specific to any particular hardware. They were chosen as an example to highlight the benchmarks needed for time scales before the emitter-based schemes improve upon the rates achieved without repeaters~\cite{dutt2007quantum,pfaff2013demonstration,press2008complete, thomas2022efficient,de2011ultrafast,nguyen2019quantum,Stas2022,rong2015experimental,gaetan2009observation,wilk2010entanglement}.

\begin{figure}
    \centering
    \includegraphics[scale=0.65]{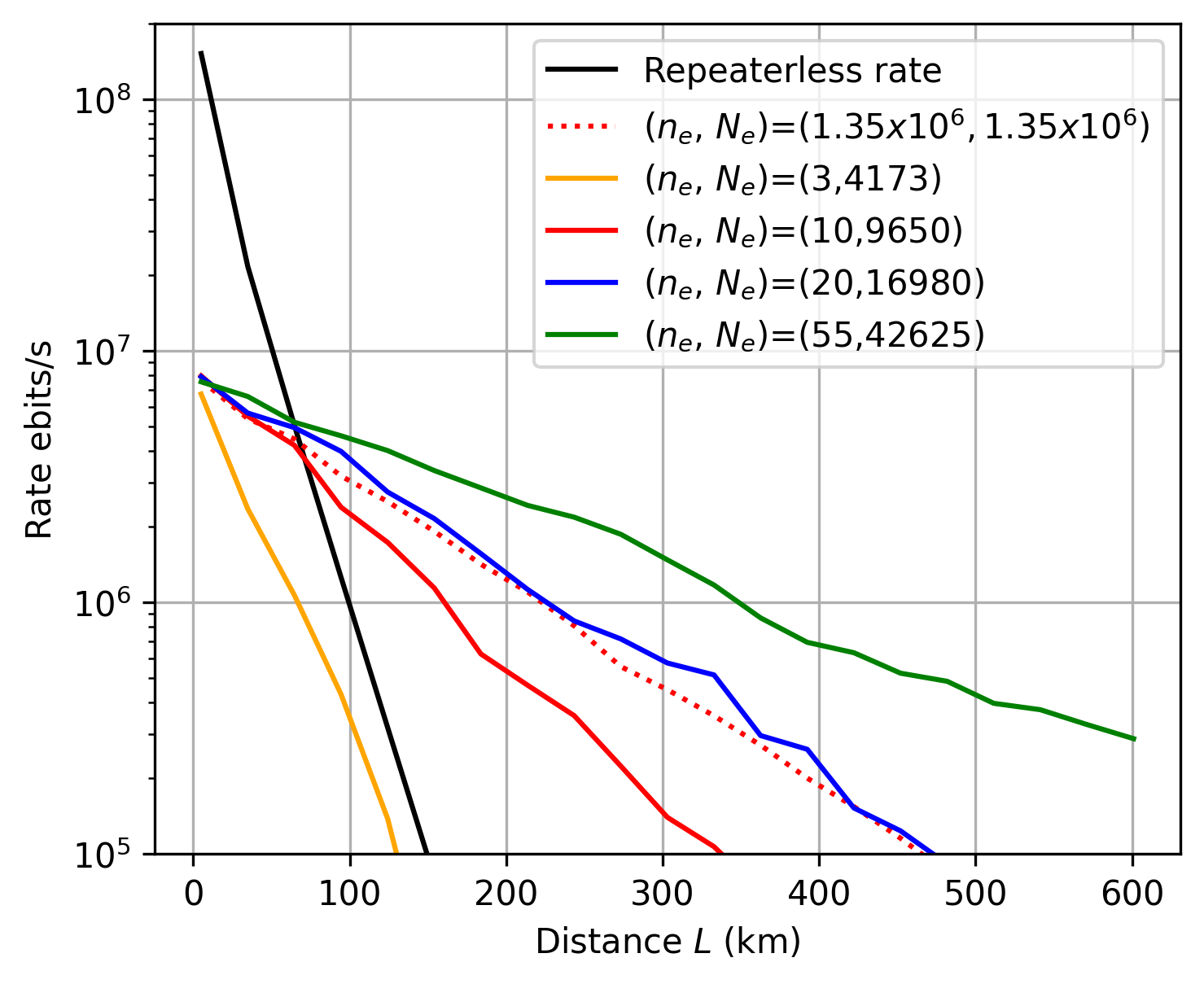}
    \caption{Rate vs distance envelopes for when the RGS with $b =7,3]$ and $m=4$ is generated using emitters and multiplexing. $n_e$ is the number of emitters used to generate one copy of the RGS, and $N_e$ is the number of emitters required per repeater to achieve the repetition rate $\tau = t_{\rm init}$. The dashed red line corresponds to RGS generated using linear optics.}
    \label{fig:ratePlot}
\end{figure}

For a given $n_e$, we first calculate the measurement wait times of all qubits in the RGS, followed by their loss probabilities as per the discussion above. The qubit emission times and measurement wait times are functions of $n_e$. Unlike the multiplexing method, the emitter-based method generates an RGS with different loss probabilities and, hence, different Pauli measurement probabilities for every logical qubit.  As a result, we cannot use Eq~\ref{eq:rate} to calculate the entanglement generation rate. Instead, we perform a Monte Carlo simulation to calculate the average rate for the emitter-based protocol. We begin by fixing $n_e$ which in turn fixes $N_e$. For each $L$, we vary the number of repeaters $n$, and calculate the entanglement generation rate using Monte Carlo simulation. Then, to each $L$ and $n_e$, we associate an entanglement generation rate maximized over $n$. In Fig~.\ref{fig:ratePlot}, for each $L$ and $n_e$, we plot the maximum entanglement generation rate -- also called the rate envelope. For the hardware parameters chosen, if the repeater generates RGS using the minimum number of emitters, given by the \textit{height function} of the state~\cite{Li2022}, the entanglement rate is less than the repeaterless rate. In other words, it is better not to use repeaters altogether than to use only $n_e = 3$ emitters in the chosen parameter regime. However, if we increase $n_e$, our protocol beats the repeaterless rate as shown in FIG.~\ref{fig:ratePlot}.

Our protocol outperforms the multiplexing-based method (dotted red line in FIG.~\ref{fig:ratePlot}.) when $n_e \geq 20$. We calculate the number of emitters used by the multiplexing method to generate the RGS with probability $\approx 1$ using the `improved multiplexing scheme' from ~\cite{Mihir2017}. Note that, for this method, quantum emitters are used as single-photons source, hence  $N_e=n_e$. For the chosen hardware parameters, our emitter-based protocol is significantly more resource-efficient as it uses 1/84th of the emitters required by the multiplexing-based method to achieve an equal entanglement rate.

\begin{table*}[ht]
    \centering
\centering
 \begin{tabular}{||p{7cm} p{2cm} p{2cm} ||} 
 \hline
 Parameter & Symbol & Value \\ 
 \hline\hline
 Fiber loss coefficient & $\alpha$ & $0.046\textrm{km}^{-1}$\\
 Speed of light in the optical fiber & $c_f$ & $2\times 10^5 $ km$/$s\\
 Chip-to-fiber coupling efficiency & $\eta_c$ & 0.99\\
 \hline
 \end{tabular}
\caption{\textbf{Assumed hardware parameters}} 
    \label{tab:rateParas}
\end{table*}

\section{Conclusions and Future Work}
In this study, we have developed an algorithm for generating multi-qubit graph states using 
$n_e$-quantum emitters, where $n_e$ is greater than the minimum number of emitters required. This algorithm delineates a sequence of Clifford operations, computational measurements, and photon emissions necessary for preparing specified graph states. The foundational elements of the algorithm are detailed in Section~\ref{sec:algo_primitives}. Our algorithm is general for any graph state.

Another primary contribution is the assessment of resource requirements for generating repeater graph states. We demonstrate a trade-off between the CNOT circuit depth and the number of emitters, as illustrated in FIG.~\ref{fig:emitter_vs_depth}. This balance is crucial for understanding the efficiency of various graph states generation protocols.

A significant focus of our work is on the entanglement generation rates in all-photonic quantum repeaters. We introduce a new protocol tailored to graph states generated from quantum emitters, aiming to minimize the waiting time between photon emission and measurement. By integrating the timing dynamics of photon release from quantum emitters with this protocol, we calculate entanglement generation rates under various parameters. In FIG.~\ref{fig:ratePlot}, we plot the envelopes the entanglement generation rates between Alice and Bob separated by a distance $L$. Our analysis emphasizes the necessity of optimizing both the graph state generation algorithm and the repeater protocol to surpass existing entanglement generation limits. Interestingly, our findings suggest that using a minimal number of emitters, such as three, performs worse than the repeaterless bound. Our protocol, given specific experimental parameters, shows improved resource efficiency compared to multiplexing protocols. In conclusion, our work lays the groundwork for enhancing quantum repeater graph state generation, keeping in mind resource efficiency and entanglement rate optimization.

Looking ahead, we identify several avenues for further optimization. The ordering of processes would be dependent on the objective function. We leave the optimality of the processes for future work. In Section~\ref{sec:circuit_depth}, we outline an algorithm to calculate the CNOT depth of the circuit obtained from the algorithm. The results are shown in Fig~\ref{fig:emitter_vs_depth}. Optimizing this algorithm for efficiency and effectiveness is a goal for subsequent research. An exciting line of work to explore would be to design an algorithm specifically for RGS that optimizes for the time of photon release such that the loss experienced by the photon before being measured is minimized. This would involve not only optimizing for the order of processes but also adding delays on the emitters in order to time the release of the photons to minimize loss.  

\appendix
\section{An abridged literature review of emitter-based preparation of photonic graph states}\label{app:review}

Refs.~\cite{Schon2005,Schon2007} proved an equivalence between $D$-dimensional matrix product states with open boundary conditions, and states that are generated sequentially and isometrically via a $D$-dimensional ancillary system which decouples in the last step. This effectively restricts the number of emitters required to generate photonic graph states according to the entanglement property of the graph. They further gave the isometries needed for certain multi-qubit states such as GHZ, W state and graph state. Ref. \cite{Rudolph2009} laid out the method for the deterministic generation of a 1D-graph state, and incorporated noise in the generation model. An experimental demonstration of such 1D graph states was presented in Refs.~\cite{Besse2020,Schwartz2016}. In Ref.~\cite{Economou2010}, the authors gave a proposal to produce 2-dimensional photonic graph states. The idea of entangled emitters emitting entangled photons has been exploited in further proposals of generation of photonic graph states in \cite{Borregaard2020,Buterakos2017,Yuan2020,Russo2018,Economou2019}, for various resource states, such as repeater graph states, and one-way quantum computing. Refs.~\cite{Yuan2020,Pichler2017} allowed for re-interference of photons with emitters after emission. In Ref.~ \cite{Hilaire2021}, the authors used emitter qubits to produce photonic graph states and analyzed the trade-off between resources needed and performance, as characterized by the achievable secret key rate per emitter qubit. In Ref.~\cite{Li2022}, the authors gave an explicit quantum circuit to generate a general photonic graph state $\ket{G}$ with the minimal number of emitters, which they characterized as the {\em height function}, $h(G)$ of the graph $G$. 

\section{Implementation timescales in SiV color center emitters}
\label{sec:apx}
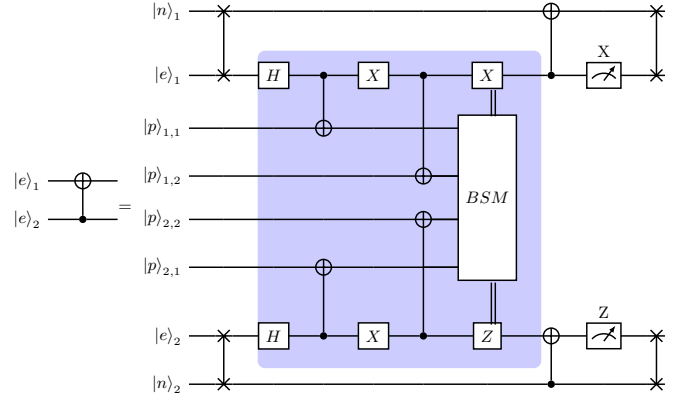
\begin{figure}
	\centering
 \resizebox{0.5\textwidth}{!}{%
    \begin{tikzpicture}
     \path [rounded corners,fill=blue!20] (-1.5, -3) rectangle (3.9, 3.05) {};
\node at (0,0) { \begin{quantikz}
      \lstick{$\ket{e}_1$} &\targ{} & \qw  \\
       \lstick{$\ket{e}_2$} &\ctrl{-1}& \qw  \\
 \end{quantikz}=
	\begin{quantikz}
	\lstick{$\ket{n}_1$} &  \swap{1} & \qw & \qw  &  \qw   & \qw & \qw & \targ{} & \qw &\swap{1}\\	
 \lstick{$\ket{e}_1$} &  \targX{} &\gate{H}  &\ctrl{1} & \gate{X}  &\ctrl{2} &  \gate{X} &\ctrl{-1}& \meter{X} & \swap{} \\
 \lstick{$\ket{p}_{1,1}$} & \qw & \qw & \targ{}  & \qw & \qw & \gate[4]{BSM}\\
 \lstick{$\ket{p}_{1,2}$} & \qw & \qw & \qw & \qw & \targ{} & \qw  \\
 \lstick{$\ket{p}_{2,2}$} & \qw &\qw  & \qw & \qw & \targ{}  & \qw  \\
  \lstick{$\ket{p}_{2,1}$} & \qw &\qw  & \targ{}  & \qw & \qw & \qw \\
\lstick{$\ket{e}_2$}& \swap{1} & \gate{H}  & \ctrl{-1}& \gate{X}  &\ctrl{-2}  &  \gate{Z}  &\targ{} &   \meter{Z} &\swap{}\\
\lstick{$\ket{n}_2$} & \targX{} & \qw  &\qw  & \qw & \qw &  \qw  &\ctrl{-1} & \qw &\swap{-1}\\ 
	\end{quantikz}};
 \draw[thick] (2.95,1.8)--(2.95,2.3);
 \draw[thick] (3.025,1.8)--(3.025,2.3);
 \draw[thick] (2.95,-1.37)--(2.95,-2.17);
 \draw[thick] (3.025,-1.37)--(3.025,-2.17);
 \end{tikzpicture}
 }
	\caption{The quantum circuit to perform CNOT between electronic spins of SiV centers, mediated through nuclear spins. The blue highlighted box represents generation of single photons using electronic spins, followed by BSM on the dual-rail photonic qubits (steps 2-7 in the text). As the photonic BSM is probabilistic, all processes in the blue box are repeated until the BSM succeeds.}
	\label{fig:enSWAP}
\end{figure}

In this section, we outline the method for applying a CNOT gate between two quantum emitters. We consider an electronic-nuclear spin system, where the nuclear spin serves as a deterministic long-lived memory qubit, such as SiV color centers. Each SiV color center consists of an electronic nuclear spin system. The electronic spin serves as the emitter and is used to emit photons of the graph state. From Section~\ref{sec:algo_primitives}, CNOT on two emitters is a necessary operation to generate a graph state. In the SiV systems, the CNOT gate cannot be applied directly between two electronic spins. Moreover, the state of the nuclear spins cannot be measured. Taking these constraints into account, we outline the steps to apply a CNOT gate between two emitters in the state $\ket{\psi}_{e_1e_2}$, mediated through the nuclear spins, along with the parameterized time scales (refer FIG.~\ref{fig:enSWAP}):
\begin{enumerate}
    \item The electronic spin states are stored in the corresponding nuclear spin states by applying a nuclear-electron swap gate \cite{Stas2022}, with time scales $t_{\rm SWAP}$. This changes the state of the system to $\ket{\psi}_{n_1n_2}\otimes\ket{?}_{e_1e_2}$, where $\ket{?}$ implies some unspecified state. 
    
    \item The electronic spin is initialized in the ground state, with the time scales given as $t_{\textrm{init}}$. This changes the state of the system to $\ket{\psi}_{n_1n_2}\otimes\ket{00}_{e_1e_2}$. 
    
    \item Apply a Hadamard gate to the electronic spin. This is equivalent to rotating the electronic spin to a superposition of a ground state $(\ket{0})$ and an excited state $(\ket{1})$. To apply a Hadamard gate, a microwave pulse is applied to the electronic spin. The time scales are $t_{\textrm{H},e}$. This changes the state of the system to $\ket{\psi}_{n_1n_2}\otimes\ket{++}_{e_1e_2}$. 
    \item Apply a laser pulse on the electron to obtain the state $\ket{\psi}_{n_1n_2}\otimes \bigotimes_{j=1,2}\left(\ket{01}_{e_jp_{j,1}}+\ket{10}_{e_jp_{j,1}}\right)$, where $j \in [1,2]$, and takes time $t_{\textrm{ex}}$. Here, $p_{j,1}$ represents the photon emitted by electronic spin $j$ and is in the single-rail encoding.
    \item Apply an $X$ gate to the electronic spin. The state of the electron-photon system is $\ket{11}_{e_jp_{j,1}}+\ket{00}_{e_jp_{j,1}}$. The time scales are $t_{X,e}$.
    \item Apply another laser pulse on the electron to obtain the following electron-photon state $\ket{110}_{e_jp_{j,1}p_{j,2}}+\ket{001}_{e_jp_{j,1}p_{j,2}}$.
    \item The state of the full system is now given as $\ket{\psi}_{n_1n_2}\bigotimes_{j=1,2}\ket{110}_{e_jp_{j,1}p_{j,2}}+\ket{001}_{e_jp_{j,1}p_{j,2}}$. We rewrite the above state as $\ket{\psi}_{n_1n_2}\bigotimes_{j=1,2}\ket{10}_{e_jp_{j_L}}+\ket{01}_{e_jp{j_L}}$, where we have used the dual rail encoding for the photonic qubits. The photons from each electronic system undergo a Bell state measurement (BSM) on the beamsplitter. Pauli $X, Z$ corrections are applied to the electronic spins conditioned on the outcomes of the BSM. Entangling photons on the beamsplitter is probabilistic and instantaneous. To combat the probabilistic process, 
    this whole process, starting from the initialization of the electronic spin is repeated till the Bell state measurement (BSM) succeeds. One attempt of photonic BSM takes time $t_{\rm ph} = t_{\textrm{init}}+t_{\textrm{H},e}+2t_{\textrm{ex}}+t_{\rm X,e}$. If $p_{\rm BSM}$ is the probability of success of the photonic BSM, the average number of trials required to get the first success is given by $1/p_{\rm BSM}$. The detector inefficiencies can be folded into $p_{\rm BSM}$~\cite{ewert20143}. The average time required to get the first BSM success is $t_{\textrm{BSM}} = t_{\rm ph}/p_{\rm BSM}+\max(t_{\rm X,e},t_{\rm Z,e})$. Here, $\max(t_{\rm X,e},t_{\rm Z,e})$ is the time taken to apply Pauli corrections on the electronic spin, and $t_{\rm Z,e}$ is time to apply Pauli-$Z$ gate on the electronic spin. If $\ket{\psi}_{n_1n_2}\otimes\ket{?}_{e_1e_2}$ is a stabilizer state, Pauli corrections affect only the phase of the state and can be tracked using classical computation. During this process, shown as the blue box in FIG.~\ref{fig:enSWAP}, the original state is stored in the nuclear spins and is not impacted by the failure of the Bell measurements. 
    \item Perform a CNOT gate on one of electron-nuclear spin pairs with nuclear spin as the control ($t_{\textrm{CNOT}_{n,e}}$) followed by $Z$ measurement of the electronic spin that takes time $t_{\textrm{meas, z}}$. 
    \item  Perform a CNOT gate on the second nuclear-electron spin pair with the electron as the control qubits, with time scale given as $t_{\textrm{CNOT}_{e,n}}$ followed by the $X$ measurement on the electronic spin, which takes $t_{\textrm{meas, x}}$.  
    \item Swap the state of the nuclear spins with the electronic spins using swap gates.
\end{enumerate}

The time taken to implement the CNOT gate between two quantum emitters $t_{\rm CNOT_{e,e}}$ using the SiV vacancy center is given by -
\begin{align*}
    t_{\rm CNOT_{e,e}}&=t_{\rm SWAP}+t_{\rm BSM}+\max(t_{\rm CNOT_{n,e}}+t_{\rm meas,z}, \\
    &t_{\rm CNOT_{e,n}}+t_{\rm meas,x})+ t_{\rm SWAP}
\end{align*}

\section{Emission Times of Photons}\label{Emission times}

After setting up the circuit for graph state generation, we assign a counter, $c_{k_e}$, to each emitter $k_e$, where $k \in [1,n]$, and $n$ is the total number of emitters used to generate the state. This counter tracks the emission time of photons. We also assign a pointer to keep track of the location in the circuit. The counters are initialized to $0$, and the pointer is initialized to the beginning of the circuit. As the pointer moves along the circuit and encounters gates, the following operations are performed based on the type of gate:

\begin{itemize}
    \item \textbf{Hadamard Gate (\( H_e \)):}  
    Let \( c_{k_e} = t \). If the pointer associated with the \( k_e^{\textrm{th}} \) emitter encounters a \( H_e \) gate, the counter is updated as:
    \[
    c_{k_e} = t + t_{H_e}.
    \]
    That is, the time required to implement the \( H_e \) gate is added to the counter \( c_{k_e} \).

    \item \textbf{Measurement of the Emitter:}  
    Let \( c_{k_e} = t \). If the pointer associated with the \( k_e^{\textrm{th}} \) emitter encounters a measurement operation, the counter is updated as:
    \[
    c_{k_e} = t + t_{\textrm{meas}} + t_{\textrm{init}}.
    \]

    \item \textbf{\( \operatorname{CNOT}_{e,e} \) Gate:}  
    If the pointer associated with the \( k_e^{\textrm{th}} \) emitter encounters a \( \operatorname{CNOT}_{e,e} \) gate involving the \( j_e^{\textrm{th}} \) emitter, the operation is delayed until both emitters are ready. Let \( c_{k_e} = t_1 \) and \( c_{j_e} = t_2 \). Define:
    \[
    t_{\textrm{max}} = \max(t_1, t_2).
    \]
    Then, the counters are updated as:
    \[
    c_{k_e} = t_{\textrm{max}} + t_{\textrm{CNOT}_{e,e}}, \quad 
    c_{j_e} = t_{\textrm{max}} + t_{\textrm{CNOT}_{e,e}}.
    \]
    The use of \( t_{\textrm{max}} \) ensures that both emitters are synchronized for the application of the \( \operatorname{CNOT}_{e,e} \) gate.

    \item \textbf{\( \operatorname{CNOT}_{e,p} \) Gate:}  
    Let \( c_{k_e} = t \). If the pointer associated with the \( k_e^{\textrm{th}} \) emitter encounters a \( \operatorname{CNOT}_{e,p} \) gate, the counter is updated as:
    \[
    c_{k_e} = t + t_{\textrm{CNOT}_{e,p}}.
    \]
    At this point, the value stored in \( c_{k_e} \) is associated with the emission time of the photon emitted by the \( k_e^{\textrm{th}} \) emitter.
\end{itemize}

The steps described above yield the emission times of photons in the graph state.

\section{$\operatorname{CNOT}_{e,e}$ depth}\label{sec:CNOT_depth}

In this appendix, we first show an example in which an increase in the number of emitters can also increase the $\operatorname{CNOT}_{e,e}$ depth. Consider the graph state in FIG.\ref{fig:example_GHZ}(a). Given two emitters, one can use the circuit shown in FIG.\ref{fig:example_GHZ}(b). The circuit has a $\operatorname{CNOT}_{e,e}$ depth of one. It is also possible to generate the graph state in FIG.\ref{fig:example_GHZ}(a) with one emitter using the circuit shown in FIG.\ref{fig:example_GHZ}(c). This circuit's $\operatorname{CNOT}_{e,e}$ depth is zero. Note that our algorithms will output these circuits given the number of emitters $n_e$. We thus see that the $\operatorname{CNOT}_{e,e}$ depth is not a monotonically decreasing function of $n_e$.

Now consider the graph state shown in FIG.\ref{fig:example}. This state can be generated using three emitters. However, in such a scenario, the emitters would be required to create one tree at a time. Thus, the $\operatorname{CNOT}_{e,e}$ gates required to generate this state from three emitters must all be applied sequentially. However, given twelve emitters, the emitters can start generating the attached trees simultaneously, thus parallelizing the process. This is evidenced by FIG.\ref{fig:example}(i), wherein we see how the usage of three emitters per tree would lead to the generation of the attached trees simultaneously. It is easy to calculate that by using three emitters, the $\operatorname{CNOT}_{e,e}$ depth of the circuit is eleven. However, if using twelve emitters, the $\operatorname{CNOT}_{e,e}$ depth of the circuit would be five. For the case of twelve emitters, the only $\operatorname{CNOT}_{e,e}$ required are for disentangling the emitters in FIG.\ref{fig:example}(i) and in FIG.\ref{fig:example}(o). The eight $\operatorname{CNOT}_{e,e}$ in FIG.\ref{fig:example}(i) can be applied in circuit depth two and in FIG.\ref{fig:example}(o) can be applied in circuit depth three. We thus see, from the aforementioned graph state, increasing the number of emitters can decrease the $\operatorname{CNOT}_{e,e}$ of the generating circuit. 

Let us now understand why increasing the number of emitters $n_e$ can increase the $\operatorname{CNOT}_{e,e}$ depth. This generally occurs when a photon that could have been absorbed by an emitter using Sec~\ref{sec:photon_absorption} is instead replaced by an emitter using Sec~\ref{sec:SFE}. This operation could add an extraneous $\operatorname{CNOT}_{e,e}$ gate in the circuit, which may increase the circuit depth. Therefore, the number of emitters should be carefully specified if the goal is to decrease the $\operatorname{CNOT}_{e,e}$ depth of the generating circuit. 

\vspace{1em}
\section{Total time taken to generate GHZ equivalent state}\label{appen: GHZ_generation}

Consider now the total time taken to generate the Clifford equivalent $m$-GHZ state. The graph representation of this state is a star graph. 

Let $T_{\operatorname{GHZ}}(n)$ be the total time taken to generate the Clifford equivalent $m$-GHZ state using $n$ emitters using the algorithms described above. Then, we can obtain the following expression:

\begin{align}
    T_{\operatorname{GHZ}}(1) &= 2t_{H_e} + t_{\textrm{meas}} + mt_{\textrm{CNOT}_{e,p}} \label{eq:GHZ_1}\\
    T_{\operatorname{GHZ}}(2) &= 2t_{H_e} + t_{\textrm{meas}} + \left \lceil \frac{m}{2} \right\rceil t_{\textrm{CNOT}_{e,p}}+t_{\textrm{CNOT}_{e,e}}\label{eq:GHZ_2}\\
    T_{\operatorname{GHZ}}(3) &= 2t_{H_e} + t_{\textrm{meas}} + \left \lceil \frac{m}{3} \right\rceil t_{\textrm{CNOT}_{e,p}}+2t_{\textrm{CNOT}_{e,e}}\label{eq:GHZ_3}\\
    T_{\operatorname{GHZ}}(n) &= 2t_{H_e} + t_{\textrm{meas}} + \left \lceil \frac{m}{n} \right\rceil t_{\textrm{CNOT}_{e,p}}+2t_{\textrm{CNOT}_{e,e}}\nonumber\\\textrm{for} &\quad n>3
\end{align}
For Eq.~\ref{eq:GHZ_1}, one can consider the circuit as an extension of Fig.~\ref{fig:example_GHZ}(b). The single emitter is used to generate $m$-GHZ state. For Eq.~\ref{eq:GHZ_2}, one can consider the extension of Fig.~\ref{fig:example_GHZ}(c). The emission of the $m$ photonic qubits is divided between the two emitters giving rise to the factor $\left \lceil \frac{m}{2} \right\rceil$. The $\operatorname{CNOT}_{e,e}$ between the emitters introduces the factor $t_{\textrm{CNOT}_{e,e}}$. For Eq.~\ref{eq:GHZ_3}, the emission of the $m$ photonic qubits is now divided between three emitters giving rise to the factor $\left \lceil \frac{m}{3} \right\rceil$. The emitters also need to be entangled. This can be done in circuit depth two, introducing a factor $2t_{\textrm{CNOT}_{e,e}}$. We note further that for $n>3$, where $n$ is the number of emitters used for generation, the entanglement of the emitters can be achieved in circuit depth two. However, the $m$ photonic qubit emission is now divided between $n$ emitters, giving rise to $\left \lceil \frac{m}{n} \right\rceil$.

We now ask the following question: given the metric as the total time taken to generate the $m$-GHZ equivalent graph state, for what values of $m$ is it better to use an extra emitter?

Using some simple algebra, we find that
\begin{itemize}
    \item Using two emitters would be better than one emitter if $\left \lfloor \frac{m}{2}  \right \rfloor \geq \frac{t_{\textrm{CNOT}_{e,e}}}{t_{\textrm{CNOT}_{e,p}}}$. 
    \item Using three emitters would be better than two emitters if $\left \lceil \frac{m}{2}  \right \rceil - \left \lceil \frac{m}{3}  \right \rceil \geq \frac{t_{\textrm{CNOT}_{e,e}}}{t_{\textrm{CNOT}_{e,p}}}$. 
    \item For all $p \geq \left \lceil \frac{m}{2}  \right \rceil - \left \lceil \frac{m}{3}  \right \rceil \geq \frac{t_{\textrm{CNOT}_{e,e}}}{t_{\textrm{CNOT}_{e,p}}}$, it is better to use a $p$ emitters. In other words, for these values, it is always better to use an extra emitter. The reason to see this jump can be attributed to the fact that coefficient for $t_{\textrm{CNOT}_{e,e}}$ in the expression of $T_{\operatorname{GHZ}}(n)$ is fixed for $n\geq 3$.
\end{itemize}
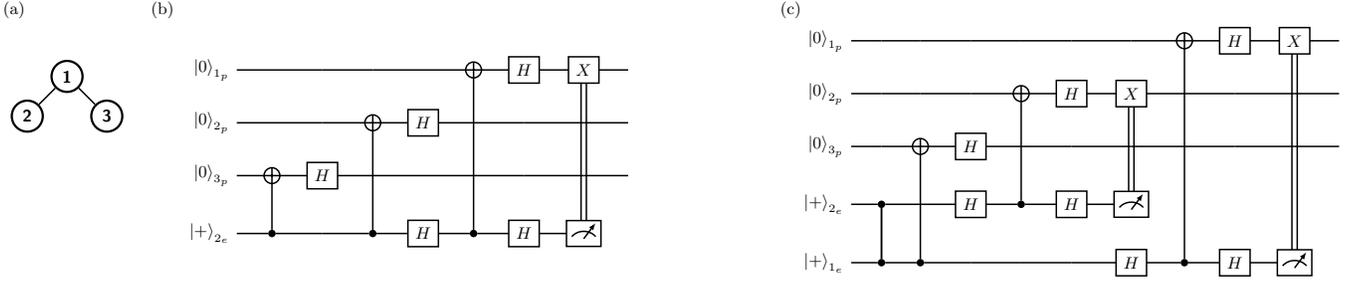
\begin{figure*}[htb!]
    \centering
    \resizebox{\textwidth}{!}{%
    \begin{tikzpicture}[roundnode/.style={circle, draw = black, very thick,  minimum size=1mm,font=\sffamily\small\bfseries},edgestyle/.style={draw=black, thick}]
    \def \n {10}
\node[roundnode]  (1)  at (-10+0.1,0+4) {1};
\node[roundnode]  (2)  at (-10.75+0.1,-.75+4) {2};
\node[roundnode]  (3)  at (-9.25+0.1,-.75+4) {3};

\draw[edgestyle] (1) edge (2);
\draw[edgestyle] (1) edge (3);

\node at (-10.9,5.25){(a)};
    \node at (-8.1,5.25){(b)};
    \node at (-6.25+\n,5.25){(c)};
        \node at (-3.5,2.3) {
        \begin{quantikz}
         \lstick{$\ket{0}_{1_p}$} & \qw & \qw & \qw & \qw & \targ{} & \gate{H}& \gate{X}  &\qw  \\
    \lstick{$\ket{0}_{2_p}$}  & \qw & \qw &\targ{} & \gate{H}  & \qw & \qw & \qw &\qw  \\
    \lstick{$\ket{0}_{3_p}$} &\targ{} & \gate{H} & \qw & \qw & \qw & \qw & \qw & \qw \\
    \lstick{$\ket{+}_{2_e}$} & \ctrl{-1} & \qw & \ctrl{-2} & \gate{H} &\ctrl{-3} & \gate{H}& \meter{} \vcw{-3}  \\
        \end{quantikz}
        };

        \node at (\n-1,2.3) {
        \begin{quantikz}
        \lstick{$\ket{0}_{1_p}$} & \qw & \qw & \qw & \qw & \qw & \qw & \targ{} & \gate{H}& \gate{X}&\qw   \\
    \lstick{$\ket{0}_{2_p}$} & \qw  & \qw & \qw &\targ{} & \gate{H} & \gate{X} & \qw & \qw & \qw &\qw \\
    \lstick{$\ket{0}_{3_p}$} & \qw & \targ{} & \gate{H} & \qw & \qw & \qw & \qw & \qw & \qw   &\qw\\
    \lstick{$\ket{+}_{2_e}$} & \ctrl{1} & \qw & \gate{H} & \ctrl{-2} & \gate{H}  & \meter{} \vcw{-2}  \\
    \lstick{$\ket{+}_{1_e}$} & \ctrl{-1} & \ctrl{-2} & \qw & \qw & \qw & \gate{H} & \ctrl{-4} & \gate{H} & \meter{}\vcw{-4} \\
        \end{quantikz}
        };
    \end{tikzpicture}}
    \caption{Quantum circuit for graph state generation (a) Clifford equivalent GHZ state, (b) Quantum circuit for generation of the graph state using one emitter, (c) Quantum circuit for generation of graph state  using two emitter}

    \label{fig:example_GHZ}
\end{figure*}

Consider the following parameter regime $\frac{t_{\textrm{CNOT}_{e,e}}}{t_{\textrm{CNOT}_{e,p}}} = 10$. Then, using the above analyses, we find, $m\leq 10$, a single emitter yields the optimal total time. For $10<m\leq 60$, use of two emitters yields the optimal time. For $m\geq 60$, it is beneficial to use up to $m$ emitters. 

\section{Justification for Algorithmic building blocks} \label{app:stabilizer_form}
\subsection{Graph State and Pauli Operators}

Consider a graph $G = (V, E)$, where $V$ represents the set of vertices (nodes) and $E$ represents the set of edges (connections between vertices). To define a \textit{graph state}, we associate each vertex with a qubit prepared in the state $\ket{+}$, defined as:

$$
\ket{+} = \frac{\ket{0} + \ket{1}}{\sqrt{2}}.
$$

The edges in the graph correspond to entangling operations between the qubits, specifically, the application of controlled-phase ($\operatorname{CZ}$) gates. The $\operatorname{CZ}$ gate is a two-qubit gate that introduces a phase shift when both qubits are in the $\ket{1}$ state.

Formally, the graph state $\ket{G}$ is defined as:

\begin{equation}
\ket{G} = \prod_{(i,j) \in E} \operatorname{CZ}_{i,j} \ket{+}^{\otimes |V|},
\end{equation}

where $\operatorname{CZ}_{i,j}$ represents a controlled-phase gate acting on qubits $i$ and $j$, and $\ket{+}^{\otimes |V|}$ denotes the tensor product of $\ket{+}$ states for all qubits in $V$. The product over $\operatorname{CZ}_{i,j}$ is taken over all edges $(i, j)$ in the set $E$. 

Alternatively, a graph state can be described using the stabilizer formalism. For each vertex $j$, we define a stabilizer operator $S_j$ as:

\begin{align}
S_j = X_j \prod_{k \in \mathcal{N}(j)} Z_k,
\end{align}

where $X_j$ and $Z_k$ are the Pauli $X$ and $Z$ operators acting on the $j$-th and $k$-th qubits, respectively. $\mathcal{N}(j)$ denotes the \textit{neighborhood} of vertex $j$ in the graph $G$, i.e., the set of vertices connected to $j$ by an edge.

The graph state $\ket{G}$ is the simultaneous eigenstate with eigenvalue $+1$ of all the stabilizer operators $\{ S_j \}_{j \in V}$. That is:

$$
S_j \ket{G} = \ket{G}, \quad \forall j \in V.
$$

This property uniquely defines the graph state in terms of its stabilizers.

Consider an $m'$-qubit graph state $\ket{G'}$, comprising $m_e' \geq 0$ emitter qubits and $m_p' > 1$ photon qubits. The graph state is characterized by stabilizer generators that define the correlations among its qubits.

We use $P_{j_p}$ to denote a Pauli operator $P \in \{I, X, Y, Z\}$ acting on the $j_p^{\text{th}}$ photon qubit, where $j \in [1, m_p']$. Similarly, $P_{k_e}$ denotes a Pauli operator acting on the $k_e^{\text{th}}$ emitter qubit, where $k \in [1, m_e']$.

To describe operators acting on all qubits except the $j_p^{\textrm{th}}$ qubit, we use the following notation:
\begin{align}
\mathbf{Q}_{V \setminus \{j_p\}} = &P_{1_p} \otimes P_{2_p} \otimes \cdots \otimes P_{(j-1)_p} \otimes P_{(j+1)_p} \otimes \cdots \otimes\nonumber \\
&P_{m_p'} \otimes P_{1_e} \otimes \cdots \otimes P_{m_e'},
\end{align}
where $V$ represents the set of all qubits in the graph state. Similarly, $\mathbf{I}_{V \setminus \{j_p\}}$ represents the identity operator acting on all qubits except the $j_p^{\text{th}}$ photon.

\subsection{Swapping with a Free Emitter}
Consider a free emitter qubit $e$ initialized in the state $\ket{0}_e$. Let $P_e$ denote a Pauli operator acting on this emitter qubit. The combined initial state of the system is given by $\ket{G'} \otimes \ket{0}_e$.

The stabilizer formalism describes the initial state using the following stabilizer generators:
\begin{align*}
\mathbf{Q}_{V \setminus \{j_p\}} \otimes Z_{j_p} \otimes I_e, \\
\tilde{\mathbf{Q}}_{V \setminus \{j_p\}} \otimes X_{j_p} \otimes I_e, \\
\mathbf{I}_{V \setminus \{j_p\}} \otimes I_{j_p} \otimes Z_e,
\end{align*}
where $\tilde{\mathbf{Q}}_{V \setminus \{j_p\}}$ represents a different set of Pauli operators acting on the qubits in $V \setminus \{j_p\}$. Note here that the stabilizer operators in the stabilizer generators of the initial state $\ket{G'} \otimes \ket{0}_e$ can be described as above.

The process of swapping the state of the free emitter $e$ with the $j_p^{\text{th}}$ photon involves the following sequence of operations:

\begin{enumerate}
\item \textbf{Hadamard Gate on the Free Emitter ($H_e$):}\\
   Applying the Hadamard gate on the emitter qubit transforms the third stabilizer generator as $Z_e \mapsto X_e$. The stabilizers after this operation are:
   \begin{align*}
   \mathbf{Q}_{V \setminus \{j_p\}} \otimes Z_{j_p} \otimes I_e, \\
   \tilde{\mathbf{Q}}_{V \setminus \{j_p\}} \otimes X_{j_p} \otimes I_e, \\
   \mathbf{I}_{V \setminus \{j_p\}} \otimes I_{j_p} \otimes X_e.
   \end{align*}

\item \textbf{Controlled-NOT Gate ($\operatorname{CNOT}_{e \rightarrow j_p}$):}\\
   Applying the controlled-NOT gate with $e$ as the control and $j_p$ as the target modifies the stabilizers as follows:
   \begin{align*}
   Z_{j_p} I_{e} &\mapsto Z_e Z_{j_p}, \\
   I_{j_p} X_e &\mapsto X_e X_{j_p}.
   \end{align*}
   The stabilizers now read:
   \begin{align*}
   \mathbf{Q}_{V \setminus \{j_p\}} \otimes Z_{j_p} \otimes Z_e, \\
   \tilde{\mathbf{Q}}_{V \setminus \{j_p\}} \otimes X_{j_p} \otimes I_e, \\
   \mathbf{I}_{V \setminus \{j_p\}} \otimes X_{j_p} \otimes X_e.
   \end{align*}

\item \textbf{Hadamard Gates on $e$ and $j_p$ ($H_e \otimes H_{j_p}$):}\\
   Applying Hadamard gates swaps the $X$ and $Z$ operators on $e$ and $j_p$, resulting in:
   \begin{align*}
   \mathbf{Q}_{V \setminus \{j_p\}} \otimes X_{j_p} \otimes X_e, \\
   \tilde{\mathbf{Q}}_{V \setminus \{j_p\}} \otimes Z_{j_p} \otimes I_e, \\
   \mathbf{I}_{V \setminus \{j_p\}} \otimes Z_{j_p} \otimes Z_e.
   \end{align*}

\item \textbf{Second Controlled-NOT Gate ($\operatorname{CNOT}_{e \rightarrow j_p}$):}\\
   The second controlled-NOT gate completes the swap, leaving the stabilizers as:
   \begin{align*}
   \mathbf{Q}_{V \setminus \{j_p\}} \otimes I_{j_p} \otimes X_e, \\
   \tilde{\mathbf{Q}}_{V \setminus \{j_p\}} \otimes Z_{j_p} \otimes Z_e, \\
   \mathbf{I}_{V \setminus \{j_p\}} \otimes Z_{j_p} \otimes I_e.
   \end{align*}

\item \textbf{Final Hadamard Gate on $e$ ($H_e$):}\\
   A final Hadamard gate on $e$ swaps the remaining $X$ and $Z$ operators, resulting in:
   \begin{align*}
   \mathbf{Q}_{V \setminus \{j_p\}} \otimes I_{j_p} \otimes Z_e, \\
   \tilde{\mathbf{Q}}_{V \setminus \{j_p\}} \otimes Z_{j_p} \otimes X_e, \\
   \mathbf{I}_{V \setminus \{j_p\}} \otimes Z_{j_p} \otimes I_e.
   \end{align*}
\end{enumerate}

This sequence of operations effectively swaps the quantum state of the free emitter $e$ with the $j_p^{\text{th}}$ photon, as reflected in the updated stabilizer generators. This sequence of operations can be combined to give the time-forward circuit in Table~\ref{tab:SFE_cases}. 

\subsection{Absorption by an entangled emitter}
Consider an $m'$ qubit graph state $\ket{G'}$ between $m_e'\geq 1$ emitters and $m_p' \geq 1$ photons. Consider $k_e, j_p \in V(G')$, where $k_e$ represents an emitter qubit and $j_p$ represents a photonic qubit, $k\in[1,{m_e'}]$ and $j\in[1,m_{p'}]$.

In this section, $\mathbf{Q}_{V \setminus \{j_p,k_e\}}$, $\mathbf{\tilde{Q}}_{V \setminus \{j_p,k_e\}}$, and $\mathbf{\bar{Q}}_{V \setminus \{j_p,k_e\}}$ denotes the tensor product of Pauli operators acting on all qubits in the graph state except $k_e$ and $j_p$.
$\mathbf{I}_{V \setminus \{j_p,k_e\}}$ represents the identity operator acting on all qubits in $V \setminus \{j_p,k_e\}$.
Other terms follow the same notation as in the stabilizer formalism.

\subsubsection{Case I}
 Assume the neighborhood of $k_e$ is given by $\mathbf{N}_{G'}(k_e) = \{j_p\}$, meaning that the emitter qubit $k_e$ shares an edge only to the photonic qubit $j_p$. 

We apply a Hadamard gate on $k_e$, followed by a $\operatorname{CNOT}_{k_e,j_p}$ operation. After these operations, the transformed graph state is denoted as $G'_1$. 

\begin{enumerate}
\item \textbf{Initial Stabilizers:}\\
The stabilizer generators before the transformations are:
\begin{align*}
    \mathbf{I}_{V \setminus\{j_p,k_e\}} \otimes X_{k_e} \otimes Z_{j_p}, \\
    \mathbf{Q}_{V \setminus\{j_p,k_e\}} \otimes Z_{k_e} \otimes X_{j_p}, \\
    \tilde{\mathbf{Q}}_{V \setminus\{j_p,k_e\}} \otimes I_{k_e} \otimes Z_{j_p}.
\end{align*}
In the initial state, the emitter qubit $k_e$ has an edge only with a photonic qubit $j_p$, while $j_p$ may also share edges with other qubits in the graph. Consequently, the stabilizers of the initial state can be described in the form above. 

\item \textbf{After Applying the Hadamard Gate on $k_e$ ($H_{k_e}$):}\\
The Hadamard gate swaps the $X$ and $Z$ operators on $k_e$, resulting in:
\begin{align*}
    \mathbf{I}_{V \setminus\{j_p,k_e\}} \otimes Z_{k_e} \otimes Z_{j_p}, \\
    \mathbf{Q}_{V \setminus\{j_p,k_e\}} \otimes X_{k_e} \otimes X_{j_p}, \\
    \tilde{\mathbf{Q}}_{V \setminus\{j_p,k_e\}} \otimes I_{k_e} \otimes Z_{j_p}.
\end{align*}

\item \textbf{After Applying the Controlled-NOT Gate ($\operatorname{CNOT}_{k_e,j_p}$):}\\
The controlled-NOT gate with $k_e$ as the control and $j_p$ as the target updates the stabilizers as follows:
\begin{align*}
    \mathbf{I}_{V \setminus\{j_p,k_e\}} \otimes I_{k_e} \otimes Z_{j_p}, \\
    \mathbf{Q}_{V \setminus\{j_p,k_e\}} \otimes X_{k_e} \otimes I_{j_p}, \\
    \tilde{\mathbf{Q}}_{V \setminus\{j_p,k_e\}} \otimes Z_{k_e} \otimes I_{j_p}.
\end{align*}
\end{enumerate}

We see that after the transformation, the photon $j_p$ is in $\ket{0}_p$ state. The operators given above describe the time-reversed operation of Case I in Table~\ref{tab:AEE_cases}.

\subsubsection{Case 2}
Assume the neighborhood of $j_p$ is given by $\mathbf{N}_{G'}(j_p) = \{k_e\}$, meaning that the photonic qubit $j_p$ shares an edge only to the emitter qubit $k_e$.

We apply a Hadamard gate on $j_p$, followed by a $\operatorname{CNOT}_{k_e,j_p}$ operation. After these operations, the transformed graph state is denoted as $G'_1$.

\begin{enumerate}
\item \textbf{Initial Stabilizers:}\\
The stabilizer generators before the transformations are:
\begin{align*}
    \mathbf{I}_{V \setminus\{j_p,k_e\}} \otimes X_{j_p} \otimes Z_{k_e}, \\
    \mathbf{Q}_{V \setminus\{j_p,k_e\}} \otimes Z_{j_p} \otimes X_{k_e}, \\
    \tilde{\mathbf{Q}}_{V \setminus\{j_p,k_e\}} \otimes I_{j_p} \otimes Z_{k_e}.
\end{align*}
In the initial state, the photonic qubit $j_p$ has an edge only with the emitter qubit $k_e$, while $k_e$ may also share edges with other qubits in the graph. Consequently, the stabilizers of the initial state can be described in the form above.

\item \textbf{After Applying the Hadamard Gate on $j_p$ ($H_{j_p}$):}\\
The Hadamard gate swaps the $X$ and $Z$ operators on $j_p$, resulting in:
\begin{align*}
    \mathbf{I}_{V \setminus\{j_p,k_e\}} \otimes Z_{j_p} \otimes Z_{k_e}, \\
    \mathbf{Q}_{V \setminus\{j_p,k_e\}} \otimes X_{j_p} \otimes X_{k_e}, \\
    \tilde{\mathbf{Q}}_{V \setminus\{j_p,k_e\}} \otimes I_{j_p} \otimes Z_{k_e}.
\end{align*}

\item \textbf{After Applying the Controlled-NOT Gate ($\operatorname{CNOT}_{k_e,j_p}$):}\\
The controlled-NOT gate with $k_e$ as the control and $j_p$ as the target updates the stabilizers as follows:
\begin{align*}
    \mathbf{I}_{V \setminus\{j_p,k_e\}} \otimes Z_{j_p} \otimes I_{k_e}, \\
    \mathbf{Q}_{V \setminus\{j_p,k_e\}} \otimes I_{j_p} \otimes X_{k_e}, \\
    \tilde{\mathbf{Q}}_{V \setminus\{j_p,k_e\}} \otimes I_{j_p} \otimes Z_{k_e}.
\end{align*}
\end{enumerate}

We see that after the transformation, the photonic qubit $j_p$ is in the $\ket{0}_p$ state. The operators given above describe the time-reversed operation of Case 2 in Table~\ref{tab:AEE_cases}.

\subsubsection{Case 3}
Assume that the neighborhoods of $k_e$ and $j_p$ are identical, i.e., $\mathbf{N}(k_e) = \mathbf{N}(j_p)$. To transform the graph state, we first apply Hadamard gates to both $k_e$ and $j_p$, followed by a $\operatorname{CNOT}_{k_e,j_p}$ operation. This sequence of operations modifies the stabilizer group as follows.

\begin{enumerate}
\item \textbf{Initial Stabilizers:}\\
The stabilizer generators before the transformations are:
\begin{align*}
    \mathbf{Q}_{V \setminus\{j_p,k_e\}} \otimes X_{j_p} \otimes I_{k_e}, \\
    \mathbf{Q}_{V \setminus\{j_p,k_e\}} \otimes I_{j_p} \otimes X_{k_e}, \\
    \tilde{\mathbf{Q}}_{V \setminus\{j_p,k_e\}} \otimes Z_{j_p} \otimes Z_{k_e}, \\
    \bar{\mathbf{Q}}_{V \setminus\{j_p,k_e\}} \otimes I_{j_p} \otimes I_{k_e}.
\end{align*}
Since the neighborhoods of $k_e$ and $j_p$ are identical, their stabilizers are symmetric with respect to $j_p$ and $k_e$, as described above.

\item \textbf{After Applying the Hadamard Gates on $j_p$ and $k_e$ ($H_{j_p} \otimes H_{k_e}$):}\\
The Hadamard gates swap the $X$ and $Z$ operators on $j_p$ and $k_e$, resulting in:
\begin{align*}
    \mathbf{Q}_{V \setminus\{j_p,k_e\}} \otimes Z_{j_p} \otimes I_{k_e}, \\
    \mathbf{Q}_{V \setminus\{j_p,k_e\}} \otimes I_{j_p} \otimes Z_{k_e}, \\
    \tilde{\mathbf{Q}}_{V \setminus\{j_p,k_e\}} \otimes X_{j_p} \otimes X_{k_e}, \\
    \bar{\mathbf{Q}}_{V \setminus\{j_p,k_e\}} \otimes I_{j_p} \otimes I_{k_e}.
\end{align*}

\item \textbf{After Applying the Controlled-NOT Gate ($\operatorname{CNOT}_{k_e,j_p}$):}\\
The controlled-NOT gate with $k_e$ as the control and $j_p$ as the target updates the stabilizers as follows:
\begin{align*}
    \mathbf{Q}_{V \setminus\{j_p,k_e\}} \otimes Z_{j_p} \otimes Z_{k_e}, \\
    \mathbf{Q}_{V \setminus\{j_p,k_e\}} \otimes I_{j_p} \otimes Z_{k_e}, \\
    \tilde{\mathbf{Q}}_{V \setminus\{j_p,k_e\}} \otimes I_{j_p} \otimes X_{k_e}, \\
    \bar{\mathbf{Q}}_{V \setminus\{j_p,k_e\}} \otimes I_{j_p} \otimes I_{k_e}.
\end{align*}

\item \textbf{After Applying the Final Hadamard Gate on $k_e$ ($H_{k_e}$):}\\
The Hadamard gate on $k_e$ swaps the $X$ and $Z$ operators on $k_e$, resulting in:
\begin{align*}
    \mathbf{Q}_{V \setminus\{j_p,k_e\}} \otimes Z_{j_p} \otimes X_{k_e}, \\
    \mathbf{Q}_{V \setminus\{j_p,k_e\}} \otimes I_{j_p} \otimes X_{k_e}, \\
    \tilde{\mathbf{Q}}_{V \setminus\{j_p,k_e\}} \otimes I_{j_p} \otimes Z_{k_e}, \\
    \bar{\mathbf{Q}}_{V \setminus\{j_p,k_e\}} \otimes I_{j_p} \otimes I_{k_e}.  
\end{align*}

\end{enumerate}

After these transformations, the photonic qubit $j_p$ is in the $\ket{0}_{j_p}$ state, and all edges in the set $E^G(j_p)$ are removed. This effectively disentangles $j_p$ from the graph state.

\begin{acknowledgments}
This work was funded by the Army Research Office
(ARO) MURI on Quantum Network Science under grant
number W911NF2110325, and also benefited from partial funding from the National Science Foundation (NSF) Engineering Research Center (ERC) Center for Quantum Networks (CQN) awarded under grant number 1941583. EK and AP acknowledge Prajit Dhara’s insights on the SiV vacancy center.
\end{acknowledgments}



\bibliography{bibFile}

\begin{thebibliography}{59}%
\makeatletter
\providecommand \@ifxundefined [1]{%
 \@ifx{#1\undefined}
}%
\providecommand \@ifnum [1]{%
 \ifnum #1\expandafter \@firstoftwo
 \else \expandafter \@secondoftwo
 \fi
}%
\providecommand \@ifx [1]{%
 \ifx #1\expandafter \@firstoftwo
 \else \expandafter \@secondoftwo
 \fi
}%
\providecommand \natexlab [1]{#1}%
\providecommand \enquote  [1]{``#1''}%
\providecommand \bibnamefont  [1]{#1}%
\providecommand \bibfnamefont [1]{#1}%
\providecommand \citenamefont [1]{#1}%
\providecommand \href@noop [0]{\@secondoftwo}%
\providecommand \href [0]{\begingroup \@sanitize@url \@href}%
\providecommand \@href[1]{\@@startlink{#1}\@@href}%
\providecommand \@@href[1]{\endgroup#1\@@endlink}%
\providecommand \@sanitize@url [0]{\catcode `\\12\catcode `\$12\catcode `\&12\catcode `\#12\catcode `\^12\catcode `\_12\catcode `\%12\relax}%
\providecommand \@@startlink[1]{}%
\providecommand \@@endlink[0]{}%
\providecommand \url  [0]{\begingroup\@sanitize@url \@url }%
\providecommand \@url [1]{\endgroup\@href {#1}{\urlprefix }}%
\providecommand \urlprefix  [0]{URL }%
\providecommand \Eprint [0]{\href }%
\providecommand \doibase [0]{http://dx.doi.org/}%
\providecommand \selectlanguage [0]{\@gobble}%
\providecommand \bibinfo  [0]{\@secondoftwo}%
\providecommand \bibfield  [0]{\@secondoftwo}%
\providecommand \translation [1]{[#1]}%
\providecommand \BibitemOpen [0]{}%
\providecommand \bibitemStop [0]{}%
\providecommand \bibitemNoStop [0]{.\EOS\space}%
\providecommand \EOS [0]{\spacefactor3000\relax}%
\providecommand \BibitemShut  [1]{\csname bibitem#1\endcsname}%
\let\auto@bib@innerbib\@empty
\bibitem [{\citenamefont {Bartolucci}\ \emph {et~al.}(2021)\citenamefont {Bartolucci}, \citenamefont {Birchall}, \citenamefont {Bombin}, \citenamefont {Cable}, \citenamefont {Dawson}, \citenamefont {Gimeno-Segovia}, \citenamefont {Johnston}, \citenamefont {Kieling}, \citenamefont {Nickerson}, \citenamefont {Pant}, \citenamefont {Pastawski}, \citenamefont {Rudolph},\ and\ \citenamefont {Sparrow}}]{bartolucci2021fusionbased}%
  \BibitemOpen
  \bibfield  {author} {\bibinfo {author} {\bibfnamefont {S.}~\bibnamefont {Bartolucci}}, \bibinfo {author} {\bibfnamefont {P.}~\bibnamefont {Birchall}}, \bibinfo {author} {\bibfnamefont {H.}~\bibnamefont {Bombin}}, \bibinfo {author} {\bibfnamefont {H.}~\bibnamefont {Cable}}, \bibinfo {author} {\bibfnamefont {C.}~\bibnamefont {Dawson}}, \bibinfo {author} {\bibfnamefont {M.}~\bibnamefont {Gimeno-Segovia}}, \bibinfo {author} {\bibfnamefont {E.}~\bibnamefont {Johnston}}, \bibinfo {author} {\bibfnamefont {K.}~\bibnamefont {Kieling}}, \bibinfo {author} {\bibfnamefont {N.}~\bibnamefont {Nickerson}}, \bibinfo {author} {\bibfnamefont {M.}~\bibnamefont {Pant}}, \bibinfo {author} {\bibfnamefont {F.}~\bibnamefont {Pastawski}}, \bibinfo {author} {\bibfnamefont {T.}~\bibnamefont {Rudolph}}, \ and\ \bibinfo {author} {\bibfnamefont {C.}~\bibnamefont {Sparrow}},\ }\href@noop {} {\enquote {\bibinfo {title} {Fusion-based quantum computation},}\ } (\bibinfo {year} {2021}),\ \Eprint {http://arxiv.org/abs/2101.09310}
  {arXiv:2101.09310 [quant-ph]} \BibitemShut {NoStop}%
\bibitem [{\citenamefont {Kitaev}(2003)}]{Kitaev2003}%
  \BibitemOpen
  \bibfield  {author} {\bibinfo {author} {\bibfnamefont {A.}~\bibnamefont {Kitaev}},\ }\href {\doibase 10.1016/s0003-4916(02)00018-0} {\bibfield  {journal} {\bibinfo  {journal} {Annals of Physics}\ }\textbf {\bibinfo {volume} {303}},\ \bibinfo {pages} {2} (\bibinfo {year} {2003})}\BibitemShut {NoStop}%
\bibitem [{\citenamefont {Raussendorf}\ and\ \citenamefont {Briegel}(2001)}]{Briegel}%
  \BibitemOpen
  \bibfield  {author} {\bibinfo {author} {\bibfnamefont {R.}~\bibnamefont {Raussendorf}}\ and\ \bibinfo {author} {\bibfnamefont {H.~J.}\ \bibnamefont {Briegel}},\ }\href {\doibase 10.1103/PhysRevLett.86.5188} {\bibfield  {journal} {\bibinfo  {journal} {Phys. Rev. Lett.}\ }\textbf {\bibinfo {volume} {86}},\ \bibinfo {pages} {5188} (\bibinfo {year} {2001})}\BibitemShut {NoStop}%
\bibitem [{\citenamefont {Sangouard}\ \emph {et~al.}(2011)\citenamefont {Sangouard}, \citenamefont {Simon}, \citenamefont {de~Riedmatten},\ and\ \citenamefont {Gisin}}]{RevModPhys.83.33}%
  \BibitemOpen
  \bibfield  {author} {\bibinfo {author} {\bibfnamefont {N.}~\bibnamefont {Sangouard}}, \bibinfo {author} {\bibfnamefont {C.}~\bibnamefont {Simon}}, \bibinfo {author} {\bibfnamefont {H.}~\bibnamefont {de~Riedmatten}}, \ and\ \bibinfo {author} {\bibfnamefont {N.}~\bibnamefont {Gisin}},\ }\href {\doibase 10.1103/RevModPhys.83.33} {\bibfield  {journal} {\bibinfo  {journal} {Rev. Mod. Phys.}\ }\textbf {\bibinfo {volume} {83}},\ \bibinfo {pages} {33} (\bibinfo {year} {2011})}\BibitemShut {NoStop}%
\bibitem [{\citenamefont {Azuma}\ \emph {et~al.}(2015)\citenamefont {Azuma}, \citenamefont {Tamaki},\ and\ \citenamefont {Lo}}]{Azuma2015}%
  \BibitemOpen
  \bibfield  {author} {\bibinfo {author} {\bibfnamefont {K.}~\bibnamefont {Azuma}}, \bibinfo {author} {\bibfnamefont {K.}~\bibnamefont {Tamaki}}, \ and\ \bibinfo {author} {\bibfnamefont {H.-K.}\ \bibnamefont {Lo}},\ }\href {\doibase 10.1038/ncomms7787} {\bibfield  {journal} {\bibinfo  {journal} {Nature Communications}\ }\textbf {\bibinfo {volume} {6}} (\bibinfo {year} {2015}),\ 10.1038/ncomms7787}\BibitemShut {NoStop}%
\bibitem [{\citenamefont {Shettell}\ and\ \citenamefont {Markham}(2020)}]{Damian2020}%
  \BibitemOpen
  \bibfield  {author} {\bibinfo {author} {\bibfnamefont {N.}~\bibnamefont {Shettell}}\ and\ \bibinfo {author} {\bibfnamefont {D.}~\bibnamefont {Markham}},\ }\href {\doibase 10.1103/PhysRevLett.124.110502} {\bibfield  {journal} {\bibinfo  {journal} {Phys. Rev. Lett.}\ }\textbf {\bibinfo {volume} {124}},\ \bibinfo {pages} {110502} (\bibinfo {year} {2020})}\BibitemShut {NoStop}%
\bibitem [{\citenamefont {Raussendorf}\ \emph {et~al.}(2007)\citenamefont {Raussendorf}, \citenamefont {Harrington},\ and\ \citenamefont {Goyal}}]{Raussendorf2007}%
  \BibitemOpen
  \bibfield  {author} {\bibinfo {author} {\bibfnamefont {R.}~\bibnamefont {Raussendorf}}, \bibinfo {author} {\bibfnamefont {J.}~\bibnamefont {Harrington}}, \ and\ \bibinfo {author} {\bibfnamefont {K.}~\bibnamefont {Goyal}},\ }\href@noop {} {\bibfield  {journal} {\bibinfo  {journal} {New Journal of Physics}\ }\textbf {\bibinfo {volume} {9}},\ \bibinfo {pages} {199} (\bibinfo {year} {2007})}\BibitemShut {NoStop}%
\bibitem [{\citenamefont {Kieling}\ \emph {et~al.}(2007)\citenamefont {Kieling}, \citenamefont {Rudolph},\ and\ \citenamefont {Eisert}}]{Kieling2007}%
  \BibitemOpen
  \bibfield  {author} {\bibinfo {author} {\bibfnamefont {K.}~\bibnamefont {Kieling}}, \bibinfo {author} {\bibfnamefont {T.}~\bibnamefont {Rudolph}}, \ and\ \bibinfo {author} {\bibfnamefont {J.}~\bibnamefont {Eisert}},\ }\href@noop {} {\bibfield  {journal} {\bibinfo  {journal} {Phys. Rev. Lett.}\ }\textbf {\bibinfo {volume} {99}},\ \bibinfo {pages} {130501} (\bibinfo {year} {2007})}\BibitemShut {NoStop}%
\bibitem [{\citenamefont {Gimeno-Segovia}\ \emph {et~al.}(2015)\citenamefont {Gimeno-Segovia}, \citenamefont {Shadbolt}, \citenamefont {Browne},\ and\ \citenamefont {Rudolph}}]{GimenoSegovia2015}%
  \BibitemOpen
  \bibfield  {author} {\bibinfo {author} {\bibfnamefont {M.}~\bibnamefont {Gimeno-Segovia}}, \bibinfo {author} {\bibfnamefont {P.}~\bibnamefont {Shadbolt}}, \bibinfo {author} {\bibfnamefont {D.~E.}\ \bibnamefont {Browne}}, \ and\ \bibinfo {author} {\bibfnamefont {T.}~\bibnamefont {Rudolph}},\ }\href@noop {} {\bibfield  {journal} {\bibinfo  {journal} {Phys. Rev. Lett.}\ }\textbf {\bibinfo {volume} {115}},\ \bibinfo {pages} {020502} (\bibinfo {year} {2015})}\BibitemShut {NoStop}%
\bibitem [{\citenamefont {Pant}\ \emph {et~al.}(2019)\citenamefont {Pant}, \citenamefont {Towsley}, \citenamefont {Englund},\ and\ \citenamefont {Guha}}]{Pant2019}%
  \BibitemOpen
  \bibfield  {author} {\bibinfo {author} {\bibfnamefont {M.}~\bibnamefont {Pant}}, \bibinfo {author} {\bibfnamefont {D.}~\bibnamefont {Towsley}}, \bibinfo {author} {\bibfnamefont {D.}~\bibnamefont {Englund}}, \ and\ \bibinfo {author} {\bibfnamefont {S.}~\bibnamefont {Guha}},\ }\href@noop {} {\bibfield  {journal} {\bibinfo  {journal} {Nat. Commun.}\ }\textbf {\bibinfo {volume} {10}},\ \bibinfo {pages} {1070} (\bibinfo {year} {2019})}\BibitemShut {NoStop}%
\bibitem [{\citenamefont {Silverstone}\ \emph {et~al.}(2016)\citenamefont {Silverstone}, \citenamefont {Bonneau}, \citenamefont {O’Brien},\ and\ \citenamefont {Thompson}}]{silverstone2016silicon}%
  \BibitemOpen
  \bibfield  {author} {\bibinfo {author} {\bibfnamefont {J.~W.}\ \bibnamefont {Silverstone}}, \bibinfo {author} {\bibfnamefont {D.}~\bibnamefont {Bonneau}}, \bibinfo {author} {\bibfnamefont {J.~L.}\ \bibnamefont {O’Brien}}, \ and\ \bibinfo {author} {\bibfnamefont {M.~G.}\ \bibnamefont {Thompson}},\ }\href@noop {} {\bibfield  {journal} {\bibinfo  {journal} {IEEE Journal of Selected Topics in Quantum Electronics}\ }\textbf {\bibinfo {volume} {22}},\ \bibinfo {pages} {390} (\bibinfo {year} {2016})}\BibitemShut {NoStop}%
\bibitem [{\citenamefont {Meyer-Scott}\ \emph {et~al.}(2020)\citenamefont {Meyer-Scott}, \citenamefont {Silberhorn},\ and\ \citenamefont {Migdall}}]{meyer2020single}%
  \BibitemOpen
  \bibfield  {author} {\bibinfo {author} {\bibfnamefont {E.}~\bibnamefont {Meyer-Scott}}, \bibinfo {author} {\bibfnamefont {C.}~\bibnamefont {Silberhorn}}, \ and\ \bibinfo {author} {\bibfnamefont {A.}~\bibnamefont {Migdall}},\ }\href@noop {} {\bibfield  {journal} {\bibinfo  {journal} {Review of Scientific Instruments}\ }\textbf {\bibinfo {volume} {91}},\ \bibinfo {pages} {041101} (\bibinfo {year} {2020})}\BibitemShut {NoStop}%
\bibitem [{\citenamefont {Mosley}\ \emph {et~al.}(2008)\citenamefont {Mosley}, \citenamefont {Lundeen}, \citenamefont {Smith}, \citenamefont {Wasylczyk}, \citenamefont {U’Ren}, \citenamefont {Silberhorn},\ and\ \citenamefont {Walmsley}}]{mosley2008heralded}%
  \BibitemOpen
  \bibfield  {author} {\bibinfo {author} {\bibfnamefont {P.~J.}\ \bibnamefont {Mosley}}, \bibinfo {author} {\bibfnamefont {J.~S.}\ \bibnamefont {Lundeen}}, \bibinfo {author} {\bibfnamefont {B.~J.}\ \bibnamefont {Smith}}, \bibinfo {author} {\bibfnamefont {P.}~\bibnamefont {Wasylczyk}}, \bibinfo {author} {\bibfnamefont {A.~B.}\ \bibnamefont {U’Ren}}, \bibinfo {author} {\bibfnamefont {.~f.~C.}\ \bibnamefont {Silberhorn}}, \ and\ \bibinfo {author} {\bibfnamefont {I.~A.}\ \bibnamefont {Walmsley}},\ }\href@noop {} {\bibfield  {journal} {\bibinfo  {journal} {Physical Review Letters}\ }\textbf {\bibinfo {volume} {100}},\ \bibinfo {pages} {133601} (\bibinfo {year} {2008})}\BibitemShut {NoStop}%
\bibitem [{\citenamefont {Alexander}\ \emph {et~al.}(2024)\citenamefont {Alexander}, \citenamefont {Bahgat}, \citenamefont {Benyamini}, \citenamefont {Black}, \citenamefont {Bonneau}, \citenamefont {Burgos}, \citenamefont {Burridge}, \citenamefont {Campbell}, \citenamefont {Catalano}, \citenamefont {Ceballos} \emph {et~al.}}]{alexander2024manufacturable}%
  \BibitemOpen
  \bibfield  {author} {\bibinfo {author} {\bibfnamefont {K.}~\bibnamefont {Alexander}}, \bibinfo {author} {\bibfnamefont {A.}~\bibnamefont {Bahgat}}, \bibinfo {author} {\bibfnamefont {A.}~\bibnamefont {Benyamini}}, \bibinfo {author} {\bibfnamefont {D.}~\bibnamefont {Black}}, \bibinfo {author} {\bibfnamefont {D.}~\bibnamefont {Bonneau}}, \bibinfo {author} {\bibfnamefont {S.}~\bibnamefont {Burgos}}, \bibinfo {author} {\bibfnamefont {B.}~\bibnamefont {Burridge}}, \bibinfo {author} {\bibfnamefont {G.}~\bibnamefont {Campbell}}, \bibinfo {author} {\bibfnamefont {G.}~\bibnamefont {Catalano}}, \bibinfo {author} {\bibfnamefont {A.}~\bibnamefont {Ceballos}},  \emph {et~al.},\ }\href@noop {} {\bibfield  {journal} {\bibinfo  {journal} {arXiv preprint arXiv:2404.17570}\ } (\bibinfo {year} {2024})}\BibitemShut {NoStop}%
\bibitem [{\citenamefont {Senellart}\ \emph {et~al.}(2017)\citenamefont {Senellart}, \citenamefont {Solomon},\ and\ \citenamefont {White}}]{senellart2017high}%
  \BibitemOpen
  \bibfield  {author} {\bibinfo {author} {\bibfnamefont {P.}~\bibnamefont {Senellart}}, \bibinfo {author} {\bibfnamefont {G.}~\bibnamefont {Solomon}}, \ and\ \bibinfo {author} {\bibfnamefont {A.}~\bibnamefont {White}},\ }\href@noop {} {\bibfield  {journal} {\bibinfo  {journal} {Nature nanotechnology}\ }\textbf {\bibinfo {volume} {12}},\ \bibinfo {pages} {1026} (\bibinfo {year} {2017})}\BibitemShut {NoStop}%
\bibitem [{\citenamefont {Aharonovich}\ \emph {et~al.}(2016)\citenamefont {Aharonovich}, \citenamefont {Englund},\ and\ \citenamefont {Toth}}]{aharonovich2016solid}%
  \BibitemOpen
  \bibfield  {author} {\bibinfo {author} {\bibfnamefont {I.}~\bibnamefont {Aharonovich}}, \bibinfo {author} {\bibfnamefont {D.}~\bibnamefont {Englund}}, \ and\ \bibinfo {author} {\bibfnamefont {M.}~\bibnamefont {Toth}},\ }\href@noop {} {\bibfield  {journal} {\bibinfo  {journal} {Nature photonics}\ }\textbf {\bibinfo {volume} {10}},\ \bibinfo {pages} {631} (\bibinfo {year} {2016})}\BibitemShut {NoStop}%
\bibitem [{\citenamefont {Knill}\ \emph {et~al.}(2001)\citenamefont {Knill}, \citenamefont {Laflamme},\ and\ \citenamefont {Milburn}}]{Knill2001}%
  \BibitemOpen
  \bibfield  {author} {\bibinfo {author} {\bibfnamefont {E.}~\bibnamefont {Knill}}, \bibinfo {author} {\bibfnamefont {R.}~\bibnamefont {Laflamme}}, \ and\ \bibinfo {author} {\bibfnamefont {G.~J.}\ \bibnamefont {Milburn}},\ }\href@noop {} {\bibfield  {journal} {\bibinfo  {journal} {Nature}\ }\textbf {\bibinfo {volume} {409}},\ \bibinfo {pages} {46} (\bibinfo {year} {2001})}\BibitemShut {NoStop}%
\bibitem [{\citenamefont {Pant}\ \emph {et~al.}(2017)\citenamefont {Pant}, \citenamefont {Krovi}, \citenamefont {Englund},\ and\ \citenamefont {Guha}}]{Mihir2017}%
  \BibitemOpen
  \bibfield  {author} {\bibinfo {author} {\bibfnamefont {M.}~\bibnamefont {Pant}}, \bibinfo {author} {\bibfnamefont {H.}~\bibnamefont {Krovi}}, \bibinfo {author} {\bibfnamefont {D.}~\bibnamefont {Englund}}, \ and\ \bibinfo {author} {\bibfnamefont {S.}~\bibnamefont {Guha}},\ }\href {\doibase 10.1103/PhysRevA.95.012304} {\bibfield  {journal} {\bibinfo  {journal} {Phys. Rev. A}\ }\textbf {\bibinfo {volume} {95}},\ \bibinfo {pages} {012304} (\bibinfo {year} {2017})}\BibitemShut {NoStop}%
\bibitem [{\citenamefont {Wang}\ and\ \citenamefont {Fang}(2020)}]{Wang2020}%
  \BibitemOpen
  \bibfield  {author} {\bibinfo {author} {\bibfnamefont {Y.}~\bibnamefont {Wang}}\ and\ \bibinfo {author} {\bibfnamefont {K.}~\bibnamefont {Fang}},\ }\href@noop {} {\bibfield  {journal} {\bibinfo  {journal} {Phys. Rev. A}\ }\textbf {\bibinfo {volume} {102}},\ \bibinfo {pages} {052601} (\bibinfo {year} {2020})}\BibitemShut {NoStop}%
\bibitem [{\citenamefont {Varnava}\ \emph {et~al.}(2008)\citenamefont {Varnava}, \citenamefont {Browne},\ and\ \citenamefont {Rudolph}}]{Varnava_2008}%
  \BibitemOpen
  \bibfield  {author} {\bibinfo {author} {\bibfnamefont {M.}~\bibnamefont {Varnava}}, \bibinfo {author} {\bibfnamefont {D.~E.}\ \bibnamefont {Browne}}, \ and\ \bibinfo {author} {\bibfnamefont {T.}~\bibnamefont {Rudolph}},\ }\href {\doibase 10.1103/PhysRevLett.100.060502} {\bibfield  {journal} {\bibinfo  {journal} {Phys. Rev. Lett.}\ }\textbf {\bibinfo {volume} {100}},\ \bibinfo {pages} {060502} (\bibinfo {year} {2008})}\BibitemShut {NoStop}%
\bibitem [{\citenamefont {Li}\ \emph {et~al.}(2015)\citenamefont {Li}, \citenamefont {Humphreys}, \citenamefont {Mendoza},\ and\ \citenamefont {Benjamin}}]{Ying2015}%
  \BibitemOpen
  \bibfield  {author} {\bibinfo {author} {\bibfnamefont {Y.}~\bibnamefont {Li}}, \bibinfo {author} {\bibfnamefont {P.~C.}\ \bibnamefont {Humphreys}}, \bibinfo {author} {\bibfnamefont {G.~J.}\ \bibnamefont {Mendoza}}, \ and\ \bibinfo {author} {\bibfnamefont {S.~C.}\ \bibnamefont {Benjamin}},\ }\href {\doibase 10.1103/PhysRevX.5.041007} {\bibfield  {journal} {\bibinfo  {journal} {Phys. Rev. X}\ }\textbf {\bibinfo {volume} {5}},\ \bibinfo {pages} {041007} (\bibinfo {year} {2015})}\BibitemShut {NoStop}%
\bibitem [{Note1()}]{Note1}%
  \BibitemOpen
  \bibinfo {note} {There have been studies on recycling graph states generated from failed fusion attempts and optimizing the space-time multiplexing for fully-LO preparation of photonic graph states~\cite {Gimeno-Segovia2017}.}\BibitemShut {Stop}%
\bibitem [{\citenamefont {Lindner}\ and\ \citenamefont {Rudolph}(2009{\natexlab{a}})}]{lindner2009proposal}%
  \BibitemOpen
  \bibfield  {author} {\bibinfo {author} {\bibfnamefont {N.~H.}\ \bibnamefont {Lindner}}\ and\ \bibinfo {author} {\bibfnamefont {T.}~\bibnamefont {Rudolph}},\ }\href@noop {} {\bibfield  {journal} {\bibinfo  {journal} {Physical review letters}\ }\textbf {\bibinfo {volume} {103}},\ \bibinfo {pages} {113602} (\bibinfo {year} {2009}{\natexlab{a}})}\BibitemShut {NoStop}%
\bibitem [{\citenamefont {Li}\ \emph {et~al.}(2022)\citenamefont {Li}, \citenamefont {Economou},\ and\ \citenamefont {Barnes}}]{Li2022}%
  \BibitemOpen
  \bibfield  {author} {\bibinfo {author} {\bibfnamefont {B.}~\bibnamefont {Li}}, \bibinfo {author} {\bibfnamefont {S.~E.}\ \bibnamefont {Economou}}, \ and\ \bibinfo {author} {\bibfnamefont {E.}~\bibnamefont {Barnes}},\ }\href {\doibase 10.1038/s41534-022-00522-6} {\bibfield  {journal} {\bibinfo  {journal} {npj Quantum Information}\ }\textbf {\bibinfo {volume} {8}} (\bibinfo {year} {2022}),\ 10.1038/s41534-022-00522-6}\BibitemShut {NoStop}%
\bibitem [{\citenamefont {Sch\"on}\ \emph {et~al.}(2005)\citenamefont {Sch\"on}, \citenamefont {Solano}, \citenamefont {Verstraete}, \citenamefont {Cirac},\ and\ \citenamefont {Wolf}}]{Schon2005}%
  \BibitemOpen
  \bibfield  {author} {\bibinfo {author} {\bibfnamefont {C.}~\bibnamefont {Sch\"on}}, \bibinfo {author} {\bibfnamefont {E.}~\bibnamefont {Solano}}, \bibinfo {author} {\bibfnamefont {F.}~\bibnamefont {Verstraete}}, \bibinfo {author} {\bibfnamefont {J.~I.}\ \bibnamefont {Cirac}}, \ and\ \bibinfo {author} {\bibfnamefont {M.~M.}\ \bibnamefont {Wolf}},\ }\href {\doibase 10.1103/PhysRevLett.95.110503} {\bibfield  {journal} {\bibinfo  {journal} {Phys. Rev. Lett.}\ }\textbf {\bibinfo {volume} {95}},\ \bibinfo {pages} {110503} (\bibinfo {year} {2005})}\BibitemShut {NoStop}%
\bibitem [{\citenamefont {Sch\"on}\ \emph {et~al.}(2007)\citenamefont {Sch\"on}, \citenamefont {Hammerer}, \citenamefont {Wolf}, \citenamefont {Cirac},\ and\ \citenamefont {Solano}}]{Schon2007}%
  \BibitemOpen
  \bibfield  {author} {\bibinfo {author} {\bibfnamefont {C.}~\bibnamefont {Sch\"on}}, \bibinfo {author} {\bibfnamefont {K.}~\bibnamefont {Hammerer}}, \bibinfo {author} {\bibfnamefont {M.~M.}\ \bibnamefont {Wolf}}, \bibinfo {author} {\bibfnamefont {J.~I.}\ \bibnamefont {Cirac}}, \ and\ \bibinfo {author} {\bibfnamefont {E.}~\bibnamefont {Solano}},\ }\href {\doibase 10.1103/PhysRevA.75.032311} {\bibfield  {journal} {\bibinfo  {journal} {Phys. Rev. A}\ }\textbf {\bibinfo {volume} {75}},\ \bibinfo {pages} {032311} (\bibinfo {year} {2007})}\BibitemShut {NoStop}%
\bibitem [{\citenamefont {Stas}\ \emph {et~al.}(2022)\citenamefont {Stas}, \citenamefont {Huan}, \citenamefont {Machielse}, \citenamefont {Knall}, \citenamefont {Suleymanzade}, \citenamefont {Pingault}, \citenamefont {Sutula}, \citenamefont {Ding}, \citenamefont {Knaut}, \citenamefont {Assumpcao}, \citenamefont {Wei}, \citenamefont {Bhaskar}, \citenamefont {Riedinger}, \citenamefont {Sukachev}, \citenamefont {Park}, \citenamefont {Lončar}, \citenamefont {Levonian},\ and\ \citenamefont {Lukin}}]{Stas2022}%
  \BibitemOpen
  \bibfield  {author} {\bibinfo {author} {\bibfnamefont {P.-J.}\ \bibnamefont {Stas}}, \bibinfo {author} {\bibfnamefont {Y.~Q.}\ \bibnamefont {Huan}}, \bibinfo {author} {\bibfnamefont {B.}~\bibnamefont {Machielse}}, \bibinfo {author} {\bibfnamefont {E.~N.}\ \bibnamefont {Knall}}, \bibinfo {author} {\bibfnamefont {A.}~\bibnamefont {Suleymanzade}}, \bibinfo {author} {\bibfnamefont {B.}~\bibnamefont {Pingault}}, \bibinfo {author} {\bibfnamefont {M.}~\bibnamefont {Sutula}}, \bibinfo {author} {\bibfnamefont {S.~W.}\ \bibnamefont {Ding}}, \bibinfo {author} {\bibfnamefont {C.~M.}\ \bibnamefont {Knaut}}, \bibinfo {author} {\bibfnamefont {D.~R.}\ \bibnamefont {Assumpcao}}, \bibinfo {author} {\bibfnamefont {Y.-C.}\ \bibnamefont {Wei}}, \bibinfo {author} {\bibfnamefont {M.~K.}\ \bibnamefont {Bhaskar}}, \bibinfo {author} {\bibfnamefont {R.}~\bibnamefont {Riedinger}}, \bibinfo {author} {\bibfnamefont {D.~D.}\ \bibnamefont {Sukachev}}, \bibinfo {author} {\bibfnamefont {H.}~\bibnamefont {Park}}, \bibinfo {author}
  {\bibfnamefont {M.}~\bibnamefont {Lončar}}, \bibinfo {author} {\bibfnamefont {D.~S.}\ \bibnamefont {Levonian}}, \ and\ \bibinfo {author} {\bibfnamefont {M.~D.}\ \bibnamefont {Lukin}},\ }\href {\doibase 10.1126/science.add9771} {\bibfield  {journal} {\bibinfo  {journal} {Science}\ }\textbf {\bibinfo {volume} {378}},\ \bibinfo {pages} {557–560} (\bibinfo {year} {2022})}\BibitemShut {NoStop}%
\bibitem [{\citenamefont {Dhara}\ \emph {et~al.}(2023)\citenamefont {Dhara}, \citenamefont {Englund},\ and\ \citenamefont {Guha}}]{Dhara2023}%
  \BibitemOpen
  \bibfield  {author} {\bibinfo {author} {\bibfnamefont {P.}~\bibnamefont {Dhara}}, \bibinfo {author} {\bibfnamefont {D.}~\bibnamefont {Englund}}, \ and\ \bibinfo {author} {\bibfnamefont {S.}~\bibnamefont {Guha}},\ }\href@noop {} {\bibfield  {journal} {\bibinfo  {journal} {Phys. Rev. Res.}\ }\textbf {\bibinfo {volume} {5}},\ \bibinfo {pages} {033149} (\bibinfo {year} {2023})}\BibitemShut {NoStop}%
\bibitem [{\citenamefont {Inlek}\ \emph {et~al.}(2017)\citenamefont {Inlek}, \citenamefont {Crocker}, \citenamefont {Lichtman}, \citenamefont {Sosnova},\ and\ \citenamefont {Monroe}}]{inlek2017multispecies}%
  \BibitemOpen
  \bibfield  {author} {\bibinfo {author} {\bibfnamefont {I.~V.}\ \bibnamefont {Inlek}}, \bibinfo {author} {\bibfnamefont {C.}~\bibnamefont {Crocker}}, \bibinfo {author} {\bibfnamefont {M.}~\bibnamefont {Lichtman}}, \bibinfo {author} {\bibfnamefont {K.}~\bibnamefont {Sosnova}}, \ and\ \bibinfo {author} {\bibfnamefont {C.}~\bibnamefont {Monroe}},\ }\href@noop {} {\bibfield  {journal} {\bibinfo  {journal} {Physical review letters}\ }\textbf {\bibinfo {volume} {118}},\ \bibinfo {pages} {250502} (\bibinfo {year} {2017})}\BibitemShut {NoStop}%
\bibitem [{\citenamefont {Tan}\ \emph {et~al.}(2015)\citenamefont {Tan}, \citenamefont {Gaebler}, \citenamefont {Lin}, \citenamefont {Wan}, \citenamefont {Bowler}, \citenamefont {Leibfried},\ and\ \citenamefont {Wineland}}]{tan2015multi}%
  \BibitemOpen
  \bibfield  {author} {\bibinfo {author} {\bibfnamefont {T.~R.}\ \bibnamefont {Tan}}, \bibinfo {author} {\bibfnamefont {J.~P.}\ \bibnamefont {Gaebler}}, \bibinfo {author} {\bibfnamefont {Y.}~\bibnamefont {Lin}}, \bibinfo {author} {\bibfnamefont {Y.}~\bibnamefont {Wan}}, \bibinfo {author} {\bibfnamefont {R.}~\bibnamefont {Bowler}}, \bibinfo {author} {\bibfnamefont {D.}~\bibnamefont {Leibfried}}, \ and\ \bibinfo {author} {\bibfnamefont {D.~J.}\ \bibnamefont {Wineland}},\ }\href@noop {} {\bibfield  {journal} {\bibinfo  {journal} {Nature}\ }\textbf {\bibinfo {volume} {528}},\ \bibinfo {pages} {380} (\bibinfo {year} {2015})}\BibitemShut {NoStop}%
\bibitem [{\citenamefont {Ghanbari}\ \emph {et~al.}(2024)\citenamefont {Ghanbari}, \citenamefont {Lin}, \citenamefont {MacLellan}, \citenamefont {Robichaud}, \citenamefont {Roztocki},\ and\ \citenamefont {Lo}}]{Ghanbari2024}%
  \BibitemOpen
  \bibfield  {author} {\bibinfo {author} {\bibfnamefont {S.}~\bibnamefont {Ghanbari}}, \bibinfo {author} {\bibfnamefont {J.}~\bibnamefont {Lin}}, \bibinfo {author} {\bibfnamefont {B.}~\bibnamefont {MacLellan}}, \bibinfo {author} {\bibfnamefont {L.}~\bibnamefont {Robichaud}}, \bibinfo {author} {\bibfnamefont {P.}~\bibnamefont {Roztocki}}, \ and\ \bibinfo {author} {\bibfnamefont {H.-K.}\ \bibnamefont {Lo}},\ }\href@noop {} {\bibfield  {journal} {\bibinfo  {journal} {Phys. Rev. A (Coll. Park.)}\ }\textbf {\bibinfo {volume} {110}},\ \bibinfo {pages} {052605} (\bibinfo {year} {2024})}\BibitemShut {NoStop}%
\bibitem [{\citenamefont {Buterakos}\ \emph {et~al.}(2017)\citenamefont {Buterakos}, \citenamefont {Barnes},\ and\ \citenamefont {Economou}}]{Buterakos2017}%
  \BibitemOpen
  \bibfield  {author} {\bibinfo {author} {\bibfnamefont {D.}~\bibnamefont {Buterakos}}, \bibinfo {author} {\bibfnamefont {E.}~\bibnamefont {Barnes}}, \ and\ \bibinfo {author} {\bibfnamefont {S.~E.}\ \bibnamefont {Economou}},\ }\href {\doibase 10.1103/PhysRevX.7.041023} {\bibfield  {journal} {\bibinfo  {journal} {Phys. Rev. X}\ }\textbf {\bibinfo {volume} {7}},\ \bibinfo {pages} {041023} (\bibinfo {year} {2017})}\BibitemShut {NoStop}%
\bibitem [{\citenamefont {Pirandola}\ \emph {et~al.}(2017)\citenamefont {Pirandola}, \citenamefont {Laurenza}, \citenamefont {Ottaviani},\ and\ \citenamefont {Banchi}}]{Pirandola2017}%
  \BibitemOpen
  \bibfield  {author} {\bibinfo {author} {\bibfnamefont {S.}~\bibnamefont {Pirandola}}, \bibinfo {author} {\bibfnamefont {R.}~\bibnamefont {Laurenza}}, \bibinfo {author} {\bibfnamefont {C.}~\bibnamefont {Ottaviani}}, \ and\ \bibinfo {author} {\bibfnamefont {L.}~\bibnamefont {Banchi}},\ }\href {\doibase 10.1038/ncomms15043} {\bibfield  {journal} {\bibinfo  {journal} {Nature Communications}\ }\textbf {\bibinfo {volume} {8}} (\bibinfo {year} {2017}),\ 10.1038/ncomms15043}\BibitemShut {NoStop}%
\bibitem [{\citenamefont {Varnava}\ \emph {et~al.}(2007)\citenamefont {Varnava}, \citenamefont {Browne},\ and\ \citenamefont {Rudolph}}]{Varnava2007}%
  \BibitemOpen
  \bibfield  {author} {\bibinfo {author} {\bibfnamefont {M.}~\bibnamefont {Varnava}}, \bibinfo {author} {\bibfnamefont {D.~E.}\ \bibnamefont {Browne}}, \ and\ \bibinfo {author} {\bibfnamefont {T.}~\bibnamefont {Rudolph}},\ }\href {\doibase 10.1088/1367-2630/9/6/203} {\bibfield  {journal} {\bibinfo  {journal} {New Journal of Physics}\ }\textbf {\bibinfo {volume} {9}},\ \bibinfo {pages} {203} (\bibinfo {year} {2007})}\BibitemShut {NoStop}%
\bibitem [{\citenamefont {Sinclair}\ \emph {et~al.}(2014)\citenamefont {Sinclair}, \citenamefont {Saglamyurek}, \citenamefont {Mallahzadeh}, \citenamefont {Slater}, \citenamefont {George}, \citenamefont {Ricken}, \citenamefont {Hedges}, \citenamefont {Oblak}, \citenamefont {Simon}, \citenamefont {Sohler},\ and\ \citenamefont {Tittel}}]{Sinclair2014}%
  \BibitemOpen
  \bibfield  {author} {\bibinfo {author} {\bibfnamefont {N.}~\bibnamefont {Sinclair}}, \bibinfo {author} {\bibfnamefont {E.}~\bibnamefont {Saglamyurek}}, \bibinfo {author} {\bibfnamefont {H.}~\bibnamefont {Mallahzadeh}}, \bibinfo {author} {\bibfnamefont {J.~A.}\ \bibnamefont {Slater}}, \bibinfo {author} {\bibfnamefont {M.}~\bibnamefont {George}}, \bibinfo {author} {\bibfnamefont {R.}~\bibnamefont {Ricken}}, \bibinfo {author} {\bibfnamefont {M.~P.}\ \bibnamefont {Hedges}}, \bibinfo {author} {\bibfnamefont {D.}~\bibnamefont {Oblak}}, \bibinfo {author} {\bibfnamefont {C.}~\bibnamefont {Simon}}, \bibinfo {author} {\bibfnamefont {W.}~\bibnamefont {Sohler}}, \ and\ \bibinfo {author} {\bibfnamefont {W.}~\bibnamefont {Tittel}},\ }\href {\doibase 10.1103/PhysRevLett.113.053603} {\bibfield  {journal} {\bibinfo  {journal} {Phys. Rev. Lett.}\ }\textbf {\bibinfo {volume} {113}},\ \bibinfo {pages} {053603} (\bibinfo {year} {2014})}\BibitemShut {NoStop}%
\bibitem [{\citenamefont {Varnava}\ \emph {et~al.}(2006)\citenamefont {Varnava}, \citenamefont {Browne},\ and\ \citenamefont {Rudolph}}]{varnava2006loss}%
  \BibitemOpen
  \bibfield  {author} {\bibinfo {author} {\bibfnamefont {M.}~\bibnamefont {Varnava}}, \bibinfo {author} {\bibfnamefont {D.~E.}\ \bibnamefont {Browne}}, \ and\ \bibinfo {author} {\bibfnamefont {T.}~\bibnamefont {Rudolph}},\ }\href@noop {} {\bibfield  {journal} {\bibinfo  {journal} {Physical review letters}\ }\textbf {\bibinfo {volume} {97}},\ \bibinfo {pages} {120501} (\bibinfo {year} {2006})}\BibitemShut {NoStop}%
\bibitem [{\citenamefont {Takeoka}\ \emph {et~al.}(2014)\citenamefont {Takeoka}, \citenamefont {Guha},\ and\ \citenamefont {Wilde}}]{Takeoka2014}%
  \BibitemOpen
  \bibfield  {author} {\bibinfo {author} {\bibfnamefont {M.}~\bibnamefont {Takeoka}}, \bibinfo {author} {\bibfnamefont {S.}~\bibnamefont {Guha}}, \ and\ \bibinfo {author} {\bibfnamefont {M.~M.}\ \bibnamefont {Wilde}},\ }\href {\doibase 10.1038/ncomms6235} {\bibfield  {journal} {\bibinfo  {journal} {Nature Communications}\ }\textbf {\bibinfo {volume} {5}} (\bibinfo {year} {2014}),\ 10.1038/ncomms6235}\BibitemShut {NoStop}%
\bibitem [{\citenamefont {Wilde}\ \emph {et~al.}(2017)\citenamefont {Wilde}, \citenamefont {Tomamichel},\ and\ \citenamefont {Berta}}]{Wilde2017}%
  \BibitemOpen
  \bibfield  {author} {\bibinfo {author} {\bibfnamefont {M.~M.}\ \bibnamefont {Wilde}}, \bibinfo {author} {\bibfnamefont {M.}~\bibnamefont {Tomamichel}}, \ and\ \bibinfo {author} {\bibfnamefont {M.}~\bibnamefont {Berta}},\ }\href {\doibase 10.1109/TIT.2017.2648825} {\bibfield  {journal} {\bibinfo  {journal} {IEEE Transactions on Information Theory}\ }\textbf {\bibinfo {volume} {63}},\ \bibinfo {pages} {1792} (\bibinfo {year} {2017})}\BibitemShut {NoStop}%
\bibitem [{\citenamefont {Dutt}\ \emph {et~al.}(2007)\citenamefont {Dutt}, \citenamefont {Childress}, \citenamefont {Jiang}, \citenamefont {Togan}, \citenamefont {Maze}, \citenamefont {Jelezko}, \citenamefont {Zibrov}, \citenamefont {Hemmer},\ and\ \citenamefont {Lukin}}]{dutt2007quantum}%
  \BibitemOpen
  \bibfield  {author} {\bibinfo {author} {\bibfnamefont {M.~G.}\ \bibnamefont {Dutt}}, \bibinfo {author} {\bibfnamefont {L.}~\bibnamefont {Childress}}, \bibinfo {author} {\bibfnamefont {L.}~\bibnamefont {Jiang}}, \bibinfo {author} {\bibfnamefont {E.}~\bibnamefont {Togan}}, \bibinfo {author} {\bibfnamefont {J.}~\bibnamefont {Maze}}, \bibinfo {author} {\bibfnamefont {F.}~\bibnamefont {Jelezko}}, \bibinfo {author} {\bibfnamefont {A.}~\bibnamefont {Zibrov}}, \bibinfo {author} {\bibfnamefont {P.}~\bibnamefont {Hemmer}}, \ and\ \bibinfo {author} {\bibfnamefont {M.}~\bibnamefont {Lukin}},\ }\href@noop {} {\bibfield  {journal} {\bibinfo  {journal} {Science}\ }\textbf {\bibinfo {volume} {316}},\ \bibinfo {pages} {1312} (\bibinfo {year} {2007})}\BibitemShut {NoStop}%
\bibitem [{\citenamefont {Pfaff}\ \emph {et~al.}(2013)\citenamefont {Pfaff}, \citenamefont {Taminiau}, \citenamefont {Robledo}, \citenamefont {Bernien}, \citenamefont {Markham}, \citenamefont {Twitchen},\ and\ \citenamefont {Hanson}}]{pfaff2013demonstration}%
  \BibitemOpen
  \bibfield  {author} {\bibinfo {author} {\bibfnamefont {W.}~\bibnamefont {Pfaff}}, \bibinfo {author} {\bibfnamefont {T.~H.}\ \bibnamefont {Taminiau}}, \bibinfo {author} {\bibfnamefont {L.}~\bibnamefont {Robledo}}, \bibinfo {author} {\bibfnamefont {H.}~\bibnamefont {Bernien}}, \bibinfo {author} {\bibfnamefont {M.}~\bibnamefont {Markham}}, \bibinfo {author} {\bibfnamefont {D.~J.}\ \bibnamefont {Twitchen}}, \ and\ \bibinfo {author} {\bibfnamefont {R.}~\bibnamefont {Hanson}},\ }\href@noop {} {\bibfield  {journal} {\bibinfo  {journal} {Nature Physics}\ }\textbf {\bibinfo {volume} {9}},\ \bibinfo {pages} {29} (\bibinfo {year} {2013})}\BibitemShut {NoStop}%
\bibitem [{\citenamefont {Press}\ \emph {et~al.}(2008)\citenamefont {Press}, \citenamefont {Ladd}, \citenamefont {Zhang},\ and\ \citenamefont {Yamamoto}}]{press2008complete}%
  \BibitemOpen
  \bibfield  {author} {\bibinfo {author} {\bibfnamefont {D.}~\bibnamefont {Press}}, \bibinfo {author} {\bibfnamefont {T.~D.}\ \bibnamefont {Ladd}}, \bibinfo {author} {\bibfnamefont {B.}~\bibnamefont {Zhang}}, \ and\ \bibinfo {author} {\bibfnamefont {Y.}~\bibnamefont {Yamamoto}},\ }\href@noop {} {\bibfield  {journal} {\bibinfo  {journal} {Nature}\ }\textbf {\bibinfo {volume} {456}},\ \bibinfo {pages} {218} (\bibinfo {year} {2008})}\BibitemShut {NoStop}%
\bibitem [{\citenamefont {Thomas}\ \emph {et~al.}(2022)\citenamefont {Thomas}, \citenamefont {Ruscio}, \citenamefont {Morin},\ and\ \citenamefont {Rempe}}]{thomas2022efficient}%
  \BibitemOpen
  \bibfield  {author} {\bibinfo {author} {\bibfnamefont {P.}~\bibnamefont {Thomas}}, \bibinfo {author} {\bibfnamefont {L.}~\bibnamefont {Ruscio}}, \bibinfo {author} {\bibfnamefont {O.}~\bibnamefont {Morin}}, \ and\ \bibinfo {author} {\bibfnamefont {G.}~\bibnamefont {Rempe}},\ }\href@noop {} {\bibfield  {journal} {\bibinfo  {journal} {Nature}\ }\textbf {\bibinfo {volume} {608}},\ \bibinfo {pages} {677} (\bibinfo {year} {2022})}\BibitemShut {NoStop}%
\bibitem [{\citenamefont {De~Greve}\ \emph {et~al.}(2011)\citenamefont {De~Greve}, \citenamefont {McMahon}, \citenamefont {Press}, \citenamefont {Ladd}, \citenamefont {Bisping}, \citenamefont {Schneider}, \citenamefont {Kamp}, \citenamefont {Worschech}, \citenamefont {H{\"o}fling}, \citenamefont {Forchel} \emph {et~al.}}]{de2011ultrafast}%
  \BibitemOpen
  \bibfield  {author} {\bibinfo {author} {\bibfnamefont {K.}~\bibnamefont {De~Greve}}, \bibinfo {author} {\bibfnamefont {P.~L.}\ \bibnamefont {McMahon}}, \bibinfo {author} {\bibfnamefont {D.}~\bibnamefont {Press}}, \bibinfo {author} {\bibfnamefont {T.~D.}\ \bibnamefont {Ladd}}, \bibinfo {author} {\bibfnamefont {D.}~\bibnamefont {Bisping}}, \bibinfo {author} {\bibfnamefont {C.}~\bibnamefont {Schneider}}, \bibinfo {author} {\bibfnamefont {M.}~\bibnamefont {Kamp}}, \bibinfo {author} {\bibfnamefont {L.}~\bibnamefont {Worschech}}, \bibinfo {author} {\bibfnamefont {S.}~\bibnamefont {H{\"o}fling}}, \bibinfo {author} {\bibfnamefont {A.}~\bibnamefont {Forchel}},  \emph {et~al.},\ }\href@noop {} {\bibfield  {journal} {\bibinfo  {journal} {Nature Physics}\ }\textbf {\bibinfo {volume} {7}},\ \bibinfo {pages} {872} (\bibinfo {year} {2011})}\BibitemShut {NoStop}%
\bibitem [{\citenamefont {Nguyen}\ \emph {et~al.}(2019)\citenamefont {Nguyen}, \citenamefont {Sukachev}, \citenamefont {Bhaskar}, \citenamefont {Machielse}, \citenamefont {Levonian}, \citenamefont {Knall}, \citenamefont {Stroganov}, \citenamefont {Riedinger}, \citenamefont {Park}, \citenamefont {Lon{\v{c}}ar} \emph {et~al.}}]{nguyen2019quantum}%
  \BibitemOpen
  \bibfield  {author} {\bibinfo {author} {\bibfnamefont {C.}~\bibnamefont {Nguyen}}, \bibinfo {author} {\bibfnamefont {D.}~\bibnamefont {Sukachev}}, \bibinfo {author} {\bibfnamefont {M.}~\bibnamefont {Bhaskar}}, \bibinfo {author} {\bibfnamefont {B.}~\bibnamefont {Machielse}}, \bibinfo {author} {\bibfnamefont {D.}~\bibnamefont {Levonian}}, \bibinfo {author} {\bibfnamefont {E.}~\bibnamefont {Knall}}, \bibinfo {author} {\bibfnamefont {P.}~\bibnamefont {Stroganov}}, \bibinfo {author} {\bibfnamefont {R.}~\bibnamefont {Riedinger}}, \bibinfo {author} {\bibfnamefont {H.}~\bibnamefont {Park}}, \bibinfo {author} {\bibfnamefont {M.}~\bibnamefont {Lon{\v{c}}ar}},  \emph {et~al.},\ }\href@noop {} {\bibfield  {journal} {\bibinfo  {journal} {Physical review letters}\ }\textbf {\bibinfo {volume} {123}},\ \bibinfo {pages} {183602} (\bibinfo {year} {2019})}\BibitemShut {NoStop}%
\bibitem [{\citenamefont {Rong}\ \emph {et~al.}(2015)\citenamefont {Rong}, \citenamefont {Geng}, \citenamefont {Shi}, \citenamefont {Liu}, \citenamefont {Xu}, \citenamefont {Ma}, \citenamefont {Kong}, \citenamefont {Jiang}, \citenamefont {Wu},\ and\ \citenamefont {Du}}]{rong2015experimental}%
  \BibitemOpen
  \bibfield  {author} {\bibinfo {author} {\bibfnamefont {X.}~\bibnamefont {Rong}}, \bibinfo {author} {\bibfnamefont {J.}~\bibnamefont {Geng}}, \bibinfo {author} {\bibfnamefont {F.}~\bibnamefont {Shi}}, \bibinfo {author} {\bibfnamefont {Y.}~\bibnamefont {Liu}}, \bibinfo {author} {\bibfnamefont {K.}~\bibnamefont {Xu}}, \bibinfo {author} {\bibfnamefont {W.}~\bibnamefont {Ma}}, \bibinfo {author} {\bibfnamefont {F.}~\bibnamefont {Kong}}, \bibinfo {author} {\bibfnamefont {Z.}~\bibnamefont {Jiang}}, \bibinfo {author} {\bibfnamefont {Y.}~\bibnamefont {Wu}}, \ and\ \bibinfo {author} {\bibfnamefont {J.}~\bibnamefont {Du}},\ }\href@noop {} {\bibfield  {journal} {\bibinfo  {journal} {Nature communications}\ }\textbf {\bibinfo {volume} {6}},\ \bibinfo {pages} {8748} (\bibinfo {year} {2015})}\BibitemShut {NoStop}%
\bibitem [{\citenamefont {Ga{\"e}tan}\ \emph {et~al.}(2009)\citenamefont {Ga{\"e}tan}, \citenamefont {Miroshnychenko}, \citenamefont {Wilk}, \citenamefont {Chotia}, \citenamefont {Viteau}, \citenamefont {Comparat}, \citenamefont {Pillet}, \citenamefont {Browaeys},\ and\ \citenamefont {Grangier}}]{gaetan2009observation}%
  \BibitemOpen
  \bibfield  {author} {\bibinfo {author} {\bibfnamefont {A.}~\bibnamefont {Ga{\"e}tan}}, \bibinfo {author} {\bibfnamefont {Y.}~\bibnamefont {Miroshnychenko}}, \bibinfo {author} {\bibfnamefont {T.}~\bibnamefont {Wilk}}, \bibinfo {author} {\bibfnamefont {A.}~\bibnamefont {Chotia}}, \bibinfo {author} {\bibfnamefont {M.}~\bibnamefont {Viteau}}, \bibinfo {author} {\bibfnamefont {D.}~\bibnamefont {Comparat}}, \bibinfo {author} {\bibfnamefont {P.}~\bibnamefont {Pillet}}, \bibinfo {author} {\bibfnamefont {A.}~\bibnamefont {Browaeys}}, \ and\ \bibinfo {author} {\bibfnamefont {P.}~\bibnamefont {Grangier}},\ }\href@noop {} {\bibfield  {journal} {\bibinfo  {journal} {Nature Physics}\ }\textbf {\bibinfo {volume} {5}},\ \bibinfo {pages} {115} (\bibinfo {year} {2009})}\BibitemShut {NoStop}%
\bibitem [{\citenamefont {Wilk}\ \emph {et~al.}(2010)\citenamefont {Wilk}, \citenamefont {Ga{\"e}tan}, \citenamefont {Evellin}, \citenamefont {Wolters}, \citenamefont {Miroshnychenko}, \citenamefont {Grangier},\ and\ \citenamefont {Browaeys}}]{wilk2010entanglement}%
  \BibitemOpen
  \bibfield  {author} {\bibinfo {author} {\bibfnamefont {T.}~\bibnamefont {Wilk}}, \bibinfo {author} {\bibfnamefont {A.}~\bibnamefont {Ga{\"e}tan}}, \bibinfo {author} {\bibfnamefont {C.}~\bibnamefont {Evellin}}, \bibinfo {author} {\bibfnamefont {J.}~\bibnamefont {Wolters}}, \bibinfo {author} {\bibfnamefont {Y.}~\bibnamefont {Miroshnychenko}}, \bibinfo {author} {\bibfnamefont {P.}~\bibnamefont {Grangier}}, \ and\ \bibinfo {author} {\bibfnamefont {A.}~\bibnamefont {Browaeys}},\ }\href@noop {} {\bibfield  {journal} {\bibinfo  {journal} {Physical review letters}\ }\textbf {\bibinfo {volume} {104}},\ \bibinfo {pages} {010502} (\bibinfo {year} {2010})}\BibitemShut {NoStop}%
\bibitem [{\citenamefont {Lindner}\ and\ \citenamefont {Rudolph}(2009{\natexlab{b}})}]{Rudolph2009}%
  \BibitemOpen
  \bibfield  {author} {\bibinfo {author} {\bibfnamefont {N.~H.}\ \bibnamefont {Lindner}}\ and\ \bibinfo {author} {\bibfnamefont {T.}~\bibnamefont {Rudolph}},\ }\href {\doibase 10.1103/PhysRevLett.103.113602} {\bibfield  {journal} {\bibinfo  {journal} {Phys. Rev. Lett.}\ }\textbf {\bibinfo {volume} {103}},\ \bibinfo {pages} {113602} (\bibinfo {year} {2009}{\natexlab{b}})}\BibitemShut {NoStop}%
\bibitem [{\citenamefont {Besse}\ \emph {et~al.}(2020)\citenamefont {Besse}, \citenamefont {Reuer}, \citenamefont {Collodo}, \citenamefont {Wulff}, \citenamefont {Wernli}, \citenamefont {Copetudo}, \citenamefont {Malz}, \citenamefont {Magnard}, \citenamefont {Akin}, \citenamefont {Gabureac}, \citenamefont {Norris}, \citenamefont {Cirac}, \citenamefont {Wallraff},\ and\ \citenamefont {Eichler}}]{Besse2020}%
  \BibitemOpen
  \bibfield  {author} {\bibinfo {author} {\bibfnamefont {J.-C.}\ \bibnamefont {Besse}}, \bibinfo {author} {\bibfnamefont {K.}~\bibnamefont {Reuer}}, \bibinfo {author} {\bibfnamefont {M.~C.}\ \bibnamefont {Collodo}}, \bibinfo {author} {\bibfnamefont {A.}~\bibnamefont {Wulff}}, \bibinfo {author} {\bibfnamefont {L.}~\bibnamefont {Wernli}}, \bibinfo {author} {\bibfnamefont {A.}~\bibnamefont {Copetudo}}, \bibinfo {author} {\bibfnamefont {D.}~\bibnamefont {Malz}}, \bibinfo {author} {\bibfnamefont {P.}~\bibnamefont {Magnard}}, \bibinfo {author} {\bibfnamefont {A.}~\bibnamefont {Akin}}, \bibinfo {author} {\bibfnamefont {M.}~\bibnamefont {Gabureac}}, \bibinfo {author} {\bibfnamefont {G.~J.}\ \bibnamefont {Norris}}, \bibinfo {author} {\bibfnamefont {J.~I.}\ \bibnamefont {Cirac}}, \bibinfo {author} {\bibfnamefont {A.}~\bibnamefont {Wallraff}}, \ and\ \bibinfo {author} {\bibfnamefont {C.}~\bibnamefont {Eichler}},\ }\href {\doibase 10.1038/s41467-020-18635-x} {\bibfield  {journal} {\bibinfo  {journal} {Nature
  Communications}\ }\textbf {\bibinfo {volume} {11}} (\bibinfo {year} {2020}),\ 10.1038/s41467-020-18635-x}\BibitemShut {NoStop}%
\bibitem [{\citenamefont {Schwartz}\ \emph {et~al.}(2016)\citenamefont {Schwartz}, \citenamefont {Cogan}, \citenamefont {Schmidgall}, \citenamefont {Don}, \citenamefont {Gantz}, \citenamefont {Kenneth}, \citenamefont {Lindner},\ and\ \citenamefont {Gershoni}}]{Schwartz2016}%
  \BibitemOpen
  \bibfield  {author} {\bibinfo {author} {\bibfnamefont {I.}~\bibnamefont {Schwartz}}, \bibinfo {author} {\bibfnamefont {D.}~\bibnamefont {Cogan}}, \bibinfo {author} {\bibfnamefont {E.~R.}\ \bibnamefont {Schmidgall}}, \bibinfo {author} {\bibfnamefont {Y.}~\bibnamefont {Don}}, \bibinfo {author} {\bibfnamefont {L.}~\bibnamefont {Gantz}}, \bibinfo {author} {\bibfnamefont {O.}~\bibnamefont {Kenneth}}, \bibinfo {author} {\bibfnamefont {N.~H.}\ \bibnamefont {Lindner}}, \ and\ \bibinfo {author} {\bibfnamefont {D.}~\bibnamefont {Gershoni}},\ }\href {\doibase 10.1126/science.aah4758} {\bibfield  {journal} {\bibinfo  {journal} {Science}\ }\textbf {\bibinfo {volume} {354}},\ \bibinfo {pages} {434} (\bibinfo {year} {2016})}\BibitemShut {NoStop}%
\bibitem [{\citenamefont {Economou}\ \emph {et~al.}(2010)\citenamefont {Economou}, \citenamefont {Lindner},\ and\ \citenamefont {Rudolph}}]{Economou2010}%
  \BibitemOpen
  \bibfield  {author} {\bibinfo {author} {\bibfnamefont {S.~E.}\ \bibnamefont {Economou}}, \bibinfo {author} {\bibfnamefont {N.}~\bibnamefont {Lindner}}, \ and\ \bibinfo {author} {\bibfnamefont {T.}~\bibnamefont {Rudolph}},\ }\href {\doibase 10.1103/PhysRevLett.105.093601} {\bibfield  {journal} {\bibinfo  {journal} {Phys. Rev. Lett.}\ }\textbf {\bibinfo {volume} {105}},\ \bibinfo {pages} {093601} (\bibinfo {year} {2010})}\BibitemShut {NoStop}%
\bibitem [{\citenamefont {Borregaard}\ \emph {et~al.}(2020)\citenamefont {Borregaard}, \citenamefont {Pichler}, \citenamefont {Schr\"oder}, \citenamefont {Lukin}, \citenamefont {Lodahl},\ and\ \citenamefont {S\o{}rensen}}]{Borregaard2020}%
  \BibitemOpen
  \bibfield  {author} {\bibinfo {author} {\bibfnamefont {J.}~\bibnamefont {Borregaard}}, \bibinfo {author} {\bibfnamefont {H.}~\bibnamefont {Pichler}}, \bibinfo {author} {\bibfnamefont {T.}~\bibnamefont {Schr\"oder}}, \bibinfo {author} {\bibfnamefont {M.~D.}\ \bibnamefont {Lukin}}, \bibinfo {author} {\bibfnamefont {P.}~\bibnamefont {Lodahl}}, \ and\ \bibinfo {author} {\bibfnamefont {A.~S.}\ \bibnamefont {S\o{}rensen}},\ }\href {\doibase 10.1103/PhysRevX.10.021071} {\bibfield  {journal} {\bibinfo  {journal} {Phys. Rev. X}\ }\textbf {\bibinfo {volume} {10}},\ \bibinfo {pages} {021071} (\bibinfo {year} {2020})}\BibitemShut {NoStop}%
\bibitem [{\citenamefont {Zhan}\ and\ \citenamefont {Sun}(2020)}]{Yuan2020}%
  \BibitemOpen
  \bibfield  {author} {\bibinfo {author} {\bibfnamefont {Y.}~\bibnamefont {Zhan}}\ and\ \bibinfo {author} {\bibfnamefont {S.}~\bibnamefont {Sun}},\ }\href {\doibase 10.1103/PhysRevLett.125.223601} {\bibfield  {journal} {\bibinfo  {journal} {Phys. Rev. Lett.}\ }\textbf {\bibinfo {volume} {125}},\ \bibinfo {pages} {223601} (\bibinfo {year} {2020})}\BibitemShut {NoStop}%
\bibitem [{\citenamefont {Russo}\ \emph {et~al.}(2018)\citenamefont {Russo}, \citenamefont {Barnes},\ and\ \citenamefont {Economou}}]{Russo2018}%
  \BibitemOpen
  \bibfield  {author} {\bibinfo {author} {\bibfnamefont {A.}~\bibnamefont {Russo}}, \bibinfo {author} {\bibfnamefont {E.}~\bibnamefont {Barnes}}, \ and\ \bibinfo {author} {\bibfnamefont {S.~E.}\ \bibnamefont {Economou}},\ }\href {\doibase 10.1103/PhysRevB.98.085303} {\bibfield  {journal} {\bibinfo  {journal} {Phys. Rev. B}\ }\textbf {\bibinfo {volume} {98}},\ \bibinfo {pages} {085303} (\bibinfo {year} {2018})}\BibitemShut {NoStop}%
\bibitem [{\citenamefont {Gimeno-Segovia}\ \emph {et~al.}(2019)\citenamefont {Gimeno-Segovia}, \citenamefont {Rudolph},\ and\ \citenamefont {Economou}}]{Economou2019}%
  \BibitemOpen
  \bibfield  {author} {\bibinfo {author} {\bibfnamefont {M.}~\bibnamefont {Gimeno-Segovia}}, \bibinfo {author} {\bibfnamefont {T.}~\bibnamefont {Rudolph}}, \ and\ \bibinfo {author} {\bibfnamefont {S.~E.}\ \bibnamefont {Economou}},\ }\href {\doibase 10.1103/PhysRevLett.123.070501} {\bibfield  {journal} {\bibinfo  {journal} {Phys. Rev. Lett.}\ }\textbf {\bibinfo {volume} {123}},\ \bibinfo {pages} {070501} (\bibinfo {year} {2019})}\BibitemShut {NoStop}%
\bibitem [{\citenamefont {Pichler}\ \emph {et~al.}(2017)\citenamefont {Pichler}, \citenamefont {Choi}, \citenamefont {Zoller},\ and\ \citenamefont {Lukin}}]{Pichler2017}%
  \BibitemOpen
  \bibfield  {author} {\bibinfo {author} {\bibfnamefont {H.}~\bibnamefont {Pichler}}, \bibinfo {author} {\bibfnamefont {S.}~\bibnamefont {Choi}}, \bibinfo {author} {\bibfnamefont {P.}~\bibnamefont {Zoller}}, \ and\ \bibinfo {author} {\bibfnamefont {M.~D.}\ \bibnamefont {Lukin}},\ }\href {\doibase 10.1073/pnas.1711003114} {\bibfield  {journal} {\bibinfo  {journal} {Proceedings of the National Academy of Sciences}\ }\textbf {\bibinfo {volume} {114}},\ \bibinfo {pages} {11362} (\bibinfo {year} {2017})}\BibitemShut {NoStop}%
\bibitem [{\citenamefont {Hilaire}\ \emph {et~al.}(2021)\citenamefont {Hilaire}, \citenamefont {Barnes},\ and\ \citenamefont {Economou}}]{Hilaire2021}%
  \BibitemOpen
  \bibfield  {author} {\bibinfo {author} {\bibfnamefont {P.}~\bibnamefont {Hilaire}}, \bibinfo {author} {\bibfnamefont {E.}~\bibnamefont {Barnes}}, \ and\ \bibinfo {author} {\bibfnamefont {S.~E.}\ \bibnamefont {Economou}},\ }\href {\doibase 10.22331/q-2021-02-15-397} {\bibfield  {journal} {\bibinfo  {journal} {Quantum}\ }\textbf {\bibinfo {volume} {5}},\ \bibinfo {pages} {397} (\bibinfo {year} {2021})}\BibitemShut {NoStop}%
\bibitem [{\citenamefont {Ewert}\ and\ \citenamefont {van Loock}(2014)}]{ewert20143}%
  \BibitemOpen
  \bibfield  {author} {\bibinfo {author} {\bibfnamefont {F.}~\bibnamefont {Ewert}}\ and\ \bibinfo {author} {\bibfnamefont {P.}~\bibnamefont {van Loock}},\ }\href@noop {} {\bibfield  {journal} {\bibinfo  {journal} {Physical review letters}\ }\textbf {\bibinfo {volume} {113}},\ \bibinfo {pages} {140403} (\bibinfo {year} {2014})}\BibitemShut {NoStop}%
\bibitem [{\citenamefont {Gimeno-Segovia}\ \emph {et~al.}(2017)\citenamefont {Gimeno-Segovia}, \citenamefont {Cable}, \citenamefont {Mendoza}, \citenamefont {Shadbolt}, \citenamefont {Silverstone}, \citenamefont {Carolan}, \citenamefont {Thompson}, \citenamefont {O~Brien},\ and\ \citenamefont {Rudolph}}]{Gimeno-Segovia2017}%
  \BibitemOpen
  \bibfield  {author} {\bibinfo {author} {\bibfnamefont {M.}~\bibnamefont {Gimeno-Segovia}}, \bibinfo {author} {\bibfnamefont {H.}~\bibnamefont {Cable}}, \bibinfo {author} {\bibfnamefont {G.~J.}\ \bibnamefont {Mendoza}}, \bibinfo {author} {\bibfnamefont {P.}~\bibnamefont {Shadbolt}}, \bibinfo {author} {\bibfnamefont {J.~W.}\ \bibnamefont {Silverstone}}, \bibinfo {author} {\bibfnamefont {J.}~\bibnamefont {Carolan}}, \bibinfo {author} {\bibfnamefont {M.~G.}\ \bibnamefont {Thompson}}, \bibinfo {author} {\bibfnamefont {J.~L.}\ \bibnamefont {O~Brien}}, \ and\ \bibinfo {author} {\bibfnamefont {T.}~\bibnamefont {Rudolph}},\ }\href@noop {} {\bibfield  {journal} {\bibinfo  {journal} {New J. Phys.}\ }\textbf {\bibinfo {volume} {19}},\ \bibinfo {pages} {063013} (\bibinfo {year} {2017})}\BibitemShut {NoStop}%
\end{thebibliography}%

\end{document}